\documentclass[11pt, letterpaper]{book}
\pdfoutput=1

\usepackage[utf8]{inputenc}
\usepackage{appendix}
\usepackage{pst-all}
\usepackage{pstcol}
\usepackage[sc]{mathpazo} 

\usepackage[format=hang,indention=-1.0cm,margin=10pt]{caption}
\usepackage{graphicx}
\usepackage{lettrine}
\usepackage{amsfonts}
\usepackage{amsmath}
\usepackage{amssymb}
\usepackage{bbm}
\usepackage{fancyhdr}                    
\usepackage[hmarginratio=1:1,textwidth=12.5cm,totalheight=20.5cm,includehead]{geometry}
\usepackage{subfigure}
\usepackage{setspace}

\makeatletter
\def\cleardoublepage{\clearpage\if@twoside \ifodd\c@page\else%
    \hbox{}%
    \thispagestyle{empty}
    \newpage%
    \if@twocolumn\hbox{}\newpage\fi\fi\fi}
\makeatother

\renewenvironment{thebibliography}[1]{\begin{oldthebibliography}{#1}%
    \setlength{\parskip}{0cm}\setlength{\itemsep}{.2cm}}%
{\end{oldthebibliography}}

\providecommand{\openone}{ {\leavevmode\hbox{\small1\kern-3.8pt\normalsize1}}}

\def\>{\rangle}
\def\<{\langle}

\def\d{\text{d}}

\newcommand{\etal}{ {\it et al.} }
\newcommand{\Bell}{{\ensuremath{|\mathrm{Bell}\>}}}
\newcommand{\mcO}{\mathcal{O}}
\newcommand{\mcU}{\mathcal{U}}
\newcommand{\mcH}{\mathcal{H}}
\newcommand{\rmH}{\mathrm{H}}
\newcommand{\rmqm}{\mathrm{qm}}
\newcommand{\rmd}{\mathrm{d}}
\newcommand{\rmc}{\mathrm{c}}

\newcommand{\rmRMT}{\mathrm{RMT}}
\newcommand{\rms}{\mathrm{s}}
\newcommand{\rmZ}{\mathrm{Z}}
\newcommand{\rme}{\mathrm{e}} \newcommand{\e}{\rme}
\newcommand{\rmi}{\mathrm{i}} \newcommand{\mimath}{\rmi}
\newcommand{\tr}{\mathop{\mathrm{tr}}\nolimits}
\newcommand{\Or}{\mathord{\mathrm{O}}} 

\newcommand{\eref}[1]{eq.~(\ref{#1})}
\newcommand{\sref}[1]{sec.~\ref{#1}}
\newcommand{\fref}[1]{fig.~\ref{#1}}

\newcommand{\Eref}[1]{Eq.~(\ref{#1})}

\newcommand{\Fref}[1]{Fig.~\ref{#1}}

\newcommand{\bm}{\boldsymbol}

\newcommand{\into}{\int_0^t \rmd \tau \int_0^t \rmd \tau'}
\newcommand{\intoh}{\int_0^t \rmd \tau \int_0^\tau \rmd \tau'}

\newhsbcolor{huezero}{0 1 1}
\newhsbcolor{huepone}{0.1 1 1}
\newhsbcolor{huepfour}{0.4 1 1}
\newhsbcolor{huepsix}{0.6 1 1}
\newhsbcolor{huepseven}{0.7 1 1}
\newhsbcolor{hueoneosix}{0.1666 1 1}
\newhsbcolor{hueoneothree}{0.3333 1 1}
\newhsbcolor{hueoneotwo}{0.5 1 1}
\newhsbcolor{huetwoothree}{0.6666 1 1}
\newhsbcolor{huefiveosix}{0.8333 1 1}

\newcommand{\ie}{{\it i.e.} }
\newcommand{\Ie}{{\it I.e.} }
\newcommand{\eg}{e.~g. }
\graphicspath{{eps/}{RMTconcurrence_eps/}}
\begin{document}

\begin{titlepage}
\begin{center}
\includegraphics[width=.3\textwidth]{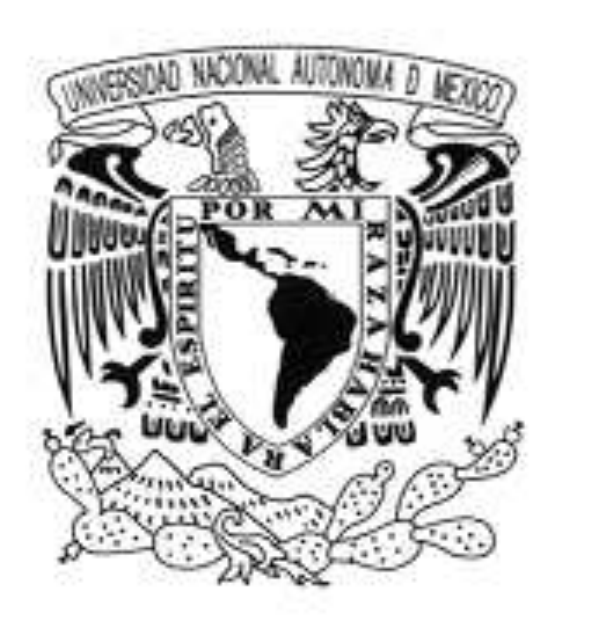}\hfill\includegraphics[width=.3\textwidth]{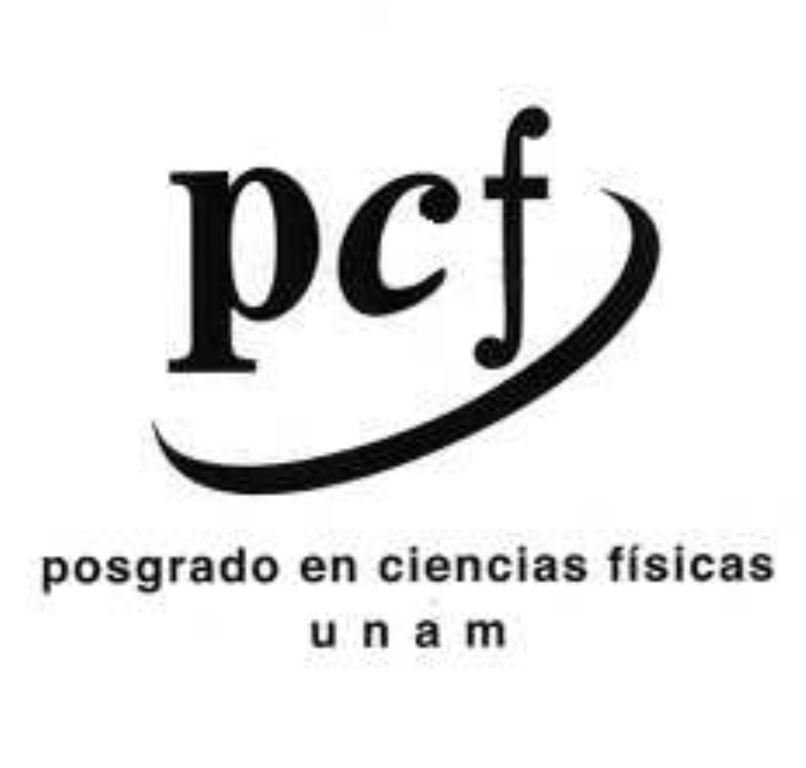}\\
\large UNIVERSIDAD NACIONAL AUT\'ONOMA DE M\'EXICO\\
POSGRADO EN CIENCIAS F\'ISICAS\\[1cm]
\huge One, Two, and $n$ Qubit Decoherence\\[.5cm]\large TESIS\\[1.2cm]

\begin{tabular}{rl}
 Que para obtener el grado de: & Doctor en Ciencias (F\'isica)\\
 Presenta: & Carlos Francisco Pineda Zorrilla\\
 Directores de tesis:& Dr.~Toma\v{z} Prosen \\
 & Dr. Thomas H. Seligman \\
 \end{tabular}

\vspace*{\stretch{1}}
{\small Miembros del Comit\'e Tutoral: \\
Dr. Jorge Flores, Dr. Toma\v{z} Prosen,  y Dr. Thomas H. Seligman}\\
\includegraphics[width=.33\textwidth]{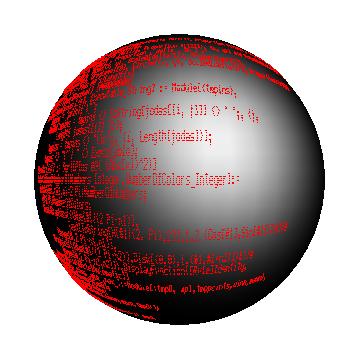}\includegraphics[width=.33\textwidth]{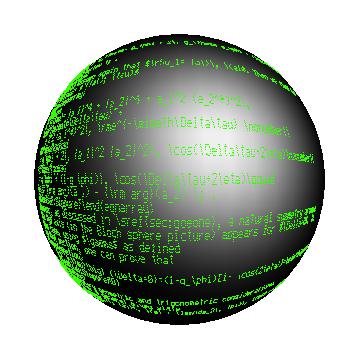}\includegraphics[width=.33\textwidth]{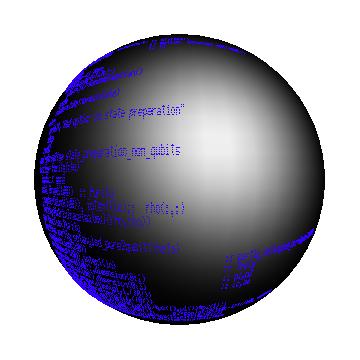}
M\'exico D. F., 2007
\end{center}
\cleardoublepage
\pagestyle{empty}
\begin{center}
\includegraphics[width=.3\textwidth]{logo_unam}\hfill\includegraphics[width=.3\textwidth]{logo_pcf}\\
\large UNIVERSIDAD NACIONAL AUT\'ONOMA DE M\'EXICO\\
POSGRADO EN CIENCIAS F\'ISICAS\\[1cm]
\huge One, Two, and $n$ Qubit Decoherence\\[.5cm]\large THESIS\\[1.2cm]

\begin{tabular}{rl}
 To obtain the degree: & Doctor en Ciencias (F\'isica)\\
 Presents: & Carlos Francisco Pineda Zorrilla\\
 Directors:& Dr. Toma\v{z} Prosen \\
 & Dr. Thomas H. Seligman \\
 \end{tabular}

\vspace*{\stretch{1}}
{\small Members of the Tutorial Committee: \\
Dr. Jorge Flores, Dr. Toma\v{z} Prosen,  and Dr. Thomas H. Seligman}\\
\includegraphics[width=.33\textwidth]{Bola1.jpg}\includegraphics[width=.33\textwidth]{Bola2.jpg}\includegraphics[width=.33\textwidth]{Bola3.jpg}
M\'exico D. F., 2007
\end{center}
\end{titlepage}

\cleardoublepage

\setcounter{page}{1}\pagenumbering{roman} \setcounter{page}{1}
\vspace*{\stretch{1}}
\begin{flushright}
{\em !`No contaban con mi astucia!}\\El Chapul\'in Colorado
\end{flushright}
\vspace*{\stretch{1}}
\thispagestyle{empty}
\cleardoublepage

{ \setlength\parindent{0cm} \setlength\parskip{0.1cm}

\begin{center} \Huge Gracias \end{center}
A Thomas por su paciencia.  A Toma\v z y Fran\c{c}ois por su
invaluable ayuda para crecer profesionalmente y también por brindarme
su amistad.  Carolina Spinel, Pier M., Thomas G., Chrisomalis,
Rolando C., Carlos B.,  Rocío J., Ivette F. y Manuel T.
influyeron (positivamente) en mi desarrollo
profesional durante mi doctorado.
  
Desde un punto de vista personal debo agradecer a demasiadas
personas.  A mis padres, Thomas Seligman (de nuevo), Edna (por darme
todo su amor), mis hermanas Maca y Nena, Emi y a toda
la familia paterna y materna.  

A los amigos atemporales: Alf, Jose, Betty, Camilo,
Mi Perro, Ignacio F., Ceci L.M. y Sapoiguanacaimán (seguro alguien se me olvido
acá).

En México agradezco a mis primeros amigos: Horacio, Eduardo, Félix,
Fico, Emilio, Yanalté, Luz Alliete, Rebescua y Mírimix.  A los compas del
Gato Macho en especial a Giovanni, Holbert y John Jairo.  A Jose
Nicolás, MariaC, Yari (más Javo, Aleja, SICME y Julián) y  {\tt
rv\_k292}.  A Elisa, Angie y Fabiola.  A Angelina y Guillermina. A Caro
Af., {\tt dagato}, Evelin, Karen.

A varios amigos de la uni, en particular a Elías, Enrique, Belinka,
Olivia U., {\tt gay\_thama},  Verónica, Reyes, Blas y Carlos Natorro.  Al
grupo de investigación:   Luis, Mau, Rakel, David, Marc, Olivier,
Sergey, Gursoy, Pablo, Choker, Rafael M., Claudia, Steffan, Luqi y
Jorge Flores.  A la gente de los eventos en Cuernavaca, DF, Dresden,
París, Les Houches y Buzios.  A los que me han brindado hospitalidad
en Los Alamos, Ljubljana, Freiburgh, Brecia, New York, Innsbruck y Bratislava.

A algunos amigos en Bogotá: Ana María, Vieja tal, Maryory, Cayita,
Marta Gu, Katherine, Chocho, Maryory, Ernestina, Katherine, Carlos F.
M., y Olga Lu. En París: Veronique, Luisa S., Caroline y Julian.  En
Taxco, Luisa F. En Melgar: Melguistas. En Ljubljana: Sneza, Amir,
Osi, Marko (+ all the group) and Nade.  En BsAs: Nacho, Ceci(s) y 
Angeles.  En Montevideo: Turca y Capelanes.  En otras
partes: Rashnia, Rudi, Hartmut, Orus, Mario Z., V. Buzek,  Sole (y
Emi), Luca Bastardo, Juan Diego U. y Walter S. 

A los amigos virtuales, como {\tt maracacol}, {\tt juanita}, {\tt
lawawis}, {\tt terepoeta}, {\tt cecibl89}, {\tt 987aw0s89duf}, {\tt
marcoandue}, {\tt mzd\_78}, Audrey, {\tt terepoeta}.  A los nuevos
amigos Camilo Cardona, Sonia, Aurora, Sayab, Tzolkin, Christian,
Fernando y Yenni.

Finalmente, gracias a la música por acompañarme en momentos de alegría
soledad, tristeza, rumba y trabajo.
}
\vfill
\vfill
\thispagestyle{empty}
\cleardoublepage
\begin{center} {\bf SYNOPSIS}\\ \end{center}

\vfill
{ \setlength\parindent{0cm} \setlength\parskip{0.1cm}

We study decoherence of one, two, and $n$ non-interacting qubits.
Decoherence, measured in terms of purity, is calculated in linear
response approximation, making use of the spectator configuration.
Monte Carlo simulations illustrate the validity of this approximation
and of its extension by exponentiation.  Initially, the environment
and its interaction with the qubits are modelled by random matrices.
Purity decay of entangled and product states are qualitatively
similar though for the latter case it is slower.  

For two qubits, numerical studies reveal a one to one correspondence between
its decoherence and its internal entanglement decay.  For strong and
intermediate coupling to the environment this correspondence agrees with the
one for Werner states, for initial Bell pairs.  Using this relation we are
able to give a formula for concurrence decay. In the limit of a large
environment the evolution induces a unital channel in the two qubits,
providing a partial explanation for the relation above.

Using a kicked Ising spin network, we study  the exact evolution  of
two non-interacting  qubits in  the presence of a spin bath. Dynamics
of  this model range from integrable to chaotic and we can handle
numerics for a large number of qubits.  We find that the entanglement
(as measured by concurrence) of the two qubits has a close relation
to the purity of the pair, and closely follows an analytic relation
derived for Werner states.  As a collateral result we find that an
integrable environment causes quadratic decay of concurrence as well
as of purity, while a chaotic environment   causes linear decay.
Both quantities   display recurrences   in an integrable environment.
Good agreement with the results found using random matrix theory is
obtained. 

Finally, we analyze decoherence of a quantum register in the absence
of non-local operations  \ie $n$ non-interacting qubits coupled to an
environment.  The problem is solved in terms of a sum rule which
implies linear scaling in the number of qubits.  Each term involves a
single qubit and its entanglement with the remaining ones.  Two
conditions are essential: first decoherence must be small and second
the coupling of different qubits must be uncorrelated in the
interaction picture.  We apply the result to the random matrix model,
and illustrate its reach considering a GHZ state coupled to a spin
bath.

}
\thispagestyle{empty}
\cleardoublepage
\begin{center} {\bf RESUMEN}\\ \end{center}

\vspace{.5in}


{ \setlength\parindent{0cm} \setlength\parskip{0.2cm}
\newcommand{\ignore}[1]{}
\newcommand{\ignorenot}[1]{#1}

La decoherencia de 1, 2 y $n$ qubits no interactuantes y posiblemente
enlazados, medida en términos de la pureza, es calculada usando respuesta
lineal y el concepto de configuración de espectador. A través de
simulaciones de Monte Carlo exploramos la validez de la aproximación y su
extensión mediante exponenciación. Inicialmente, modelamos la interacción y
el medio ambiente con matrices aleatorias (MA).

Para 2 qubits, en el modelo MA, el enlazamiento interno y la decoherencia
tienen una relación uno a uno. Esta relación, para acoplamientos moderados y
fuertes, coincide con la relación correspondiente para estados de Werner, si
la condición inicial es un par de Bell. Mediante ésta, obtenemos una formula
explicita para el decaimiento del enlazamiento interno.

Introducimos un modelo de Ising pateado (MIP) para estudiar un grupo de
espines, acoplado débilmente a un baño de espines. Este modelo presenta
diná\-micas integrable, mixta y caótica para diferentes parámetros.
Observamos nuevamente la relación entre la decoherencia y el enlazamiento
interno, obtenida con el modelo MA. Inicialmente, tanto el enlazamiento como
la decoherencia decrecen cuadráticamente/linealmente para el caso 
integrable/caótico. En el caso integrable, si el acoplamiento
es a un extremo de una cadena, ambas cantidades presentan comportamientos
periódicos. Comparamos cuantitativamente el decaimiento de pureza en el
modelo MA con nuestro sistema dinámico. Los resultados positivos demuestran
la validez del modelo estocástico.

Finalmente, analizamos una memoria cuántica expuesta a decoherencia. Nuestro
resultado, que asume altas purezas e independencia de los acoplamientos en la
imagen de interacción, permite expresar la decoherencia como una suma de
términos que involucran un solo qubit y su enlazamiento con el resto de la
memoria. Aplicamos el resultado al modelo MA, generalizando nuestros
hallazgos. En el MIP, observamos que incluso en situaciones integrables,
los requerimientos se pueden cumplir.

Números PACS: 03.65.Yz, 03.65.-w, 03.65.Ud, 05.40.-a \\ 
Palabras clave: enlazamiento, enmarañamiento, RMT,
pureza, decoherencia, concurrencia, memoria cuántica, GHZ,
Bell.

} 
\thispagestyle{empty}
\cleardoublepage

\tableofcontents
\setlength\parindent{0cm} \setlength\parskip{0.3cm}
\chapter{Introduction and fundamental tools}
\label{sec:chapintro}
\pagenumbering{arabic} \setcounter{page}{1}



Studying decoherence of one, two, and $n$ qubits has a wide scope of
applications due to the huge interest in implementing ``quantum technology''.
{\it The} limiting factor for building this technology is the
sensitivity of quantum systems to undesired perturbations/coupling.
Moreover, coupling to external degrees of freedom is {\it fundamentally}
inevitable. Understanding its behavior is crucial to tame its effects. 

Since a long time the coupling of quantum systems to external degrees of
freedom has been studied (see e.g. \cite{vN55a, RevModPhys.29.454,
PhysRevD.46.5504, PhysRevD.48.3768}).  Philosophical aspects of quantum
mechanics (the measurement process and the emergence of the classical
world) are deeply connected with the problem. Some particular models
designed for specific applications have been developed, but the favorite
for general purposes (by far) is the Caldera-Leggett model
\cite{caldeiraleggett} in which the external degrees of freedom are
modeled by a set of harmonic oscillators.  Good agreement with the
experiment has been observed, e.g.  \cite{PhysRevLett.81.3523,
PhysRevLett.89.217004, Zazunov:cond-mat0702247}. 

Regarding applications to quantum information some progress has been
made. A big amount of literature exists and some aspects have been
demonstrated experimentally.  Most theoretical studies use
Caldeira-Le\-ggett like models; others explore the consequences of
using/dropping the usual Markovian approximations. Some use very
particular models either to obtain explicit analytical results or to
study specific experimental situations. A more general picture is thus
desirable to gain a deeper understanding of the physics governing
decoherence of quantum information systems.  A few comments on some
relevant papers on the field are useful to have an idea of the situation
in the literature. This list of papers is not meant to be complete, it
is just intended to give a brief overview of what people are currently
working on, in relation to this topic.  
\begin{list}{$\bullet$}%
    {\setlength{\topsep}{0cm}\setlength{\leftmargin}{0.5cm}%
    \setlength{\itemsep}{0cm}  }
\item In a series of papers \cite{yu:193306, PhysRevB.68.165322,
yu:140404} Yu and Eberly explore how two qubits (in particular their
entanglement properties) are affected by broadband bosonic reservoirs
that induce vacuum noise, phase noise, etc. They find that entanglement
after a finite time goes identically to zero. 
\item In \cite{braun:230502} a more complicated study is done using
again a traditional harmonic oscillator bath. There, the off diagonal
elements of the reduced density matrix (sometimes called ``coherences'')
are analyzed.  Interestingly, the author discovers that decoherence is
determined by a generalized Hamming distance. He introduces some
primitive spectator (see \sref{sec:twoqubitmodel}).
\item In \cite{gedik-2006-138} the author uses a spin bath as an
environment.  He studies concurrence of some Bell pairs. The results,
though interesting, are quite model dependent. In \cite{lages:026225}
the relation between integrability and decoherence is studied for a spin
bath environment, very much in the spirit of our results
\cite{pineda:012305}.  Studies of specific spin-bath environments,
aiming to understand decoherence in experimental qubit realizations are
\cite{PhysRevB.68.115322, PhysRevB.67.033301} and
\cite{simmonds:077003}.
\item In another interesting article \cite{garcia-mata:120504},
Garcia-Mata \etal use semi-classical considerations to study
multi-particle qubit entanglement. They  analyze how entanglement is
affected both by the phase space structure and by the kind of noise
applied.  
\item In \cite{Fedichkin2004} the authors propose two measures
for decoherence.  These  measures are additive, a property that is
important to express the total decoherence in terms of the decoherence
of each qubit. 
\item Some efforts to understand the implications of the
Markovian approximation are made in \cite{PhysRevA.66.012302,
lee:024301}, where it is shown that under certain circumstances that
approximation leads to very inaccurate results. 
\item In \cite{andrereview} the authors use random Hamiltonians (though
not with the minimum information properties of the classical ensembles
\cite{balian}) to study entanglement decay of $n$ qubits.  Using
Markovian approximations, they arrive to time independent Linblad
equations. They obtain multi-exponential decay. They are able to analyze
the differences between  W and GHZ states. 
\item An isolated and possibly premature (due to the interests of the
community) study is worth mentioning. In \cite{Mello1988} Mello \etal
analyze the relaxation rates of a single $1/2$ spin particle using a
random matrix model.
\end{list}

We are pioneering the use of Random Matrix Theory in the field of quantum
information theory.  Some previous work has influenced considerably our
research namely \cite{1367-2630-6-1-020, Gorin2003}, where random matrices
were used to analyze decoherence and fidelity in general quantum systems.

Joseph Emerson has developed methods to create random unitary operators (in the
spirit of the CUE) using random gates (ironically a non-trivial task to
implement efficiently) \cite{JosephEmerson12192003}. He has also explored the
utilities of such random unitary operators in fidelity and local density of
states estimation \cite{emerson:050305}. 
Other uses of randomization in quantum protocols,  like diminishing the effects
of static perturbations, have been introduced in
\cite{ShepelianzkyCorrectionRandom, 0305-4470-34-47-103} and further explored
in \cite{kern:062302}.
Some people have also used random
matrices to study fidelity decay \cite{reflosch, ShepelyanskyRMTFidelity}.

This thesis is based mainly on four publications
\cite{pinedaRMTshort, pinedalong, pineda:012305, GPS-letter}. We
shall not follow the chronological order as the logical order will
result in almost opposite.  The reasons are clear. As you gain
insight in the field, things become clearer, concepts become more
elaborate and distilled and thus more suitable for understanding. 

\paragraph{On the structure of the thesis}--

In the remaining part of the introduction we explain some concepts and
tools used along the thesis.  Some material is not new, and is not
appropriate for publication in a journal as it only contains a review of
known things.  However, for the reader a coherent presentation is always
handy.  A main concept used and studied during this thesis is
entanglement. It is understood here both as a resource (to perform quantum
information tasks) and as a cause of decoherence.  We shall define it and
discuss the two ways that we understand it. Next we explain the spectator
configuration, which is an original tool exploited during most of this
work.  Finally we introduce Random Matrix Theory (RMT). 

We next proceed to analyze decoherence of quantum systems.  We first
explore the single qubit case (\sref{sec:onequbit}). The detailed
mathematical derivation of the formulae used is given in appendix
\ref{sec:calculationSpec}.  Both the time reversal invariant (TRI) case
and the non-TRI case are analyzed in detail.  We then consider the two
qubit case (\sref{sec:twoqubit}).  Different configurations
corresponding to different physical situations are analyzed. Again TRI
and non-TRI cases are also studied.  The effect of internal entanglement
proves important and provides interesting effects. In
\sref{sec:concurrence} the relation between decoherence and entanglement
is studied. Sections \ref{sec:onequbit}, \ref{sec:twoqubit}, and
\ref{sec:concurrence} are based on \cite{pinedalong}, though the basic
idea was introduced in \cite{pinedaRMTshort}.

In \sref{sec:ki} we give an example of how some of the concepts can be
applied to a simple model: the kicked Ising spin chain.  Though this
chapter is based on \cite{pineda:012305}, major modifications have been
introduced with the aim of getting closer to the RMT models. 

Finally in \sref{sec:nqubit} we use some of the results obtained during
the thesis to analyze decoherence of an $n$ qubit register. We shall use
both the RMT model and the KI spin chain to discover and understand the
reach of the results.  This chapter is based on \cite{GPS-letter}.

In appendix \ref{sec:calculationSpec} we perform the main RMT calculation,
for a single qubit in the spectator configuration.  In appendix
\ref{sec:implementationKI} we explain how to implement numerically the kicked
Ising model in an efficient manner. The next appendix
(\ref{sec:exponentiation}) explains a way of extending some analytical
results via exponentiation. This heuristic result is tested throughout the
thesis for most Monte Carlo simulations.  We then discuss some technical
aspects of both entanglement (appendix \ref{sec:entappendix}) and random
matrix theory (appendix \ref{sec:RMTtech}). Regarding entanglement, we
discuss the definition of entanglement in more general systems than the ones
discussed in this introduction, and the physical meaning of concurrence. For
random matrix theory we mention the physical justification of the ensembles,
relations among their matrix elements, and other formulae used during the
thesis. In the last appendix we give the double integral of the form factor
for a particular ensemble and a simple proof of the Born expansion for the
echo operator.

I do hope you enjoy and have a nice time with this piece of work.

\section{Entanglement}

Though in mathematical terms entanglement is trivial to define (once
the basic tools of quantum mechanics are introduced), its consequences
challenge many deeply rooted (mis)conceptions about reality.  Here we do
not wish to discuss how the existence of entanglement affects our
understanding of reality; this is a difficult topic outside the scope of
this work, and even Einstein was puzzled by its consequences. We limit
our selves to define and discuss briefly entanglement and how to quantify
it.

A pure state of a quantum composite system is said to be entangled when
it is not the ``sum'' of its parts (technically we mean tensor product).
To be more precise, let our Hilbert space $\mcH$ be composed of two
parts: $\mcH=\mcH_A \otimes \mcH_B$. If, given a state $|\psi\> \in
\mcH$, there exist $|\psi_A\> \in \mcH_A$ and  $|\psi_B\> \in \mcH_B$
such that
\begin{equation} \label{eq:separablecondition}
	|\psi\>=|\psi_A\>\otimes |\psi_B\>
\end{equation}
it is said that $|\psi\>$ is separable or unentangled. Conversely, if
\begin{equation} \label{eq:noseparablecondition}
	|\psi\> \ne |\psi_A\>\otimes |\psi_B\>,\, 
	\forall \left(|\psi_A\> \in \mcH_A,\,|\psi_B\> \in \mcH_B\right)
\end{equation}
it is said that $|\psi\>$ is entangled.  In other words $|\psi\>$ is
entangled if and only if
\begin{equation}
  |\psi\> \<\psi | \ne \tr_A |\psi\> \<\psi | \otimes  \tr_B |\psi\> \<\psi |.
\label{eq:entnotsum}
\end{equation}
It is quite easy to show the existence of entangled states. The simplest
case can be constructed when $\dim \mcH_A=\dim \mcH_B=2$, \ie when
$\mcH_A$ and $\mcH_B$ represent qubits. Let $\{|0\>,|1\>\}$ be an
orthonormal basis in each space. The Bell state
\begin{equation} \label{eq:defBell}
  \Bell = \frac{|0\>\otimes |0\> + |1\>\otimes|1\>}{\sqrt{2}} \in \mcH 
\end{equation}
is entangled. Assuming the existence of $\alpha_A, \beta_A,\alpha_B, \beta_B
\in \mathbb{C}$, such that $\Bell=(\alpha_A |0\> + \beta_A |1\>)
\otimes(\alpha_B|0\> + \beta_B |1\>)$ results in a contradiction.  For
multipartite mixed systems a generalization of the definition of entanglement
is straightforward. See appendix \ref{sec:entappendix} for details. 

In order to get deeper insight in the entanglement properties of pure
bipartite states it is convenient to use the Schmidt decomposition
\cite{Schmidt1907,NC00a}.  Given a state $|\psi\>$ in a bipartite space
$\mcH_A \otimes\mcH_B$, there exist orthonormal states $\{|i_A\>\}$ in
$\mcH_A$ and $\{|i_B\>\}$ in $\mcH_B$ such that
\begin{equation}\label{eq:schmidtGENERAL}
	|\psi\>=\sum_{i=1}^{\min \{\dim \mcH_A,\dim \mcH_B\} }
	     \lambda_i |i_A\> \otimes |i_B\>
\end{equation}
and $0\le \lambda_i \le 1$, with $\sum_i \lambda_i^2=1$. The numbers
$\lambda_i$ are called Schmidt coefficients and play an important roll in
entanglement theory.  Consider an orthonormal (and complete) basis that
diagonalizes $\rho_A= \tr_B |\psi \> \<\psi|$.  We choose that basis to be
$|i_A\>$; its existence is guarantied by the spectral theorem.  We can
then write $|\psi\>=\sum_i |i_A\> \otimes |\tilde i_B\>$, but since
$\rho_A= \sum_i |i_A\> \< j_A|  \<\tilde i_B| \tilde j_B\> $ must be in
fact diagonal, then $\<\tilde i_B| \tilde j_B\> \propto \delta_{ij}$.
Using some $|i_B\> \propto |\tilde i_B\>$ such that $\<i_B|i_B\>=1$ and
suitably choosing its phases we can write \eref{eq:schmidtGENERAL}. The
sharp reader will notice that the Schmidt coefficients are the square
roots of the eigenvalues of the reduced density matrix of any of the two
subsystems.

The Schmidt coefficients are unique for each pure state.  From the
argumentation we can see that $\rho_A$ and
$\rho_B$ have the same eigenvalues (and with the same degeneracy) except
for $|\dim \mcH_A - \dim \mcH_B|$ zeros.  Determining whether a pure
state is entangled or not is an easy task.  From the previous paragraph
one can see that a state is not entangled if and only if one of the
Schmidt coefficients is one (implying that the others are zero).

\subsection{Decoherence as entanglement} 

Decoherence can be seen as entanglement with the environment
\cite{zurekreview, Zur91}.
\begin{quote}
Quantum correlations [in our language, entanglement] can also
disperse information throughout the degrees of freedom that are, in
effect, inaccessible to the observer \cite{Zur91}.
\end{quote}
Though a big debate has been issued since the formulation of that
para\-digm, it is now generally accepted.

We now discuss an example to explain the previous statement.  To
present the key idea it is enough to consider a central system,
composed of a single qubit,  and an environment alone; in the
original formulation a measurement apparatus was also involved to
allow the analysis of the ``collapse'' of the wave function after a
measurement process.  In this example the environment has three
characteristics: large dimension, uncontrollable dynamics, and no
possibility of being observed. We assume some interaction between the
central system and the environment.  Consider an initial state which
is (i) separable with respect to the environment and (ii) a
superposition in the central system.  \Ie
\begin{equation}
	|\psi(t=0)\>=(\alpha |0\> + \beta |1\>)\otimes |\phi\>
\label{eq:initialstatezurel}
\end{equation}
where $|0\>$ and $|1\>$ form an orthonormal basis for the qubit, $\alpha$ and
$\beta$ are complex numbers, and $|\phi\>$ is the initial state of
the environment.  Assume that the interaction depends on the state of the
qubit, \eg a controlled-$U$.  After some time, due to the interaction, the
state will be  
\begin{equation}
	|\psi({\rm big}\, t)\>=\alpha |0\>|\phi_0\> + \beta |1\>|\phi_1\>.
\label{eq:finalstatezurel}
\end{equation}
As the dimension of the environment is big, the states of the environment,
after some time scale, will be approximately orthogonal:
$\<\phi_i|\phi_j\>\approx \delta_{ij}$. Of course this last statement is
not fulfilled for an arbitrary interaction, but precisely the basis
(regarding the qubit) in which this condition is fulfilled will determine
the preferred basis which determines the pointer states. 

To quantify the degree of entanglement of a bipartite system, in a pure
state, we make use of the Schmidt coefficients. Adding  any convex function
of these coefficients is enough. We use the sum of their squares, as it
induces a very simple formula (other common choice, instead of $x^2$,  is $x
\log x$ which induces the von Neumann entropy). This measure we call {\it
purity}. Thus, for a given density matrix $\rho$, its purity is defined by
\begin{equation}
 P(\rho)=\tr \rho^2.
 \label{eq:defpurity}
\end{equation}
This quantity is 1 for pure states ($\rho=|\psi\>\<\psi|$), less than one for
mixed states, and reaches a minimum of $1/N$ (where $N$ is the dimension of
$\rho$) for the completely mixed state $\openone/N$.  If the partial trace
with respect to an environment is represented by $\tr_\rme$, a measure of
decoherence is then $P(\rho=\tr_\rme |\psi\>\<\psi|)$. An important practical
advantage of this measure is that one does not need to evaluate the Schmidt
coefficients of the density matrix $\rho$.

Other views of decoherence are common in the literature. Consider a qubit in
an initial state $|\psi\>=(|0\> + |1\>)/\sqrt{2}$.  Its corresponding density
matrix is
\begin{equation}
\rho  = 
\frac{1}{2}\begin{pmatrix}
1 & 1 \\
1 & 1
\end{pmatrix}
\label{eq:super}
\end{equation}
Assume we have pure dephasing (no amplitude damping). Typically, what will
happen is that the off diagonal elements will decay exponentially. That is,
its time evolution will be
\begin{equation}
\rho(t)  = 
\frac{1}{2}\begin{pmatrix}
1 & \rme^{-\gamma t}  \\
\rme^{-\gamma t} & 1
\end{pmatrix} .
\label{eq:puredephase}
\end{equation}
Inspired in this behavior one can relate decoherence
to the norm of the off diagonal term. We define
\begin{equation}
D (\rho ) = 4 \left| \rho_{1,2}\right|^2 .
\label{eq:deco}
\end{equation}
For states of the form \eref{eq:puredephase} we obtain the formula
$D(\rho)=P(\rho)$. However, in the general case, information about one only
gives partial information about the other; for an arbitrary one qubit density
matrix, $0 \le D(\rho) \le P(\rho)$, and thus one can have a completely pure
state with $D=0$. This quantity is used frequently  as is easy to calculate
and is related to the interference fringes shown in the very popular cat
states in phase space, see e.g. \cite{zurekreview} page 742.  A big
disadvantage of using $D$ is that it is a basis dependent quantity.  Internal
dynamics may produce a decay of the off diagonal elements of $\rho$
and thus of $D$. Moreover, if one
studies an ($n>2$)-level system the situation becomes more
complicated. Purity on the other hand works for a much wider class of
systems.

\subsection{Entanglement as a resource}

It is not difficult to understand that entanglement is a (quantum) resource,
since already classical correlations are an important (classical) resource
used extensively in classical cryptography. Entanglement, as the quantum
correlation, brings up richer possibilities.  In general, controlled
entanglement can be used for the following:
\begin{itemize}
 \item Teleportation: It is the most celebrated application, due 
 to its spectacularity and simplicity \cite{PhysRevLett.70.1895}. 
 The transfer of an unknown quantum state can be achieved using an
 entangled state, local operations, and classical communication. 
 \item Quantum computation: It is a controversial subject whether
 entanglement is {\it essential} for quantum computation, but so far it has
 been demonstrated that for an exponential speedup in pure state schemes,
 entanglement is necessary (see \cite{entanglementvsquantcom}).
 \item Communication: Both quantum and classical communication can benefit
 from entanglement. In particular, quantum key distribution extensively uses
 this resource \cite{NC00a}. 
 \item Quantum-Enhanced Metrology: It is shown that the signal/noise ratio
 can be increased qualitatively \cite{giovannetti:010401, quantummeasurment}
 if one uses entangled states.  Thus the use of highly entangled states shall
 be mandatory for precise measurements.  Generalizations of the ideas
 developed in this area can be used to build quantum positioning systems (in
 analogy to GPS), enhanced radars, and for clock synchronization. 
\end{itemize}
Still the field is quite young and ideas for exploiting entanglement are
emerging at this point. New technology is arising: some quantum random
number generators are already available as USB gadgets.  In the near future
many expected, and unexpected, technologies are going to be proposed and, no
doubt, realized. Thus it is a major concern to be able to understand,
quantify, and control internal entanglement. 

One of the first tasks of quantum information theory was to quantify the
degree of entanglement. It was soon realized that, in general, this was a
complex task. We now know, for example, that for general systems,
entanglement induces only a partial ordering. 

We now focus on the simplest possible scenario that allows entanglement, a
two qubit system.  Four conditions must be fulfilled by an entanglement
measure: (i) It must have a value between zero and one.  It is zero for
separable states and one for Bell states. (ii) Any local unitary operation
leaves entanglement unchanged.  This condition can be seen as an invariance
of the measure under a {\it local} change of basis. (iii) Local operations
plus classical communication cannot increase entanglement.  \Ie to create
entanglement we need genuine non-local quantum operations (say interaction,
skew measurements, etc).  (iv) The entanglement measure must be a convex
function.  This condition says that entanglement will not increase when
mixing ensembles.  These four conditions can be generalized to more complex
systems (multipartite or higher dimensional systems). 

Several measures of two qubit entanglement fulfill these conditions,
however these measures do not provide exactly the same ordering of states
\cite{0305-4470-34-47-329}.  In this work we shall use the concurrence.
The first reason being that it is straightforward  to compute.  Some
measures of 2 qubit mixed state entanglement require explicit maximization
over high dimensional continuous sets.  Though in the definition of
concurrence (via the entanglement of formation) a maximization is
required, the problem is solved in a general fashion, and a closed formula
is given. See appendix \ref{sec:entappendix} for details.  The second
reason being that it is used widely in the community of quantum
information, both by theoreticians and experimentalists. 

Concurrence $C$ of a two qubit density matrix $\rho$ is 
\begin{equation}
  \label{eq:concurrence}
  C(\rho)=\max \{0,\Lambda_1-\Lambda_2-\Lambda_3-\Lambda_4 \}
\end{equation}
where $\Lambda_i$ are the eigenvalues of the matrix $\sqrt{\rho (\sigma_y
\otimes \sigma_y) \rho^* (\sigma_y \otimes \sigma_y)}$ in non-increasing
order.  The superscript $^*$ denotes complex conjugation in the
computational basis and $\sigma_y$ is a Pauli matrix.  Furthermore,
concurrence fulfills all conditions of a legitimate entanglement measure
discussed at the beginning of this section.

\section{The spectator configuration}
\label{sec:spectatorintro}

One of the important contributions of this work is the concept of spectator
configuration. During the development of the thesis the concept was
discovered, and its potential is exploited here.  On one hand, it allows to
enclose all the calculations in a single one, thus simplifying greatly the
technical details.  On the other, it enables to extend easily our results
from one and two qubits to $n$ qubits. Its full potential has not yet been
exploited, but we hope that the community will take advantage of this
concept.  

The concept involves the following  Hilbert spaces:
\begin{itemize}
\item The spectator space $\mcH_\rms$. The spectator is not
      coupled to the other spaces. However, it can be correlated
      initially via entanglement with the interacting space.
\item The interacting space $\mcH_\rmi$. This subspace is dynamically
      coupled to the environment and is initially entangled with the
      spectator subspace.
\item The environment space $\mcH_\rme$. This space typically has
      a large dimension (though this is not essential at this point).
      It is initially decoupled (in the sense of entanglement) to the
      rest of the system.
\item The central system $\mcH_\rmc$. It is the tensor product of the spectator
      and the interacting space: $\mcH_\rmc=\mcH_\rms \otimes \mcH_\rmi$. 
\end{itemize}
The whole Hilbert space $\mcH$ is the tensor product of all spaces, namely
\begin{equation} \label{eq:hamiSpec}
  \mcH=\mcH_\rme\otimes \mcH_\rmc=\mcH_\rme\otimes\mcH_\rms \otimes \mcH_\rmi.
\end{equation}
Additional to the Hilbert space structure, the spectator configuration, as
explained above, has a characteristic Hamiltonian: 
\begin{align}\label{eq:genspectator}
    H & = H_\rme \otimes \openone_{\rmi,\rms} +
          H_\rms \otimes \openone_{\rme,\rmi} +
          H_\rmi \otimes \openone_{\rms,\rme} +
	  \openone_\rms \otimes W_{\rme,\rmi}\\
	&= H_\rme+H_\rms+H_\rmi+W_{\rme,\rmi}\nonumber .
\end{align}
The indices indicate the spaces in which the operators act. Where there is
no danger of confusion, the identities are dropped as in the last equality
of \eref{eq:genspectator}. The first three terms represent local
Hamiltonians in each proper subspace, and the last term is an interaction
between the interacting subspace and the environment.  The different parts
of the Hamiltonian need not to be time independent; in this work the parts
devoted to random matrix theory deal with time independent Hamiltonians.
The parts studying the kicked Ising model and the $n$-qubit chapter
use a time dependent Hamiltonian example of (\ref{eq:genspectator}).

The initial condition is always separable with
respect to the environment:
\begin{equation}
  |\psi(t=0)\> = |\psi_\rme\> \otimes |\psi_\rmc\>,\,
  |\psi_\rme\> \in \mcH_\rme,\, |\psi_\rmc\>\in \mcH_\rmc,
\label{eq:initialspectator}
\end{equation}
but not necessarily with respect to the spectator space. If there is
separability between the interacting and the spectator spaces, the problem
trivially separates and we can consider then 2 completely decoupled
problems, one in $\mcH_\rms$ and another in $\mcH_\rme \otimes \mcH_\rmi$.
The condition of the environment being pure in \eref{eq:initialspectator}
is technical.  Some calculations have been done with mixed states in the
environment(s) yielding similar results.  At this point we could
formulate, with no problem whatsoever, the model with a mixed environment,
but since for further considerations it is convenient to have a pure state
we keep it that way.

The Hamiltonians that are going to be analyzed during the thesis are not
always of the form \eref{eq:genspectator}, but do have a particular
structure due to the structure of the underlying Hilbert space. This
structure is of the form given in \eref{eq:hamiSpec}, but with the
interacting space being composed of different independent  non-interacting
groups.  For this configuration, the environment will be coupled to all
groups independently.  Under some general conditions we shall be able to
decouple this complicated problem into many spectator problems. 

Two explicit examples of a simplification of the problem using the
spectator configuration are given when we have  2 qubits as the central
system in the context of random matrix theory (\sref{sec:separateTwo} and
\sref{sec:jointTwo}). Section \ref{sec:nqubit} separates the decoherence
problem in spectator configurations in a general fashion, and exemplifies
the results with both random matrix theory and the kicked Ising spin
chain. As the reader can notice, we shall follow a line of argumentation
that will build step by step the general case. During the thesis we shall
first study the simplest case (one qubit), then the next in complexity
(two qubits) and finally explain a possible way to use the tools developed
in a more general way. 

\section{Random matrix theory: a tool} \label{sec:RMTintro}
\begin{quote} {\em All I know is I know nothing.}\\ Socrates \end{quote}

Some people in the field consider the start of Random Matrix Theory as
being the paper by John Wishart \cite{wishartRMT} (who, incidentally, died
in Acapulco). There he introduces random matrices with invariant measure
under basis transformation. His objective was to analyze multivariate
data. Others say that the landmark was placed by  \'Elie Cartan in an old
(and often forgotten) paper \cite{cartanRMT}. He introduces explicitly the
circular ensembles, with invariance properties with respect to (usually)
group operations, to generalize the integral theorem of Cauchy. Mehta
\cite{mehta} was the first to calculate many of the mathematical
properties of the classical ensembles \cite{cartanRMT}.

There was not much development of RMT outside mathematics,  until Wigner
published his famous papers \cite{Wig51, Wig55a} pioneering its use in
physics.  These papers contain two important aspects.  The first one is
the idea to study {\em statistical} properties of the resonances of
complex nuclei instead of studying its {\em particular} properties.  This
is in perfect analogy with statistical mechanics: One does not care about
the particular position of the system in its phase space but rather about
its thermodynamical (statistical) properties. The other idea was to use an
ensemble of matrices to describe the system. This is again done in analogy
with statistics mechanical, but it is conceptually quite different.
Instead of performing averages over the phase space, one does averages
over the space of systems.  Wigner was successful in describing some
experimental findings, and later evidence showed the wide scope of
applicability of his idea \cite{Brody:1981cx, guhr98random, JPA-volRMT}. 

\begin{figure}[t]
\begin{center}
	\includegraphics[width=\textwidth]{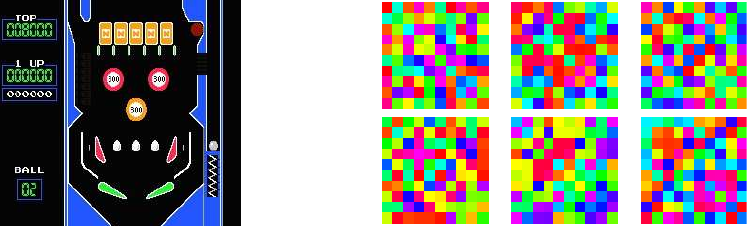}
\end{center}
\caption{Cartoon of an idea exploited in this thesis. The statistical
         properties of particular system (on the left, say a complicated
         billiard) are studied using the properties of an ensemble of random
         matrices (on the right).}
\label{fig:cartoonRMT}
\end{figure}

A next revolutionary step was marked by the papers by Casati \etal
\cite{conjectureCasati} and Bohigas \etal \cite{bohigasconjecture}.
There they conjectured that quantum systems whose classical counterparts
are chaotic have a spectrum whose fluctuations resemble those of the
appropriate classical ensembles. The revolutionary aspect of this
conjecture is that it does not require a complex composition of the
system (\ie many bodies), but only complexity in its {\it dynamics}.
Vast numerical evidence favoring this conjecture is available
\cite{guhr98random, JPA-volRMT}, but a precise understanding (\ie a
globally accepted proof) is yet outstanding. What about systems with
no classical correspondence? Defining chaoticity in this case is
cumbersome. We shall keep an oversimplified  definition: quantum 
chaotic systems are those that exhibit fluctuations in its spectrum
similar to those observed in the appropriate RMT ensemble. Alternatively 
one could say that quantum chaotic systems are those for which the correlations of most pair of observables decay to zero at large times.

In recent experiments \cite{fourparticleentanglement, photonentanglement,
antidecoherenceentanglement,NatureTomography}, it has been demonstrated
that it is possible to protect ever larger entangled quantum systems,
often arrays of qubits, ever more efficiently from decoherence.  A close
connection between the dynamics of fidelity decay and decoherence has been
shown in some instances \cite{cucchietti:045206, zurekDecoEcho,
cucchietti:210403, gpss2004}, which suggests to apply methods successful
in one field to the other.  In that context, a random matrix description
\cite{1367-2630-6-1-020, reflosch} is accessible and very effective in
describing experiments \cite{expRudi, 1367-2630-7-1-152, gorin-weaver}.
Based on this success of random matrix theory, we shall use it to model
decoherence \cite{reflosch, 1464-4266-4-4-325} of qubit systems
\cite{pinedaRMTshort}, assuming complicated dynamics in the environment,
and a complicated coupling (in the interaction picture).

Three perspectives make such a random matrix treatment particularly
attractive.  First, reduction of decoherence may, in some instances,
be achieved by isolating some ``far'' environment (including
spontaneous decay) to a degree that it can, to first approximation,
be neglected. Then it can happen that the Heisenberg time of the
relevant ``near'' environment is finite on the time scale of
decoherence.  In such a case it becomes relevant that RMT shows, in
linear response approximation, a transition from linear to quadratic
decay at times of the order of the Heisenberg time. This behavior is
seen with spin chain environments  [see \sref{sec:ki}], and is
essential for the success of the theory in describing the above
mentioned experiments of fidelity decay.  Note also that the concept
of a two stage environment has been used for basic considerations
\cite{zurekreview}.  Second, the long term goal must be to describe
in one theory the decay of fidelity that includes undesirable
deviations of the internal Hamiltonian of the central system
(already done), together with decoherence (done in this thesis).
Third, random matrix descriptions include some aspects of chaos or
mixing that are essential in the above experiments and may be
useful for application to quantum computing
\cite{0305-4470-34-47-103, ShepelyanskyRMTFidelity, Pineda01}. 

For some technical aspects of random matrix theory,
including Heisenberg time, GOE and GUE ensembles, form factor and density
of states, please go to appendix \ref{sec:RMTtech}.

\chapter{One qubit decoherence}\label{sec:onequbit}

In this chapter we analyze decoherence of a single qubit.  
We focus on weak coupling of the qubit to
an environment.  We shall use the correlation function approach proposed for
purity decay in echo-dynamics \cite{purityfidelity}, treating the coupling as
the perturbation.  The linear response approximation will be sufficient. In
this approximation the ensemble averages, which we have to take in any RMT
model, are feasible though somewhat tedious.  Exact solutions, which exist in
some instances for the decay of the fidelity amplitude \cite{shortRMT,
gorin:244105}, seem to be out of reach at present, because they require the
evaluation of four-point functions.

The general program is as follows.  Assume that the qubit is initially in a
pure state, and evolves under its own local Hamiltonian.  The qubit is  coupled
via a random matrix to a large environment in turn described by another random
matrix. The coupling to the environment gives rise to decoherence.  Averaging
both the coupling and the environment Hamiltonian over the RMT ensembles yields
the generic behavior of decoherence of the qubit. 

We present a detailed analysis  using both the Gaussian unitary (GUE) and the
Gaussian orthogonal (GOE) ensembles~\cite{cartanRMT,mehta} for the description
of the environment and the coupling.  The two ensembles correspond to time
reversal invariance (TRI) breaking and conserving dynamics respectively. 

In \sref{sec:OneQubitModel} we shall state the model,  recall the linear
response formalism for echo dynamics, and show how it can be adapted to forward
evolution. In \sref{sec:generalprogram} we discuss how to express the problem
in terms of echo dynamics.  In \sref{sec:solutionOne} we give the general
solution, arising from the calculations done in the appendix
\ref{sec:calculationSpec}.  The analysis for the GUE case is given in
\ref{sec:gueone}, whereas the one for the GOE is given in  \ref{sec:goeone}.

\section{The model} \label{sec:OneQubitModel}

We describe decoherence by considering explicitly additional degrees of
freedom (henceforth called ``environment'') which are interacting with the
qubit. The Hilbert space studied in this section is
\begin{equation}
\mcH= \mcH_1 \otimes \mcH_\e,
\label{eq:algo}
\end{equation}
where $\mcH_1$ (of dimension two) and $\mcH_\e$ (of dimension
$N_\e$) denote the Hilbert spaces of the qubit and the environment,
respectively.  The Hamiltonian is of the following form
\begin{align} \label{eq:hamiltonianqubit}
H_\lambda &=
H_1\otimes \openone_\e + \openone_1 \otimes H_\e +\lambda V_{1,\e}\\
&\equiv H_1 + H_\e + \lambda V_{1,\e} \; . 
\end{align}
Here, $H_1$  represents the Hamiltonian acting on the qubit, $H_\e$ the
Hamiltonian of the environment, and $V_{1,\e}$ the coupling between the
qubit and the environment. Notice how the indices in the operators
indicate the spaces in which they act. The real parameter $\lambda$
controls the strength of the coupling.  We shall study the time evolution
of an initially pure and separable state
\begin{equation} \label{eq:initialonequbit}
|\psi(t=0)\> = |\psi_1\>\otimes |\psi_\e\> \; ,
\end{equation}
where $|\psi_1\> \in \mcH_1$ and $|\psi_\e\>\in \mcH_\e$. At any time $t$,
the state of the whole system is thus $|\psi(t)\>=\exp(-\mimath t
H_\lambda) |\psi(0)\>$, and the state of the single qubit is $\tr_\e
|\psi(t)\> \<\psi(t)|$.  As time evolves, the qubit and the environment
get entangled, which means that after tracing out the environmental
degrees of freedom, the state of the qubit becomes mixed.

At this point we wish to compare this model with the spectator model.  We
can arrive to the one studied in this chapter from two different
directions. One is if we eliminate the spectator. The other is if we
consider a separable (with respect to the spectator) situation [\ie  if in
\eref{eq:initialspectator} we let $|\psi_\rmc\>=|\psi_\rms\> \otimes
|\psi_\rmi\>$]. 

We describe both the coupling and the dynamics in the environment within
random matrix theory. To this end, $H_\e$ and $V_{\e,1}$ are chosen both
from either the GUE or the GOE, depending on whether we wish to describe a
TRI breaking or TRI conserving situation. The Hamiltonian $H_1$ implies
another free parameter of the model, namely the level splitting $\Delta$
of the two level system representing the qubit.  While the state of the
qubit $|\psi_1\>$ implies more free parameters in our model, we assume the
state of the environment $|\psi_\e\>$ to be random.  This means that the
state is chosen from an ensemble which is invariant under unitary
transformations, and is fully consistent with our minimum information
assumption. In practice, this means that the coefficients are chosen as
complex random Gaussian variables, and subsequently the state is
normalized.

\section{Echo dynamics and linear response theory} \label{sec:generalprogram}

We shall calculate the value of purity as a function of time analytically, in a
perturbative approximation.  As we want to use the tools developed in the
appendix \ref{sec:calculationSpec} for a linear response formalism in echo
dynamics, we must state the problem in this language.  To perform this task it
is useful to consider the above Hamiltonian [\eref{eq:hamiltonianqubit}], as
composed by an unperturbed part $H_0$ and a perturbation $\lambda V$.  The
unperturbed part corresponds to the operators that act on each individual
subspace alone whereas the perturbation corresponds to the coupling among the
different subspaces; \ie $H_0=H_\e+H_1$ and $V=V_{\e,1}$.  

We write the Hamiltonian as 
\begin{equation} \label{eq:hlambda}
  H_\lambda=H_0+\lambda V, 
\end{equation} 
and introduce the evolution
operator and the echo operator defined by 
\begin{equation}\label{eq:defevecho} 
 U_\lambda(t)=e^{-\mimath H_\lambda t} , \quad
 M_\lambda(t)=U_0(t)^\dagger U_\lambda(t), 
\end{equation} respectively
($\hbar =1$ during all the thesis). The echo operator receives its name because
it evolves a state forward in time with a perturbed operator and backwards with
an unperturbed one. For the calculation of purity at a given time $t$, we
replace the forward evolution operator $U_\lambda$ by the corresponding echo
operator $M_\lambda$. Even though the resulting states are different, \ie
\begin{equation} 
  \rho(t)= \tr_{\e,\e'} U_\lambda(t) \rho U_\lambda^\dagger(t)
     \ne \rho^M(t)= \tr_{\e,\e'} M_\lambda(t) \rho M_\lambda^\dagger(t),
\end{equation} 
they are still related by the local (in the qubit and the
environment) unitary transformation $U_0(t)$.  Since local transformations do
not change the entanglement properties, it holds (exactly!) that
\begin{equation} 
  P(t)= P[\rho(t)]= P[\rho^M(t)].  
  \label{eq:purityequalecho}
\end{equation} 
This step is crucial, since the echo operator admits a series
expansion with much larger range of validity (both, in time and perturbation
strength).  However the numerical simulations are all done with forward
evolution alone as they require less computational effort.

The Born expansion of the echo operator up to second order reads
\begin{equation} \label{eq:bornexpansion} 
  M_\lambda(t)= \openone -\mimath \lambda I(t) - \lambda^2 J(t) 
                       + \Or(\lambda^3), 
\end{equation} 
with
\begin{equation} 
  I(t)= \int_0^t \rmd \tau \tilde V(\tau), 
    \quad J(t)= \intoh \tilde V(\tau) \tilde V(\tau') 
  \label{eq:expIandJ} 
\end{equation} 
and $\tilde{V}(t)=U_0(t)^\dagger V U_0(t)$ being the coupling in the
interaction picture. Using this expansion we calculate the purity of the
central system, averaged over the coupling and the Hamiltonian of the
environment.

The reader must notice that at no point we used that  the dimension of the
central system is 2. In fact, we only required the locality of the operator
$U_0$ and the fact that $M_\lambda(t)\approx \openone$. Thus, all this
reasoning (in particular eqs. \ref{eq:purityequalecho}, \ref{eq:bornexpansion},
and \ref{eq:expIandJ}) is equally valid for a completely general spectator
configuration or even more general configurations to be introduced later.

\section{The solution}\label{sec:solutionOne}

In appendix \ref{sec:calculationSpec}, we compute the average purity
$\<P(t)\>$ as a function of time in the linear response approximation
\eref{eq:bornexpansion}, following the steps outlined in
\sref{sec:generalprogram}. The average is taken with respect to the
coupling $V_{1,\e}$ [using eqs.~\ref{eq:GOEconmute} and
\ref{eq:GUEconmute}], the random initial state $|\psi_\e\>$, and the
spectrum of $H_\e$. In the limit of $N_\e\to\infty$, we
obtain~[\eref{B:pulrdef}, \eref{B:pulrires}]
\begin{equation}\label{eq:purityoneq} 
  \< P(t)\> = 1 -2\,\lambda^2 \int_0^t\rmd\tau\int_0^t\rmd\tau'\; 
    {\rm Re}\, A_{\rm JI}(\tau,\tau') + \Or(\lambda^4) , 
\end{equation} 
with 
\begin{equation}\label{eq:resajiuno}
  A_{\rm JI}(\tau,\tau')= 
    [C_1(|\tau-\tau'|)-S_1(\tau-\tau')]\bar C(|\tau-\tau'|)
                  +\chi_{\rm GOE} [1-S'_1(-\tau-\tau')], 
\end{equation} 
where $\chi_{\rm GOE} = 1$ for the TRI case, and $\chi_{\rm GOE} = 0$ for
the non-TRI case. The correlation functions $C_1(\tau), S_1(\tau),
S'_1(\tau)$, and $\bar C(\tau)$ are defined in appendix \ref{aB}.  The
first three depend on the state of the central system.  Note that
$S'_1(\tau)$  is only relevant in the case of a GOE, and curiously is not
strictly a correlation function as it contains a dependence on the {\it
sum} of both times. The last one, $\bar C(\tau)$ deserves special
attention, since it depends on the spectral properties of the environment
determined by the function
\begin{equation}\label{eq:thecorrelation} 
\frac{1}{N_\e}\left\< 
  \left| {\textstyle\sum_{j=1}^{N_\e}}\rme^{-\mimath E_j t} \right|^2
    \right\> 
    = \bar C(t) 
    = 1 + \delta(t/\tau_H) - b_2^{(\beta)}(t/\tau_H) , 
\end{equation}
(recall \eref{eq:formfactor} and subsequent equations).  Here the $E_j$'s are
the eigenenergies of $H_\e$ and $\tau_H$ is the corresponding Heisenberg
time.  Actually the validity of \eref{eq:purityoneq} is not dependent on the
environment being represented by a GOE ($\beta=1$) or GUE ($\beta=2$).  For
these the two-point form factor $b_2^{(\beta)}$ is well known ~\cite{mehta}
but any ensemble with the corresponding invariance properties will do, for
example the POE or PUE \cite{Dittes}.

We first study the GUE case with and without an internal Hamiltonian
governing the qubit.  The next step is to work out the GOE case.  There we
concentrate on the case with no internal Hamiltonian governing in the qubit
since we want to keep the discussion as simple as possible to focus on the
consequences of the weaker invariance properties of the ensemble.

\section{The GUE case} \label{sec:gueone}

We are now in the position to give an explicit formula for $\< P(t)\>$ in the
GUE case. This formula will generally depend on some properties of the initial
condition $|\psi_1\>$. We wish to write it in the most general way.  However
the symmetries involved in the problem reduce the number of parameters needed
to describe the initial state $|\psi_1\>$. 

Recall that $H=H_\e+H_1+ \lambda V$ represents an ensemble of Hamiltonians in
which $H_\e$  and $V$ are chosen from GUEs of dimension $N_\rme$ and $2N_\rme$
respectively, whereas $H_1$ together with the initial condition $|\psi_1\>$
remain fixed throughout the calculation.  The operations under which the
ensemble is invariant are local (with respect to the partitioning of the
Hilbert space into $\mcH_1$ and $\mcH_\e$), unitary (due to the invariance
properties of the GUE), and leave $H_1$ invariant. Hence the transformation
matrices must be of the form \begin{equation}\label{eq:jpjp} U\otimes
\exp(\mimath \alpha H_1) \end{equation} with $\alpha$ a real number and $U$ a
unitary operator acting on $\mcH_\e$.  The solution must also be invariant
under that transformation.

This freedom allows to choose a convenient basis to solve the problem. On
the one hand, it allows to write $H_0$ in diagonal form (as done during
the discussion of appendix \ref{sec:calculationSpec}), and on the other
hand, we can use it to represent the initial state of the qubit in such a
way that there is no phase shift between the two components of the qubit.
This can be achieved by appropriately choosing $\alpha$ in \eref{eq:jpjp}.
We thus write, without
loosing generality 
\begin{equation}\label{eq:initialOneGUE} 
|\psi_1\>=\cos\phi |0\>+ \sin\phi |1\> \; ,
\end{equation}
where $|0\>$ and $|1\>$ are eigenstates of $H_1$. Notice that if
$\phi\in\{0,\pi/2\}$, $|\psi_1\>$ is an eigenstate of $H_1$.  Finally, we
choose the origin of the energy scale in such a way that the Hamiltonian
of the qubit can be written as
$H_1=(\Delta/2)|0\>\<0|-(\Delta/2)|1\>\<1|$. 

We obtain the average purity from the general expression in
\eref{eq:purityoneq} and \eref{eq:resajiuno}. For a pure initial state
$\rho_1= |\psi_1\> \<\psi_1|$ the relevant correlation functions ${\rm
Re}\, C_1(\tau)$, $S_1(\tau)$, and $\bar C(\tau)$ are given in
\eref{aB:ReC1}, \eref{aB:S1pure}, and \eref{B:Cbardef}, respectively.
Using the symmetry of the resulting integrand with respect to the exchange
of $\tau$ and $\tau'$, we find
\begin{equation}\label{eq:genGUEone} \< P(t)\>= 1
-4\lambda^2\int_0^t\rmd\tau\int_0^\tau\rmd\tau'\; \bar C(\tau')\;
\big [\, 1- g_\phi\; (1- \cos\Delta\tau')\, \big ]
+ \Or(\lambda^4,N_{\rm e}^{-1})
\end{equation}
with
\begin{equation}\label{eq:defg}
g_\phi=\cos^4\phi+\sin^4\phi = \frac{3+\cos(4\phi)}{4}
\end{equation}
quantifying the ``distance'' between $|\psi\>$ and an eigenbasis of $H_1$.
 
Let us consider the following two limits for $H_1$. The ``degenerate
limit'', where the level splitting $\Delta$ is much smaller than the mean
level spacing $d_\e= 2\pi/\tau_H$ of the environmental Hamiltonian, and
the ``fast limit'', where the level splitting is much larger. In the
latter case, the  internal evolution of the qubit is fast compared with
the evolution in the environment.  (We shall refer to these limits also in
later sections.)

The degenerate limit leads to the known formula~\cite{pinedaRMTshort}
\begin{equation}\label{eq:DegenerateOne}
P_{\rm D}(t)=1-\lambda^2 f_{\tau_\rmH}(t),
\end{equation}
with
\begin{equation}\label{eq:deff}
f_{\tau_\rmH}(t)=
2t\max\{t,\tau_\rmH\}+\frac{2}{3\tau_\rmH}(\min\{t,\tau_\rmH\})^3.
\end{equation}
The result does not depend on the initial state of the qubit.  Due to the
degeneracy all states are eigenstates of $H_1$ and thus equivalent.  The
leading term of the purity decay is linear before the Heisenberg time and
quadratic after the Heisenberg time. Similar features were already
observed in fidelity decay and purity decay in other contexts
\cite{reflosch}.

In the fast limit ($\Delta\gg d_\e$), purity is obtained from
\eref{eq:genGUEone} by replacing $\cos\Delta\tau'$  by 1 when it is
multiplied with the $\delta$ function [see \eref{eq:thecorrelation}], and by
zero everywhere else. For finite $N_\e$ care must be taken, since we are
assuming Zeno time (which is given by the ``width of the $\delta$-function'')
to be much smaller than all other time scales, such that $\Delta \ll N_\e\,
d_\e$. The resulting expression is \begin{equation}\label{eq:FastOne} P_{\rm
F}(t)=1-\lambda^2 [ (1-g_\phi)f_{\tau_\rmH}(t)+2 g_\phi t \tau_\rmH]\;.
\end{equation} Typically (unless $|\psi_1\>$ is an eigenstate), this formula
displays a dominantly linear decay below the Heisenberg time, and a dominantly
quadratic decay above, similar to \eref{eq:DegenerateOne}.

\begin{figure} \centering
\includegraphics{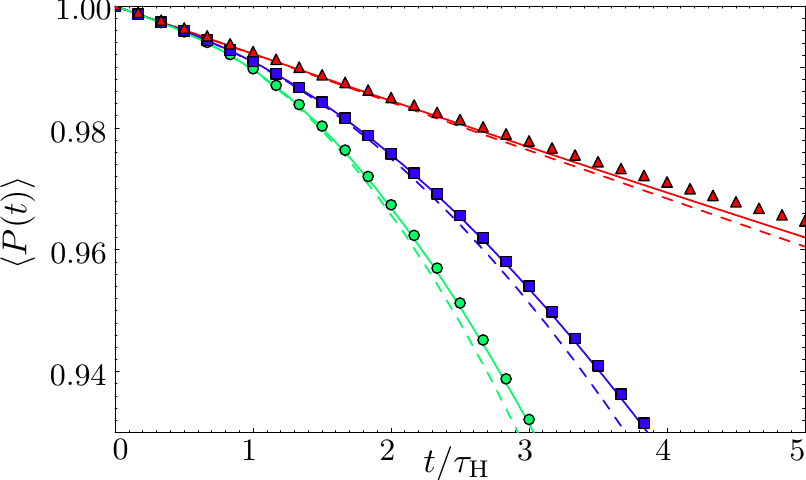} 
\caption{Numerical simulations for the average purity of one qubit
using non-TRI Hamiltonians, as a function of time in units of the
  Heisenberg time $\tau_H$ of the environment.  For the coupling
  strength $\lambda= 0.01$ and $N_\e=2048$, we show the dependence of
  $\< P(t)\>$ on $\Delta$ (the level splitting) and $\phi$
  (characterizing the initial state) of the internal qubit: $\Delta=
  0$ (green circles), $\Delta= 8$, $\phi= \pi/4$ (blue squares), and
  $\Delta= 8$, $\phi= 0$ (red triangles). The corresponding linear
  response results (dashed lines) and exponentiated linear response
  results (solid lines) are based on \eref{eq:genGUEone} and
  \eref{eq:exponentiationone}, where $P_\infty$ is given in
  \ref{sec:exponentiation}. Note that the level splitting $\Delta$
  tends to slow down decoherence.} 
  \label{fig:holeone} 
\end{figure}

In \fref{fig:holeone} we compare numerical simulations of the average purity
$\< P(t)\>$ (symbols) with the corresponding linear response result (dashed
lines) based on eqs. \eref{eq:DegenerateOne} and \eref{eq:FastOne}.  The
numerical results are obtained from Monte Carlo simulations with 15 different
Hamiltonians and 15 different initial conditions for each Hamiltonian.  We
wish to underline two aspects. First, the energy splitting in general leads to
an attenuation of purity decay. Even though a strict inequality only holds for
the limiting cases, $P_{\rm F}(t) > P_{\rm D}(t)$ (for $t\ne 0$), we may still
say that increasing $\Delta$ tends to slow down purity decay. This result is in
agreement with earlier findings on the stability of quantum
dynamics~\cite{0305-4470-35-6-309}. Second, for the fast limit and an
eigenstate of $H_1$ ($g_\phi=1$) we find linear decay even beyond the
Heisenberg time. A similar behaviour has been obtained in~\cite{reflosch}, but
there an eigenstate of the whole Hamiltonian was required.

In \cite{pinedaRMTshort} it was shown that exponentiation of the linear
response result leads to very good agreement beyond the validity of the
original approximation. We use the formula \eref{eq:ELRextension}
\begin{equation}\label{eq:exponentiationone} P_{\rm
ELR}(t)=P_\infty+(1-P_\infty) \exp\left[- \frac{1-P_{\rm
LR}(t)}{1-P_\infty}\right].  \end{equation} where $P_{\rm LR}(t)$ is truncation
to second order in $\lambda$ of the expansion \eref{eq:genGUEone}, and
$P_\infty=1/2$ the estimated asymptotic value of purity for $t \to \infty$, see
appendix \ref{sec:exponentiation}.  From \fref{fig:holeone} we see that the
exponentiation indeed increases the accuracy of the bare linear response
approximation.

\section{The GOE case}\label{sec:goeone}

\begin{figure}[htbp] 
  \centering 
  \includegraphics[width=.5\textwidth]{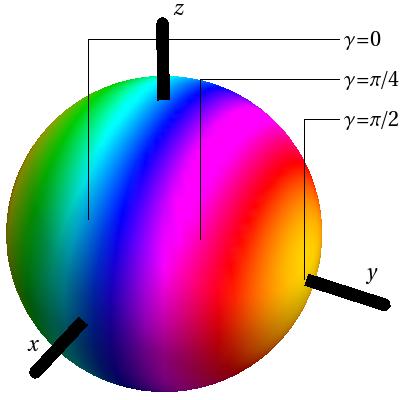} 
  \caption{Any pure initial state of the qubit can be mapped onto the
    Bloch sphere. Here, we show the angle $\gamma$ defined in
    \eref{eq:initialOneGOE} in color code. Regions of a given color
    represent subspaces which are invariant under the transformation
    $\exp(\mimath \alpha \sigma_y)$.} 
  \label{fig:blochgoe}
\end{figure}

We drop $H_1$ for a moment, leaving $H_0=H_\e$, resulting in
\begin{equation}\label{eq:noint} 
  H_\lambda=H_\e+\lambda V.  
\end{equation} 
$H_\e$ is chosen from a GOE of dimension $N_\rme$ and acts on $\mcH_\e$; $V$
is chosen from a GOE  of dimension $2N_\rme$ and acts on $\mcH_\e \otimes
\mcH_1$. The resulting ensemble of Hamiltonians is invariant under local
orthogonal transformations.  In the qubit this invariance allows rotations of
the kind $\exp(\mimath \alpha \sigma_y)\in \mcO(2)$.  If such transformations
are represented on the Bloch sphere, they become rotations around the $y$
axis. Hence, they can take any point on the Bloch sphere onto the $xy$-plane.
Supposing this point represents the initial state, it shows that we may
assume the initial state in the qubit to be of the form 
\begin{equation}\label{eq:initialOneGOE}
  |\psi_1\>=\frac{|0\>+e^{\mimath \gamma}|1\>}{\sqrt 2}.  
\end{equation} 
In this expression, $\gamma\in [-\pi/2,\pi/2]$ denotes the angle of the
vector representing the initial state  with the $xz$-plane (see
\fref{fig:blochgoe}).

In order to obtain the linear response expression for $\< P(t)\>$ we again
make use of \eref{eq:purityoneq} and \eref{eq:resajiuno}. However, apart from
the correlation functions used in the GUE case, we have now to consider in
addition $S'_1(\tau)$, as given in \eref{aB:S1ppure}. The special case
$H_1=0$ can be simply obtained by setting $\Delta =0$. This yields 
\begin{equation}\label{goeone:aji} 
  A_{\rm JI}(\tau,\tau')= \bar C(|\tau-\tau'|) + \sin^2\gamma.  
\end{equation} 
After evaluating the double integral in \eref{eq:purityoneq}, we obtain 
\begin{equation}\label{eq:purityGOEsep} 
  \<P(t)\>=1-\lambda^2 \left\{ 
             t^2 \left[3-\cos(2\gamma)\right] +2 t \tau_H - 2 B_2^{(1)}(t)  
             \right\}, 
\end{equation} 
where 
\begin{equation}\label{eq:Btwogoe} 
  B_2^{(1)}(t)= 
    2\int_0^t \rmd \tau \int_0^\tau \rmd\tau' b_2^{(1)}(\tau'/\tau_H) 
\end{equation} 
is the double integral of the form factor. It can be computed analytically,
but the resulting expression is very involved~\cite{1367-2630-6-1-020}. For
our purpose it is sufficient to note that for $t \ll \tau_\rmH$,
$B_2^{(1)}(t) \propto t^3 $ (as in the GUE case), whereas for $t \gg
\tau_\rmH$, $t- B_2^{(1)}(t)$ grows only logarithmically.  Thus it has a
similar behavior as in the GUE case, \eref{eq:finalBtwotwo}.

\begin{figure}[htbp] \centering
\includegraphics{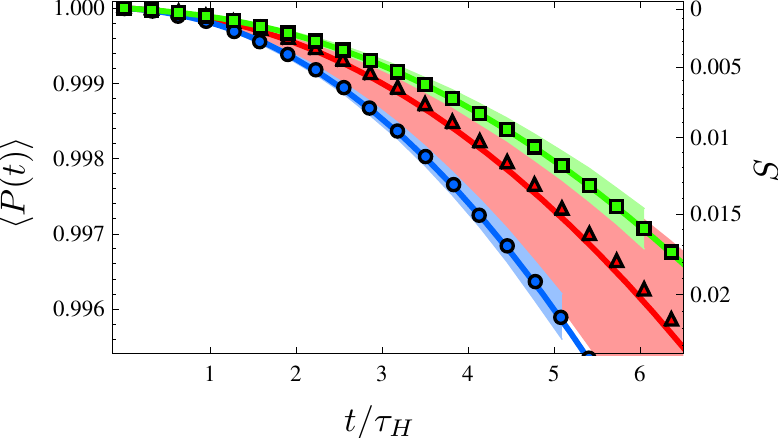}\\[1cm]
\includegraphics{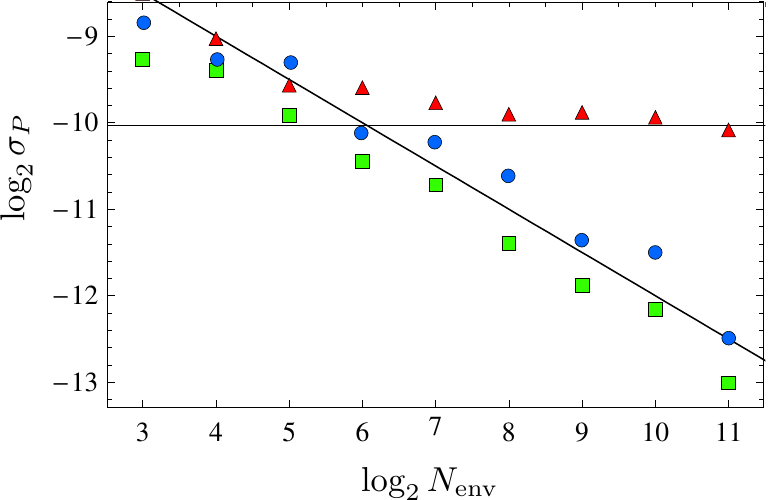}
\caption{We display the behavior of purity and von Neumann entropy,
  in different regions in the Bloch sphere connected by orthogonal
  transformations, characterized by $\gamma$ in
  \eref{eq:initialOneGOE}. On the top figure we see the enveloping
  curve after running 100 initial conditions (thick regions), their
  average (symbols) and the predicted behavior (solid curves) by
  \eref{eq:purityGOEsep}.  If $\gamma=0$ we
  use color green; if $\gamma=\pi/2$ we use blue; and if we allow
  arbitrary $\gamma$ we use red.  In the lower panel 
  $\sigma$ is plotted for $t=40$. We use
  the same coding as the upper figure.  For a fixed value of $\gamma$
  (blue and green) there is asymptotic self averaging whereas for an
  arbitrary initial condition (red) the standard deviation reaches
  the finite value predicted in \eref{eq:sigmagoe}, plotted as a
  horizontal line.  We fixed $\lambda=10^{-3}$. A line
  $\propto 1/\sqrt{N_\rme}$ is also included.}
  \label{fig:decayGOE}
\end{figure}

In \fref{fig:decayGOE} we show $\< P(t)\>$ for $\gamma= 0$ (green squares),
for $\gamma= \pi/2$ (blue circles), and for random values in the whole Bloch
sphere (red triangles).  In contrast to the GUE case in the degenerate limit,
the average purity depends on the initial state (via the angle $\gamma$).
The fastest decay of purity is observed for $\gamma= \pi/2$, where the image
under the time reversal operation becomes orthogonal to the initial state.
The slowest decay is observed for $\gamma= 0$, which characterizes states
which remain unchanged under the time reversal symmetry operation. In the
lower panel we show numerical results for the standard deviation of the
purity as a function of $N_\e$, the dimension of the Hilbert space of the
environment. We consider the same cases as on the upper panel: random initial
states $|\psi_1\>$ with fixed $\gamma=0$ (green squares), with fixed $\gamma=
\pi/2$ (blue circles) and random states $|\psi_1\>$ uniformly distributed on
the Bloch sphere (red triangles). Note that along with the random choice of
$|\psi_1\>$, also $H_\e$, $V_{1,\e}$, and $|\psi_\e\>$ are randomly chosen
from their respective ensembles. We clearly see that for those cases where
$\gamma$ is kept fixed, the standard deviation falls off like $N_\e^{-1/2}$.
By contrast, the standard deviation converges to a finite value, when
$|\psi_1\>$ chosen with no restriction.  That value can be estimated 
from the standard deviation of the function $\cos\, 2\gamma$, which yields: 
\begin{equation}\label{eq:sigmagoeprev} 
  \int_{-\pi/2}^{\pi/2} \left[
             \overline{\cos (2 \gamma)}-\cos (2 \gamma)\right]^2
	\frac{\cos (2 \gamma)}{2} \rmd \gamma=\frac{16}{45}.
\end{equation} 
Assuming that for $N_\e\to\infty$, the fluctuations of $\cos 2\gamma$, is the
only source for the fluctuations of purity, the standard deviation of the
purity is 
\begin{equation}\label{eq:sigmagoe}
  \sigma_P=
    \frac{4}{3\sqrt{5}} \lambda^2 t^2 + \mathcal{O}(\lambda^4,N_\e^{-1}).  
\end{equation}

The statements of the last two paragraphs can be directly translated to the
von Neumann entropy $S$.  For one qubit it has a one to one relation with
purity
\begin{equation} 
 S(P)=
  h\left(\frac{1+\sqrt{2P-1}}{2}\right)+h\left(\frac{1-\sqrt{2P-1}}{2}\right),
    \,h(x)=-x\log_2x.  
 \label{eq:entropur}
\end{equation} 
Observe the entropy scale on the upper panel of~\fref{fig:decayGOE}.  The
consequences of the weaker invariance properties of the GOE, and the relation
to the states will  be analyzed in detail in a general framework in a later
paper.

\chapter{Two qubit decoherence}\label{sec:twoqubit}

In this chapter, we address the question whether entanglement within a given
system affects its decoherence rate.  In particular, as the name suggests, we
are going to study two qubit decoherence.  We will use the spectator model
described in \sref{sec:spectatorintro}.  Moreover we shall consider the first
nontrivial example thereof, in the sense that the spectator space plays a
nontrivial roll. However it is still the simplest example allowing this
possibility as both the spectator and the interacting space are qubits. We
shall study two other configurations, namely when both qubits are coupled to
one or two environments. There, we shall start appreciating the power of the
spectator model; we are  going to be able to express there decoherence in terms
of the decoherence in the spectator configuration.

We will base our arguments on the calculations made in appendix
\ref{sec:calculationSpec} and the results discussed in the previous chapter.
Again we are going to study the linear response regime, and test with Monte
Carlo simulations the validity of a heuristic exponentiation.  The symmetries
of the classical ensembles will continue to play an important role to simplify
the problem. Moreover we shall find that entanglement has the property of
transporting the symmetry from one qubit to the other.

We first describe the models (configurations) we are going to study,
\sref{sec:twoqubitmodel}. Next we analyze in detail the results for the
simplest configuration, the spectator model for two qubits,
\sref{sec:spectator}. There we consider both the GUE and GOE cases. We then
generalize the result to the other configurations in \sref{sec:separateTwo} and
\sref{sec:jointTwo}. 

\section{The models}\label{sec:twoqubitmodel}

For the two qubit case, the Hilbert space structure must be slightly
more complicated than \eref{eq:algo}. We need at least to provide the
Hilbert space for a second qubit, and, in one of the models, we shall
need an additional Hilbert space for a second environment. In all cases
the initial condition is pure in the central system, but the two qubits
can share some entanglement. We shall consider three different
dynamical scenarios, all explicitly excluding any interaction between
the two qubits:

\begin{figure}[t] 
\centering 
\includegraphics[width=\textwidth]{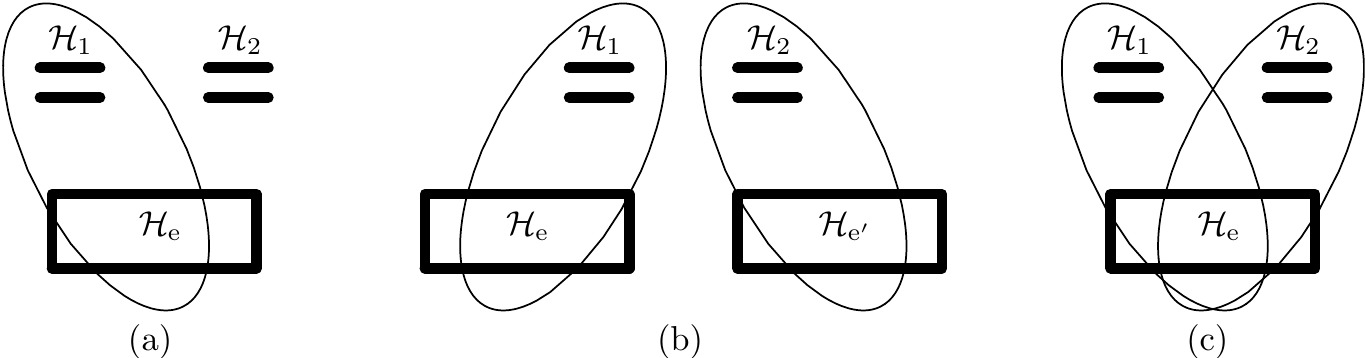}
\caption{Schematic representations  of the different dynamical
configurations studied in this article:  (a) the 2 qubit spectator
  Hamiltonian, (b) the separate, and (c) the joint environment
  Hamiltonian.}
  \label{fig:schemeconfig}
\end{figure}

\begin{itemize} \item[(a)] {\em The 2 qubit spectator Hamiltonian:}\\
The
Hilbert space structure is 
\begin{equation} 
  \mcH=\mcH_1 \otimes \mcH_2 \otimes \mcH_\rme.  
  \label{eq:hilberth1h2he} 
\end{equation} 
Both $\mcH_1$ and $\mcH_2$ are the state spaces of
the qubits ($\dim \mcH_1=\dim \mcH_2=2$) and $\mcH_\rme$ (of dimension
$N_\rme$) is the state space of an environment. The central
system is, obviously,
\begin{equation} 
  \mcH_\rmc=\mcH_1 \otimes \mcH_2.
  \label{eq:thxxx} 
\end{equation} 
Only the first qubit is coupled to an environment, and we allow for local
dynamics. The total Hamiltonian reads
\begin{equation}\label{eq:spectatorHam} 
  H_\lambda= H_1 + H_2  + H_\e  + \lambda V_{1,\e} \; , 
\end{equation} 
where $\lambda$ is again a real number modulating the strength of the
coupling.  We recall the remark in \sref{sec:spectatorintro}, namely
that the sub-indices of the operators indicate the space on which they
act.  This situation is shown schematically in
\fref{fig:schemeconfig}(a). Notice that if we choose an initial state
where the two qubits are already entangled, this provides the {\it
simplest} situation which allows to study entanglement decay.  If the
two qubits are initially not entangled, the process reduces effectively
to the one-qubit case, described previously \sref{sec:OneQubitModel}.
The special case $H_1=H_2=0$ has been considered in
Ref.~\cite{pinedaRMTshort}.

The reader must notice that this situation is an example of the
spectator configuration, discussed in \sref{sec:spectatorintro}. In
this case the interacting subspace is $\mcH_1$ and the spectator is
played by $\mcH_2$.

\item[(b)] {\em The separate environment Hamiltonian:}\\ The Hilbert space
structure is
\begin{equation} 
  \mcH=\mcH_1 \otimes \mcH_2 \otimes \mcH_\rme\otimes \mcH_{\rme'}.
  \label{eq:hilberth1h2hehep} 
\end{equation}
Besides the subspaces in \eref{eq:hilberth1h2he} (which keep their meaning), we
have an additional space $\mcH_{\rme'}$ which represents a new environment.  We
again allow for similar dynamics except that we allow the second qubit to
interact with the new environment. The two environments are assumed to be
non-interacting and uncorrelated: 
\begin{equation}\label{eq:initialconditiongeneral}
  |\psi(0)\>=|\psi_{12}\>|\psi_\e\>|\psi_{\e'}\>,\quad 
    |\psi_{12}\>\in \mcH_1 \otimes \mcH_2,\,|\psi_\e\>\in \mcH_\e,\,
     {\rm and}\, |\psi_{\e'}\>\in \mcH_{\e'}\; .  
\end{equation} 
Thus, the Hamiltonian of this model reads
\begin{equation}\label{eq:sepHam} 
  H_{\lambda_1,\lambda_2}= H_1+H_2+H_\e+H_{\e'}+
    \lambda_1\; V_{1,\e} + \lambda_2\; V_{2,\e'} \; ,
\end{equation} 
where $V_{2,\e'}$ and $H_{\e'}$ describe the coupling to -- and the dynamics
in the additional environment. Both quantities are chosen independently from
the respective random matrix ensembles, in perfect analogy with $V_{1,\e}$
and $H_\e$. The real parameters $\lambda_1$ and $\lambda_2$ fix the coupling
strengths to either environment. This model, see \fref{fig:schemeconfig}(b),
may describe two qubits that are ready to perform a distant teleportation,
where each of them is interacting only with its immediate surroundings.  It
can also represent a pair of qubits that, although close to each other,
interact with different and independent degrees of freedom.

\item[(c)] {\em The joint environment Hamiltonian:}\\ The third case, shown in
\fref{fig:schemeconfig}(c), describes a situation in which both qubits are
coupled to the same environment, even though the coupling matrices are still
independent. The Hilbert space is identical to the one for the 2 qubit
spectator configuration. The total Hamiltonian reads
\begin{equation}\label{eq:joinedHam} 
  H_{\lambda_1,\lambda_2}= H_1+H_2+H_\e+
    \lambda_1\; V_{1,\e} + \lambda_2\; V_{2,\e} \; , 
\end{equation}
where $V_{2,\e}$ describes the coupling of the second qubit to the environment.
It is chosen independently from the same random matrix ensemble as $V_{1,\e}$.
\end{itemize}

\section{The spectator Hamiltonian \label{sec:spectator}}

The first step to calculate the decoherence of the initial state
\begin{equation}\label{eq:initialcondtwoqu} 
  \varrho_0= |\psi_{12}\>\<\psi_{12}| \otimes |\psi_{\rm e}\>\<\psi_{\rm e}|, 
\end{equation} 
evolved with the Hamiltonian (\ref{eq:spectatorHam}), is to realize
that the echo operator does not contain $H_2$.  The quantum echo of
$\varrho_0$ after time $t$ is
\begin{equation}\label{eq:twoevolution} 
  \varrho^M(t)= \Bigl[ \openone_2\otimes M_\lambda(t) \Bigr] 
                \varrho_0 
		\Bigl[ \openone_2\otimes M^\dagger_\lambda(t) \Bigr].  
\end{equation} 
Since $\varrho^M(t)$ remains a pure state in $\mcH_1
\otimes \mcH_2 \otimes \mcH_{\rm e}$,
\begin{equation}\label{eq:bothpuritiesequal} 
  P(t)= \tr \rho_{\rm c}(t)^2 = \tr\rho_\e(t)^2 
\end{equation} 
with $\rho_{\rm c}(t)= \tr_\e \varrho^M(t)$ and $\rho_\e(t)= \tr_{\rm c}
\varrho^M(t)$. This simply means that, as a formality, we can calculate purity
of the central system, calculating purity of the environment.  As the echo
operator acts as the identity on the second qubit,
\begin{align}\label{eq:ahora} 
  \rho_\e(t)&= \tr_1  M_\lambda(t) ( \tr_2 \varrho_0 ) M^\dagger_\lambda(t)\\ 
    &= \tr_1  M_\lambda(t) \left( \rho_1\otimes|\psi_\e \>\<\psi_\e| \right) 
      M^\dagger_\lambda(t),\nonumber
\end{align} 
where $\rho_1= \tr_2 |\psi_{12} \>\<\psi_{12}|$.

We may therefore compute the purity of the spectator model, without ever
referring explicitly to the second qubit! Any dependence of the decay of purity
on the central system as a whole is encoded into the initial density matrix
$\rho_1$. This also implies that we can use the results obtained
in~\ref{sec:calculationSpec}, and hence \eref{eq:purityoneq} and
\eref{eq:resajiuno} remain valid. The only difference is that for the
correlation functions $C_1(\tau)$, $S_1(\tau)$, and $S'_1(\tau)$, we now have
to insert the respective expressions which apply for mixed initial states of
the first qubit. These expressions are given in~\ref{aB}. We stress, for later
reference, that this line of reasoning is not limited by the fact that the
interacting and spectator systems are qubits.

\subsection{The GUE case}\label{sec:guetwo}

We wish to write the initial condition in its simplest form. We must respect
the structure of the Hamiltonian (\ref{eq:spectatorHam}). However we can still
take advantage of all its invariance properties, when seen as an ensemble.
Given a fixed $H_1$, that ensemble of Hamiltonians is invariant under local
operations of the form 
\begin{equation} 
U_{N_\e}\otimes\exp{\mimath \alpha H_1} \otimes U_2 
\end{equation} 
where $U_{N_\e}\in\mcU(N_\e)$ is any unitary operator acting on the
environment, $\alpha$ a real number, and $U_2\in\mcU(2)$ is any unitary
operator acting on the second qubit.

\begin{figure}[t] 
\begin{center}
\includegraphics{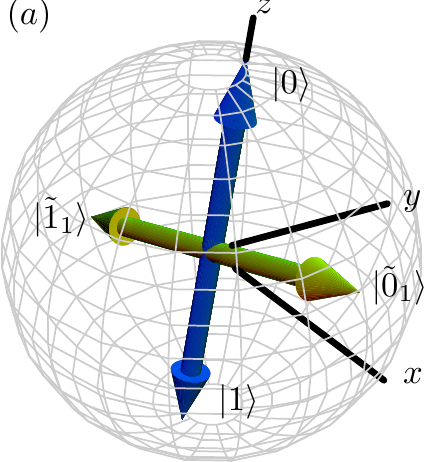}\hfill
\includegraphics{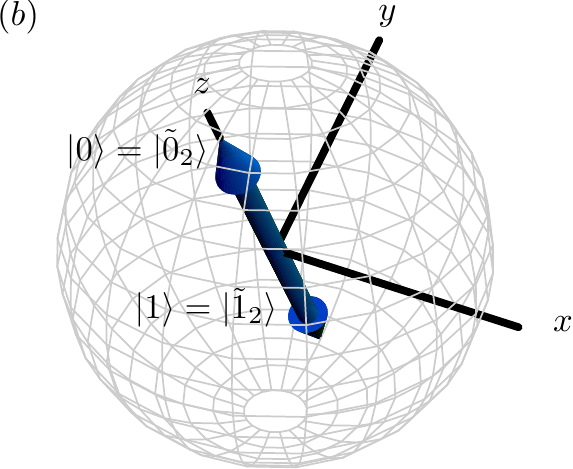} 
\end{center} 
\caption{A figure to visualize the way the initial condition is parametrized,
  using the Bloch sphere, is presented. On the left, qubit one has an
  internal Hamiltonian.  Its eigenvectors ($|0\>$ and $|1\>$) are represented
  in blue.  The $z$ axis is {\textit chosen} parallel to $|0\>$.  The $x$
  axis is chosen so that  both $|\tilde 0_1\>$ and $|\tilde
  1_1\>$ have real coefficients; \ie such that the $xz$ plane contains
  $|\tilde 0_1\>$ and $|\tilde 1_1\>$. On the right we represent the second
  qubit. We have absolute freedom to choose the basis (even if an internal
  Hamiltonian is present), and thus we choose it according to the natural
  Schmidt decomposition.}
\label{fig:initGUEdos}
\end{figure}

The freedom within the qubits allows to choose a basis $\{|0\>, |1\>\}
\otimes \{|0\>, |1\>\}$ in which the initial state [see
\eref{eq:initialcondtwoqu}] can be written as
\begin{equation} \label{eq:initialguewitness}
|\psi_{12}\>= \cos\theta(\cos\phi|0\>+\sin\phi|1\>)|0\>
             +\sin\theta(\sin\phi|0\>-\cos\phi|1\>)|1\>, 
\end{equation} 
and still, $H_1=\frac{\Delta}{2}|0\>\<0|-\frac{\Delta}{2}|1\>\<1|$ is
diagonal.  Let us prove it in a constructive manner. To find this basis we
start using the Schmidt decomposition to write 
\begin{equation}\label{eq:schmidt}
|\psi_{12}\>=\cos\theta|\tilde 0_1 \tilde 0_2\> 
          +  \sin\theta|\tilde 1_1 \tilde 1_2\> 
\end{equation} 
with $\{ |\tilde 0_i \>, |\tilde 1_i \> \}$ being an
orthonormal basis of particle $i$. For the first qubit, we fix the $z$ axis of
the Bloch sphere (containing both $|0\>$ and $|1\>$) parallel to the
eigenvectors of $H_1$, and the $y$ axis perpendicular (in the Bloch sphere) to
both the $z$ axis and $|\tilde 0_1 \>$. The states contained in the $xz$ plane
are then real superpositions of $|0\>$ and $|1\>$, which implies that
\begin{align} 
  |\tilde 0_1\> &=\cos\phi|0\>+\sin\phi|1\>,\\ |\tilde 1_1\> 
  &= \sin\phi|0\>-\cos\phi|1\>, 
\end{align} 
for some $\phi$.  In the second qubit it is enough to set 
\begin{align} 
  |\tilde 0_2\> &= |0\>,\\ 
  |\tilde 1_2\> &= |1\>.
\end{align} 
This freedom is also related to the fact that purity only depends on $\tr_2
|\psi_{12}\> \<\psi_{12}|$. A visualization of this argumentation is found in
\fref{fig:initGUEdos}. The angle $\theta \in [0,\pi/4]$ measures the
entanglement 
\begin{equation} 
  C(|\psi_{12}\>\<\psi_{12}|) = \sin2\theta
  \label{eq:relationthetaconc} 
\end{equation} 
whereas the angle $\phi \in [0,\pi/2]$ is related to an initial magnetization.

\begin{figure}[t] \centering 
\includegraphics[width=\textwidth]{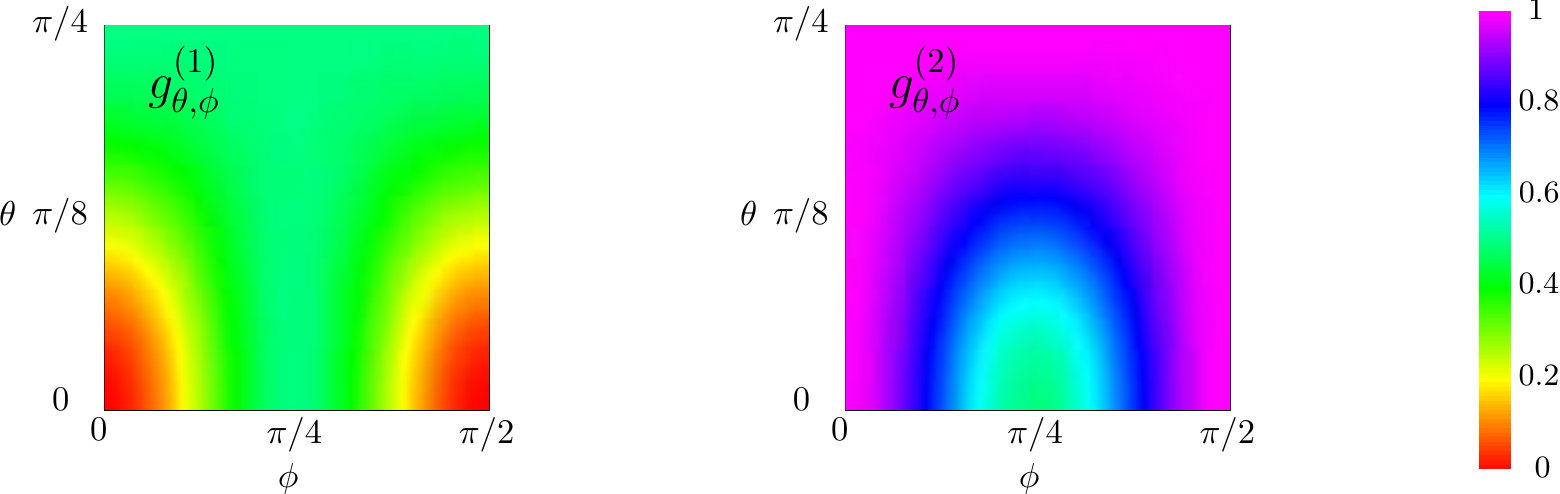}
\caption{Visualization of the geometric factors $g_{\theta,\phi}^{(1)}$, and
  $g_{\theta,\phi}^{(2)}$ from left to right respectively. For
  $g_{\theta,\phi}^{(1)}$ we see that for pure eigenstates of $H_1$ its value
  is zero. This leads to a higher qualitative stability of this kind of
  states.}
\label{fig:gsfig} 
\end{figure}

The general solution for purity using this parametrization is
\begin{equation}\label{eq:generalGUEtwo} 
  P(t)= 1 -4\lambda^2
   \int_0^t\rmd\tau\int_0^\tau\rmd\tau' \bar C(\tau')
   \big [g^{(1)}_{\theta,\phi} + g^{(2)}_{\theta,\phi} \cos\Delta\tau' \big ] 
+ \Or(\lambda^4, N_{\rm e}^{-1}), 
\end{equation} 
where the geometric factors $g^{(1)}_{\theta,\phi}\in [0,1/2]$, and
$g^{(2)}_{\theta,\phi} \in [1/2,1]$ are expressed as 
\begin{eqnarray} 
  g^{(1)}_{\theta,\phi}&= g_\theta (1-g_\phi) + g_\phi (1-g_\theta)\\ 
  g^{(2)}_{\theta,\phi}&= 2(1-g_\theta) - g_\phi(1-2g_\theta), 
\end{eqnarray} 
in terms of the functions $g_\phi$ and $g_\theta$, defined in \eref{eq:defg}.
Both geometric factors are shown in \fref{fig:gsfig}. \Eref{eq:generalGUEtwo}
is obtained from \eref{eq:purityoneq} and \eref{eq:resajiuno} by insertion of
the \eref{aB:ReC1} and \eref{aB:S1mixed} for ${\rm Re}\, C_1(\tau)$ and
$S_1(\tau)$, respectively.

We consider again two limits for $\Delta$. In the degenerate limit ($\Delta\ll
1/\tau_\rmH$) purity decay is given by
\begin{equation}\label{eq:spectatorDegenerate} 
P_{\rm D}(t)=1-\lambda^2(2-g_\theta)f_{\tau_\rmH}(t), 
\end{equation} 
where $f_{\tau_\rmH}(t)$ is defined in \eref{eq:deff}.  The result is
independent of $\phi$ since a degenerate Hamiltonian is, in this context,
equivalent to no Hamiltonian at all.  The $\theta$-dependence in this formula
shows that an entangled qubit pair is more susceptible to decoherence than a
separable one.

In the fast limit ($\Delta\gg 1/\tau_\rmH$) we get
\begin{equation}\label{eq:spectatorFast} 
P_{\rm F}(t)=1-\lambda^2\left[g^{(1)}_{\theta,\phi}f_{\tau_\rmH}(t) 
                           + 2\tau_\rmH g^{(2)}_{\theta,\phi} t \right] .  
\end{equation}

For initial states chosen as eigenstates of $H_1$ we find linear decay of
purity both below and above Heisenberg time. In order for $\rho_1$ to be an
eigenstate of $H_1$ it must, first of all, be a pure state (in $\mcH_1$).
Therefore this behavior can only occur if $\theta=0$ or $\theta= \pi/2$.  Apart
from that particular case, we observe in both limits, the fast as well as the
degenerate limit, the characteristic linear/quadratic behavior before/after the
Heisenberg time similar to the one qubit case.

\begin{figure}[t] \centering
\includegraphics[width=.6\textwidth]{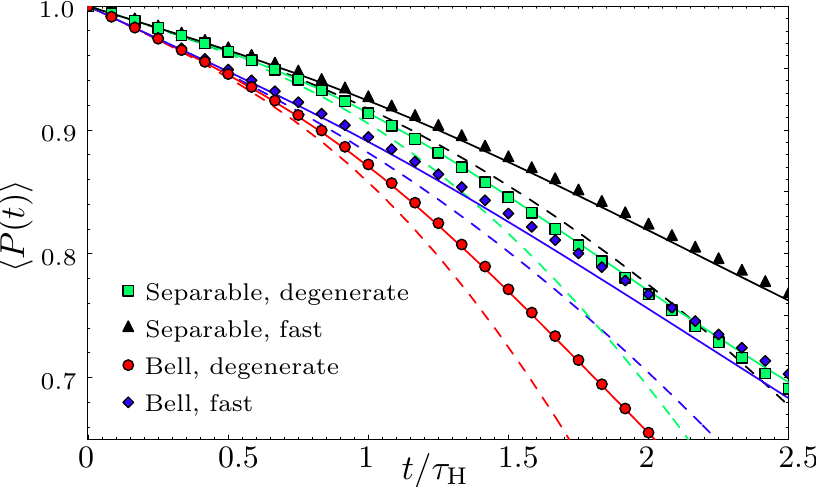} 
\caption{Numerical simulations for the average purity as a function of time in
  units of the Heisenberg time of the environment (spectator configuration, GUE
  case).  For the coupling strength $\lambda= 0.03$ we show the dependence of
  $\< P(t)\>$ on the level splitting $\Delta$ in $H_1$ and on the initial
  degree of entanglement between the two qubits (in all cases $\phi= \pi/4$).
  For separable state ($\theta= 0$) and a degenerate limit we use {\huepfour
  $\blacksquare$} whereas for the fast limit, with $\Delta=8$, we use
  $\blacktriangle$; For Bell states, the corresponding limits are encoded as
  {\huezero $\bullet$} in the degenerate limit and {\huepseven
  \protect\rotatebox{45}{\protect\scalebox{0.7071}{$\blacksquare$}}} in the
  fast limit ($\Delta=0.8$).  The corresponding linear response results (dashed
  lines) are based on \eref{eq:spectatorDegenerate}, \eref{eq:spectatorFast}
  and \eref{eq:exponentiationone}.  The exponentiated linear response results
  (solid lines) are obtained based on the results of appendix
  \ref{sec:exponentiation}.  The theoretical curves are plotted with the same
  color, as the respective numerical data.  In all cases $N_\rme=1024$.} 
\label{fig:holetwo} 
\end{figure}

In \fref{fig:holetwo} we show numerical simulations for $\< P(t)\>$.  We
average over 30 different Hamiltonians each probed with 45 different initial
conditions. We contrast Bell states ($\phi=\pi/4$, $\theta=\pi/4$) with
separable states ($\phi=\pi/4$, $\theta= 0$), and also systems with a large
level splitting ($N_\rme\gg \Delta= 8\gg 1/\tau_\rmH$) with systems having a
degenerate Hamiltonian ($\Delta= 0$). The results presented in this figure
show that entanglement generally enhances decoherence.  This can be
anticipated from \fref{fig:gsfig}, since for fixed $\phi$, increasing the
value of $\theta$ (and hence entanglement) increases both
$g^{(1)}_{\theta,\phi}$ and $g^{(2)}_{\theta,\phi}$. At the same time we find
again that increasing $\Delta$ tends to reduce the rate of decoherence. A
strict inequality only holds among the two limiting cases (just as in the one
qubit case): From $g^{(2)}_{\theta,\phi}=2-g^{(1)}_{\theta,\phi}-g_\theta$,
it follows that $(P_{\rm F}-P_{\rm D})/\lambda^2=g^{(2)}_{\theta,\phi}
[f_{\tau_\rmH}(t) -2t\tau_\rmH] \ge 0$. Therefore, for fixed initial
conditions and $t$ greater than 0, $P_{\rm F}(t) > P_{\rm D}(t)$. This is the
second aspect illustrated in \fref{fig:holetwo}.

In order to extend the formulae to longer times/smaller purities we
exponentiate them using the results of appendix \ref{sec:exponentiation}. The
numerical simulations (see \fref{fig:holetwo}) agree very well with that
heuristic exponentiated linear response formula. In one case (blue rhombus)
where the agreement is not as good, we found that it is the inaccurate
estimate $P_\infty= 1/4$ of the asymptotic value of purity, which leads to
the deviations.

\subsection{The GOE case}\label{sec:goetwo}

Let us consider the GOE average of both  $H_\e$ and $V_{1,\e}$. When
averaging $H_\e$ and $V_{1,\e}$ over the GOE, we are again confronted with
the fact that the invariance group is considerably smaller than in the GUE
case. In this context the initial entanglement between the two qubits has a
crucial importance since it ``transports'' the invariance properties from the
spectator to the coupled qubit.

For the sake of simplicity, we focus on the degenerate limit setting $H_1=0$.
Note that on the basis of the results in appendix \ref{sec:calculationSpec},
the general case can be treated similarly and the corresponding result will be
presented at the end of this subsection.

We first specify the operations under which the spectator Hamiltonian
\eref{eq:spectatorHam}, considered as a random matrix ensemble, is invariant.
As both, the internal Hamiltonian of the environment and the coupling, are
selected from the GOE the invariance operations form the group 
\begin{equation}\label{eq:invariancegoe} 
  \mcO (N_\e) \times \mcO(2) \times \mcU(2),
\end{equation} 
and have the structure $O_{N_\e} \otimes \exp (\mimath \alpha \sigma_y)
\otimes U_2$, with $O_{N_\e}$ being an orthogonal matrix (acting in
$\mcH_\e$), $\alpha$ a real number, and $U_2$ a unitary operator acting on
the spectator qubit.  The direct product structure of the invariance group
obliges us to respect the identity of each particle, but allows to analyze
each qubit separately. For instance, if we replace the random coupling
matrix $V_{1,\e}$ with one which involves both qubits, the invariance group
would be $ \mcO (N_\e) \times \mcO(4)$. As a consequence, purity decay would
become independent of the entanglement within the qubit pair: for any
entangled state one can find an orthogonal matrix which maps the state onto a
separable one \footnote{One can see a general state
$|\psi_{12}\>=\sum_{i=0}^3(a_i +\mimath b_i )|i\>$, $a_i,b_i\in \mathbb{R}$
as characterized by 2 real vectors $ \vec{a}=(a_0,a_1,a_2,a_3),
\vec{b}=O(b_0,b_1,b_2,b_3) \in \mathbb{R}^4$.  We can rotate $\vec{a}$ to a
vector having only the first component $\vec{a}'=(a_0',0,0,0)$ using an
orthogonal matrix $O$. Consider the vector $\vec{b}'=O\vec{b}$ rotated by the
same transformation.  We next repeat the procedure on the last 3 axes (for
$\vec{b}'$) with an orthogonal transformation $O'$, to obtain a vector having
only the first two components. We thus have, an orthogonal transformation
that takes $|\psi_{12}\>$ to $O'O|\psi_{12}\>=\sum_{i=0}^1(a_i^{''} +\mimath
b_i{''}  )|i\> = |0_1\>(\alpha |0_2\>+\beta |1_2\>$, a separable state.}.

We write the initial condition $|\psi_{12}\>$ as in \eref{eq:schmidt}. For the
coupled particle follows the same analysis made in \sref{sec:goeone}.  We can
thus write $|\tilde{0}_1\>=2^{-1/2}[|0\>+ \exp(\mimath \gamma)|1\>]$ and, in
order to respect orthogonality, $|\tilde{1}_1\>=2^{-1/2}\exp(\mimath
\zeta)[|0\>- \exp(\mimath \gamma)|1\>]$.  For the second qubit we have the same
complete freedom as in \eref{sec:guetwo}.  We thus select $|\tilde{0}_2\>=|0\>$
and $|\tilde{1}_2\>=\exp(-\mimath \zeta)|1\>$ to erase the relative phase in
the first qubit and finally write the initial state as 
\begin{equation}
  |\psi_{12}\>= \frac{
    \cos \theta (|0\>+ e^{\mimath \gamma} |1\>) |0\> +
    \sin \theta  (|0\>- e^{\mimath \gamma} |1\>) |1\>}{\sqrt{2}}.  
\label{eq:psi12goe}
\end{equation} 
The average purity is still given by the double integral
expression in \eref{eq:purityoneq}. However, in the present case the mixed
initial state $\rho_1= {\rm tr}_2\, |\psi_{12}\> \<\psi_{12}|$ must be used.
For $\Delta=0$, the resulting integrand reads 
\begin{equation} 
  A_{\rm JI}(\tau,\tau')= 
    (2-g_\theta)\bar C(|\tau-\tau'|)+1-g_\theta+ (2g_\theta -1)\sin^2\gamma, 
\end{equation} 
where $\bar C(|\tau-\tau'|)$ is given in
\eref{eq:thecorrelation}.  Evaluating the double integral, we obtain
\begin{equation}\label{eq:purityGOEone} 
  \< P(t)\>=1-\lambda^2 \left\{ t^2 [4 -2 \cos^2(2\theta)\cos^2 \gamma] 
        +(4-2 g_\theta) \left[ t \tau_\rmH - B_2^{(1)}(t) \right] \right\} , 
\end{equation} 
where $B_2^{(1)}(t)$ is given in \eref{eq:Btwogoe}.  As in the GUE case, this
result depends on the entanglement of the initial state and, as in the
one-qubit GOE case, it also depends on $\gamma$.  Again it turns out that Bell
states are more susceptible to decoherence than separable states. Note however,
that purity as a function of $\theta$ is not monotonous. Hence, a finite
increase of entanglement does not guarantee that the purity decreases
everywhere in time. For separable states, $g_\theta= 1$, we retrieve formula
\eref{eq:purityGOEsep}. However, for completely entangled states, $\theta=
\pi/4$ the dependence on $\gamma$ is lost. This is understood from a physical
point of view, noticing that any local unitary operation  on a Bell state can
be reduced to a single local unitary operation acting on a single qubit. In
other words, given  $U_{1,2}=U_1 \otimes U_2$, there exists a unitary $U_1'$
such that 
\begin{equation}\label{eq:bilocaltolocal} 
  U_1 \otimes U_2 |{\rm Bell}\>= U_1' \otimes \openone |{\rm Bell}\> 
\end{equation} 
($|\rm{Bell}\>$ is any 2 qubit pure state with
$C=1$, e.g. $|00\>+|11\>$).  We can then say that the invariance properties in
the coupled qubit are inherited from the spectator qubit via entanglement.

Let us obtain the standard deviation for the different possible initial
conditions in the qubits. We want to analyze the situation separately for a
fixed value of concurrence. Then, as the invariant measure of the ensemble of
initial conditions, and fixing the amount of entanglement, we shall use the
tensor product of the invariant measures in each of the qubits. Since there is
no dependence of \eref{eq:purityGOEone} on the second qubit, the appropriate
invariant measure is trivially inherited from the invariant measure for a
single qubit.  The resulting value for the standard deviation is
\begin{equation} 
  \sigma_P=\frac{4 \lambda^2 t^2 \cos^2(2\theta)}{3 \sqrt{5}}.
  \label{eq:sigmados}
\end{equation}

Based on the appendix \ref{sec:calculationSpec} we can also obtain the average
purity for $\Delta\ne 0$. The parametrization of the initial states is more
complicated since two preferred directions arise, one from the eigenvectors of
the internal Hamiltonian and the other from the invariance group.  The result
can be expressed in the form given in \eref{eq:purityoneq}, with
\begin{multline} 
  {\rm Re}\, A_{\rm JI}(\tau,\tau')= \bar C(|\tau-\tau'|)\; \big
[\, g^{(1)}_{\theta,\phi} + g^{(2)}_{\theta,\phi}\; \cos\Delta(\tau-\tau')\,
\big ] \nonumber\\ + g_{\theta,\phi}^{(1)} - (1-g_{\theta,\phi}^{(2)})\;
\cos[\Delta (\tau+\tau') -2\eta]  + \Or(\lambda^4, N_{\rm e}^{-1})\; .
\end{multline} 
The angle $\eta$ is related to a phase shift between the components of any of
the eigenvectors of the initial density matrix $\rho_1$.  Here we can again see
the term containing a sum of times, \ie a term that is not a correlation
function. However, this term is small, and adequate values must me chosen to
see its effect. Still it is observable in numerical simulations, with moderate
effort.

\section{The separate environment Hamiltonian} \label{sec:separateTwo} We
proceed to study purity decay with other configurations of the environment.
Consider the separate environment configuration, pictured in
\fref{fig:schemeconfig}(b).  The corresponding uncoupled Hamiltonian is
\begin{equation}\label{eq:sehamone} 
  H_0 = H_1+H_2+H_{\e}+H_{\e'}
\end{equation}
and the coupling is 
\begin{equation}\label{eq:sehamdos} 
  \lambda V=\lambda_1 V_{\e,1}+\lambda_2 V_{\e',2}. 
\end{equation} 
From now on we assume that the
internal Hamiltonians of the environment and the couplings are chosen from the
GUE. The generalization for the GOE can be obtained along the same lines using
the corresponding results of \sref{sec:goetwo}.  The initial condition has a
separable structure with respect to both environments, see
\eref{eq:initialconditiongeneral}. The coupling in the interaction picture
separates into two parts acting on different subspaces $\lambda \tilde
V=\lambda_1 \tilde V^{(1)}+ \lambda_2 \tilde V^{(2)}$, where
\begin{equation}\label{eq:twopersep} 
  \tilde V^{(1)}=\rme^{\mimath (H_1 +H_{\e})} V_{\e,1} 
     \rme^{-\mimath (H_1 +H_{\e})},\quad 
   \tilde V^{(2)}=\rme^{\mimath (H_2 +H_{\e'})} V_{\e',2} 
     \rme^{-\mimath (H_2 +H_{\e'})}.
\end{equation} 
Notice that $\tilde V^{(1)}$ ($\tilde V^{(2)}$) does not depend
on $H_{\e'}$ ($H_{\e}$). Since $V^{(1)}$ and $V^{(2)}$ are uncorrelated,
quadratic averages separate as 
\begin{equation}\label{eq:serpperint} 
  \lambda^2 \<\tilde V_{ij} \tilde V_{kl}\> =
     \lambda_1^2 \<\tilde V^{(1)}_{ij} \tilde V^{(1)}_{kl}\> 
    +\lambda_2^2 \<\tilde V^{(2)}_{ij} \tilde V^{(2)}_{kl}\>.
\end{equation} 
This leads to a natural separation of each of the contributions
to purity 
\begin{equation} 
  1- \< P(t)\> = 1- P_{\rm spec}^{(1)}(t) + 1 - P_{\rm spec}^{(2)}(t) , 
\end{equation} 
where $P_{\rm spec}^{(i)}(t)$ denotes the
average purity with particle $i$ being a spectator, as given in
\sref{sec:spectator}. In this way, the problem formally reduces to that of the
spectator model. The respective expressions in \sref{sec:spectator} may be
used. For instance, if we assume broken TRI, we obtain from
\eref{eq:generalGUEtwo} 
\begin{equation}\label{eq:sepenvfull} 
\< P(t)\>=1-4 \sum_{i=1}^{2} 
  \lambda_i^2\intoh \left[g^{(1)}_{\theta,\phi_i}
    +g^{(2)}_{\theta,\phi_i}\cos\Delta_i\tau'\right] \bar C_i(\tau') +
  \Or(\lambda^4,N_\e^{-1}), 
\end{equation} 
where $\bar C_1$ and $\bar C_2$ are
the correlation functions of the corresponding environments defined in exact
correspondence with \eref{eq:thecorrelation}, for $H_\e$ and $H_{\e'}$
respectively.  If in one or both of the qubits, the level splitting in the
internal Hamiltonians is very large/small compared to the Heisenberg time in
the corresponding environment (denoted by $\tau_\e$ and $\tau_{\e'}$ for $H_\e$
and $H_{\e'}$, respectively) the degenerate and/or fast approximations may be
used.  As an example, if $\Delta_1 \ll 1/\tau_\e$ and $\Delta_2 \ll
1/\tau_{\e'}$ we find 
\begin{equation}\label{eq:DeglimitSep} 
  P_{\rm D}(t)=
    1-(2-g_\theta)(\lambda_1^2 f_{\tau_\e}(t)+\lambda_2^2 f_{\tau_{\e'}}(t)),
\end{equation} 
whereas if $\Delta_1 \gg 1/\tau_\e$ and $\Delta_2 \gg 1/\tau_{\e'}$ 
\begin{equation}\label{eq:FastlimitSep} 
  P_{\rm F}(t)=1- 
  \lambda_1^2\left[g^{(1)}_{\theta,\phi_1}f_{\tau_\e}(t)+
     2\tau_\e g^{(2)}_{\theta,\phi_1} t \right] -
  \lambda_2^2\left[g^{(1)}_{\theta,\phi_2}f_{\tau_{\e'}}(t)
    +2\tau_{\e'} g^{(2)}_{\theta,\phi_2} t \right].  
\end{equation} 
It is interesting to note that if we have two separate but equivalent
environments (\ie both Heisenberg times are equal), we get exactly the same
result as for a single environment.  Also notice that the Hamiltonian of the
entire system separates and thus the total entanglement of the two subsystems
($\mcH_1 \otimes \mcH_\e$ and $\mcH_2 \otimes \mcH_{\e'}$) becomes time
independent.

\section{The joint environment configuration} \label{sec:jointTwo} 

The last configuration we shall consider is the one of joint environment; see
\fref{fig:schemeconfig}(c).  Its uncoupled Hamiltonian is
\begin{equation}\label{eq:johamu} 
  H_0 = H_1+H_2+H_{\e} 
\end{equation} 
whereas
the coupling is given by 
\begin{equation}\label{eq:johamc} 
  \lambda V=\lambda_1 V_{\e,1}+\lambda_2 V_{\e,2}.  
\end{equation} 
Notice the similarity with
\eref{eq:sehamone} and \eref{eq:sehamdos}.  However, as discussed in the
introduction, they represent very different physical situations. The coupling
in the interaction picture can again be split $\lambda \tilde V=\lambda_1
\tilde V^{(1)}+ \lambda_2 \tilde V^{(2)}$, where 
\begin{equation} 
  \tilde V^{(1)}=
    \rme^{\mimath (H_1 +H_{\e})} V_{\e,1} \rme^{-\mimath (H_1 +H_{\e})},
  \quad 
  \tilde V^{(2)}=
    \rme^{\mimath (H_2 +H_{\e})} V_{\e,2}\rme^{-\mimath (H_2 +H_{\e})}.  
\end{equation} 
Note the slight difference with
\eref{eq:twopersep}. However $V^{(1)}$ and $V^{(2)}$ are still uncorrelated,
enabling us to write again \eref{eq:serpperint}.

From  now  on,  the  calculation  is  formally the  same  as  in  the separate
environment  case.  Hence  we  can  inherit  the  result \eref{eq:sepenvfull}
directly, taking into  account that since they  come from the  same
environmental Hamiltonian, the  two correlation functions are the same. In case
any of the Hamiltonians fulfills the fast or degenerate limit conditions, the
corresponding expressions to \eref{eq:DeglimitSep} and \eref{eq:FastlimitSep}
can be written. As an example, if the first qubit  has no internal Hamiltonian,
and the second  one has a big energy difference, the resulting expression for
purity decay is 
\begin{equation}\label{eq:joinedFast} 
  \< P(t)\> =1-
    \lambda_1^2 (2-g_\theta)f_{\tau_\rmH}(t) -
    \lambda_2^2\left[g^{(1)}_{\theta,\phi_2}f_{\tau_\rmH}(t) 
           +2\tau_\rmH g^{(2)}_{\theta,\phi_2} t \right] 
    + \Or(\lambda^4,N_\e^{-1}), 
\end{equation}
where $\tau_\rmH$ is the Heisenberg time of the joint environment.  Monte Carlo
simulations  showing the  validity of  the result  were done  with satisfactory
results, comparable to those  obtained in \fref{fig:holetwo}. The parameter
range checked was similar to that in the figure.

%
%
%

\chapter{Entanglement decay}\label{sec:concurrence}


In the previous chapter we studied purity decay of two qubits.  Purity
measures entanglement with the environment, but we can wonder how decoherence
affects the internal quantum mechanical properties of the central system.
Possibly the most important quantum mechanical property of a multi-particle
system is its internal entanglement.  As discussed in appendix
\ref{sec:entappendix}, a simple and meaningful measure of two qubit
entanglement is concurrence ($C$).  Since concurrence is defined in terms of
the eigenvalues of a Hermitian $4\times4$-matrix, an analytical treatment,
even in linear response approximation, is much more involved than in the case
of purity.  A study along the same lines followed in the last chapters, is
out of reach for the time being.

We shall first explore a relation (first found in \cite{pineda:012305}, 
partly explained in \cite{ziman:052325}, and further studied in
\cite{pinedaRMTshort}) between concurrence and purity. We show that this
relation is valid in a wide parameter range (\sref{sec:cpplane}).  Combining
it with an appropriate formula for purity decay,  we obtain an analytic
prediction for concurrence decay in \sref{sec:concdecay}. We compare
our prediction with Monte Carlo simulations. It is essential
that the two qubits do not interact; otherwise the coupling between the
qubits would act as an additional sink (or source) for internal entanglement
-- a complication we wish to avoid.  In the last part of the section we
extend our ideas to non-Bell states.

\section{The $CP$ plane}\label{sec:cpplane} 

\begin{figure}
\centering \includegraphics[height=4cm]{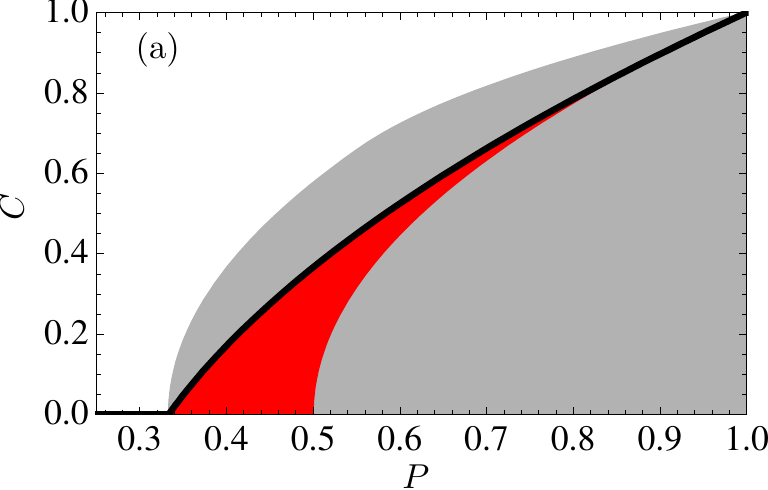}\hfill
\includegraphics[height=4cm]{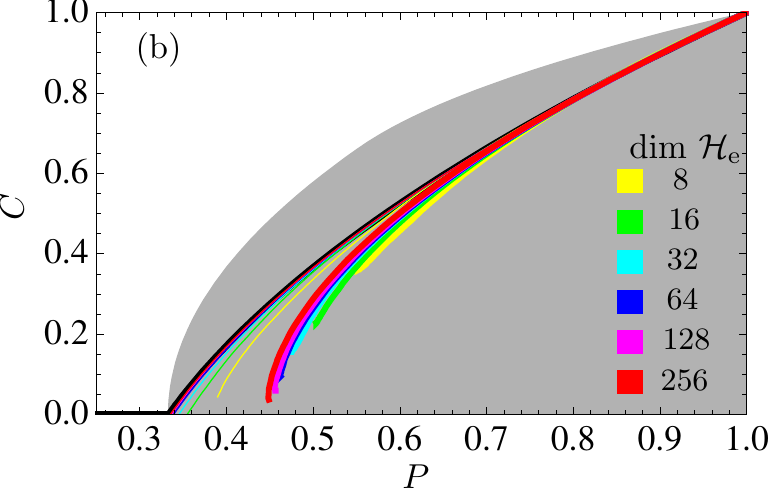}
\caption{We present the $CP$ plane. In (a) we show the area of
  concurrence-purity ($CP$) combinations which are allowed for
  physical states (gray area plus the set $\{(0,P),
  P\in[1/4,1/3]\}$). The image of a Bell state under the set of local
  unital operations defines the red area.  The Werner curve
  \eref{eq:goodexponwernerCtime}  is shown as a thick black solid
  line.  In (b) we show curves $(\<C(t)\>,\<P(t)\>)$ as obtained from
  numerical simulations of the spectator Hamiltonian. GUE matrices
  are used during all numerical experiments in this section. We
  choose $\Delta=1$ and vary $\dim(\mcH_\e)$ for two different
  coupling strengths $\lambda=0.02$ (thick lines) and $\lambda=0.14$
  (thin lines). The resulting curves are plotted with different
  colors, according to $\dim(\mcH_\e)$ as indicated in the figure
  legend.}
\label{fig:exampleacuuWerner_one}
\end{figure}

We study the relation between concurrence and purity using the $CP$-plane,
where we plot concurrence against purity with time as a parameter.  This
plane is plotted in \fref{fig:exampleacuuWerner_one}. The gray area indicates
the region of physically admissible states \cite{munro:030302}.  The upper
boundary of this region is given by the maximally entangled mixed states.
This set depends on the entanglement measures chosen \cite{wei:022110}. Large
purity, \ie low entanglement with the environment, is required in order to
have large concurrence, \ie entanglement within the pair. This is a
consequence of the monogamy of entanglement.  Note also that concurrence
becomes identically zero before purity reaches its minimum at $1/4$.  Another
region of interest, plotted in red in \fref{fig:exampleacuuWerner_one}(a),
corresponds to those states (density matrices), which form the image of a
Bell state under the set of local (\ie acting separately on each qubit)
unital operations \cite{ziman:052325}.  Unital operations are those which
preserve the identity \cite{Keyl2002}.  Single qubit unital operations
include bit flip and phase flip, whereas an example of a non-unital operation
is amplitude damping \cite{NC00a}.  Finally, the Werner states $\rho_{\rm
W}=\alpha \frac{\openone}{4}+ (1-\alpha)|\rm{Bell}\>\<\rm{Bell}|$, $0\le
\alpha\le 1$, define a smooth curve on the $CP$-plane (black solid line).
The analytic form of this curve is \cite{pinedaRMTshort}
\begin{equation}\label{eq:goodexponwernerCtime}
C_{\rm W}(P)=\max\left\{0, \frac{\sqrt{12P-3}-1}{2}\right\},
\end{equation}
and will be referred to as the Werner curve. Note that states mapped to the
Werner curve are not necessarily Werner states.

In \fref{fig:exampleacuuWerner_one}(b), we perform numerical simulations in
the spectator configuration assuming broken time reversal symmetry (GUE
case).  We compute the average concurrence for a given interval of purity
using 15 different Hamiltonians and 15 different initial states for each
Hamiltonian. We fix the level splitting in the coupled qubit to $\Delta=1$
and consider two different values $\lambda=0.02$ and $0.14$ for the coupling
to the environment.  \Fref{fig:exampleacuuWerner_one}(b) shows the resulting
$CP$-curves for different dimensions of $\mcH_\e$.  Observe that for both
values of $\lambda$, the curves converge to a certain limiting curve as
$\dim(\mcH_\e)$ tends to infinity. While for $\lambda=0.02$, this curve is at
a finite distance of the Werner curve, for $\lambda=0.14$ it practically
coincides with $C_{\rm W}(P)$.  Varying the configuration, the coupling
strength, the level splitting, or the ensemble (GOE/GUE), gives the same
qualitative results in the $CP$ plane, for large dimensions. In some cases we
have an accumulation towards the Werner curve, in others there is a small
variation, but staying in the unital area.

\begin{figure}
\begin{center} \includegraphics{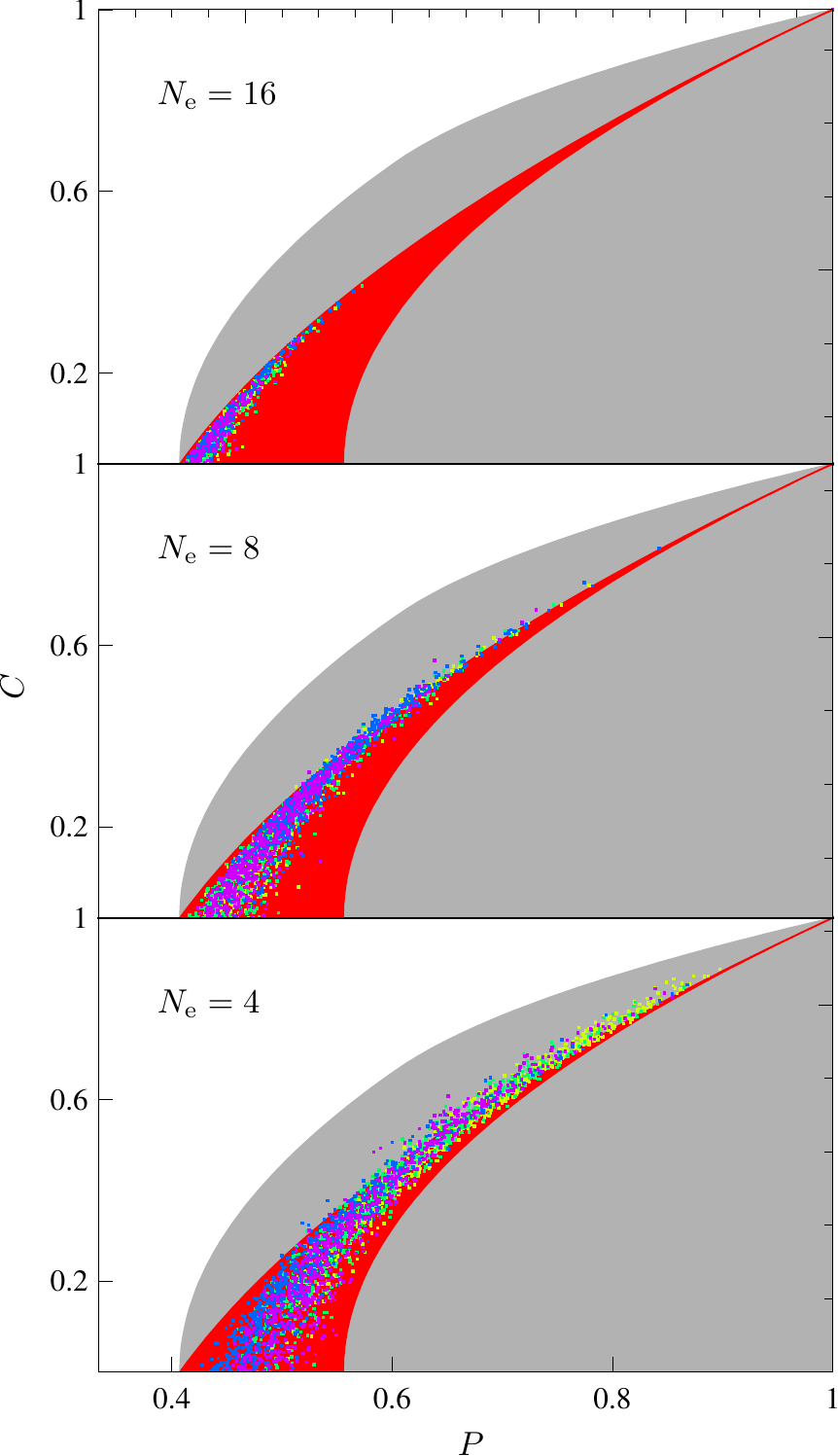} \end{center}
\caption{We show the points reached in the $CP$ plane after the evolution
  with different sizes of the environment. Four initial conditions (with
  different colors) are studied per environment size. For very small
  environments the evolution can reach regions outside the unital region.
  Here we use the spectator configuration, $\delta=0.2$, and $\Delta=0$  with
  the GUE model.}
\label{fig:CPvaryNenv}
\end{figure}

Are the channels induced by our RMT model (RMT channels) typically unital? If
so, we could use the results of \cite{ziman:052325} to explain our numerical
findings. The fact that our RMT channels map Bell states to the unital region
only is not enough to ensure unitality.  Non-unital channels can also map
Bell states into the unital region. The amplitude damping channel acting on a
single qubit provides the simplest example. We now examine the question in
more detail.

For small dimensions the channels are clearly not unital as can be inferred
from \fref{fig:CPvaryNenv}. Bell states are mapped, after the action of our
RMT channels, to the non unital region. The probability of these events,
however, decrease rapidly with the environment size.  Within numerical
accuracy, the probability of being mapped outside the unital region, for
$\delta=0.2$ and a time range between 0 and 4000 is $\approx 10^{-5}$ already
for $N_\rme=32$.  For larger dimensions  a tighter test must be performed as
virtually all points in the $CP$ plane are in the unital region.

\begin{figure}
\centering \includegraphics{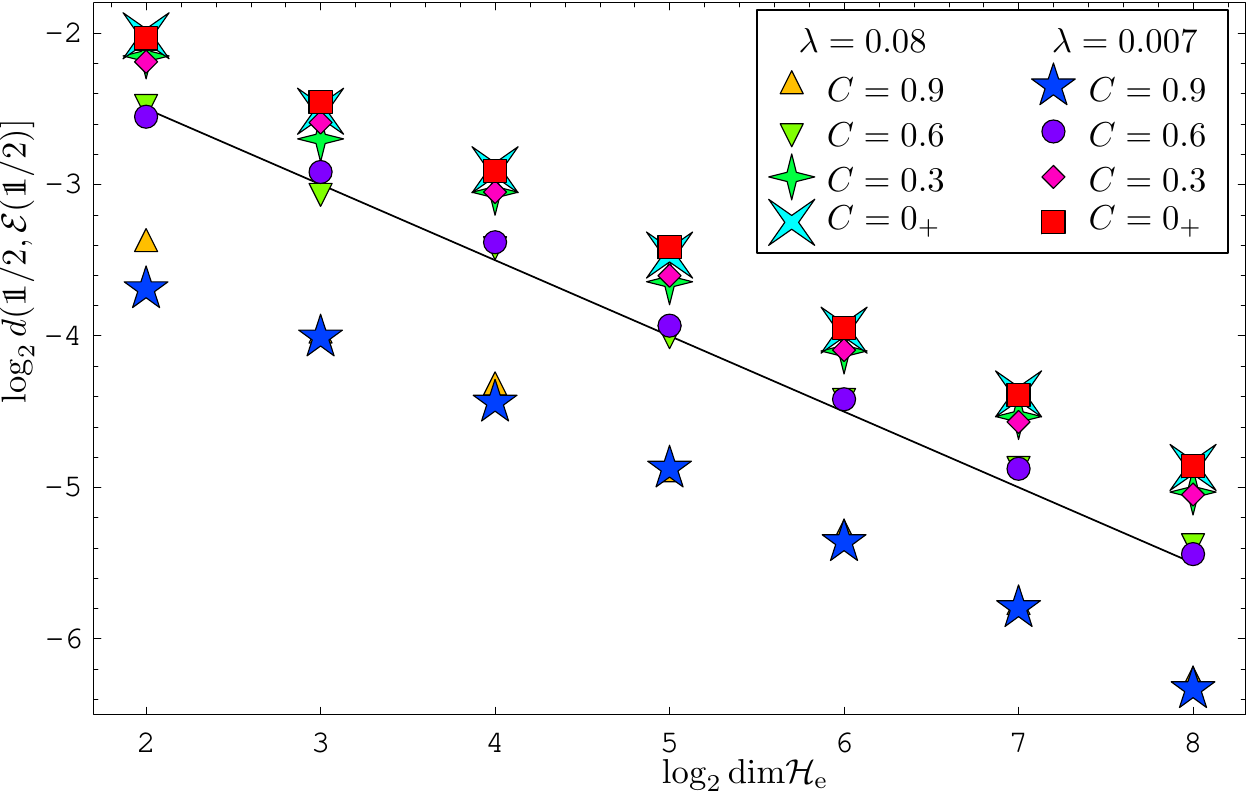}
\caption{ We evaluate the unitality condition for the spectator
  configuration (\ie on one of the qubits). We vary the size of the
  environment and test different times, that would lead approximately
  to different values of concurrence as shown in the inset. See text
  for details.  $C=0_+$ refers to the value a time step immediately
  before $C=0$. A line with slope $-1/2$ is included for comparison.}
\label{fig:theplothatgorinwanted}
\end{figure}

We want to test how close to unitality the RMT channels $\mathcal{E}_\rmRMT$
are.  The channels depend on the particular Hamiltonian chosen from the
ensemble, on time, and on the initial condition in the environment.  As we
want to test unitality of the channel on a single qubit, we can think of the
spectator configuration, or the single qubit model.  Since we want check how
closely $\mathcal{E}_\rmRMT(\openone)=\openone$, we prepare an initial pure
condition (in the qubit plus environment) that leads to a completely mixed
state in the qubit, \ie,
\begin{equation}
\frac{|0 \psi_\e^{(0)} \> + |1 \psi_\e^{(1)} \>}{\sqrt{2}}
\end{equation}
with $\< \psi_\e^{(i)} | \psi_\e^{(j)} \>=\delta_{ij}$. We let the state
evolve with a particular member of the ensemble of Hamiltonians defined in
\eref{eq:hamiltonianqubit}. Afterwards we evaluate the Euclidean distance
$d(\cdot,\cdot)$ in the Bloch sphere, of the points corresponding to the
resulting mixed state in the qubit and the fully mixed state. For unital
channels this distance should be exactly zero. The average distance is
plotted as a function of the size of the environment in
\fref{fig:theplothatgorinwanted}, for two coupling values, and various times.
Instead of reporting the times, we report the approximate value of
concurrence a Bell pair would have after the corresponding time, and thus the
area to which it would be mapped in the $CP$ plane.  We conclude form the
figure that the unitality condition is approached algebraically fast as the
size of the environment increases.

We have established convincingly that the RMT channels are nearly unital.
For large purities this allows to establish a one to one relation between
purity and concurrence, as the unital area converges rapidly to the line
$C=P$.  Another fact remains to be clarified. When/how do the curves
approach the Werner curve? 

\begin{figure}
\centering
\includegraphics{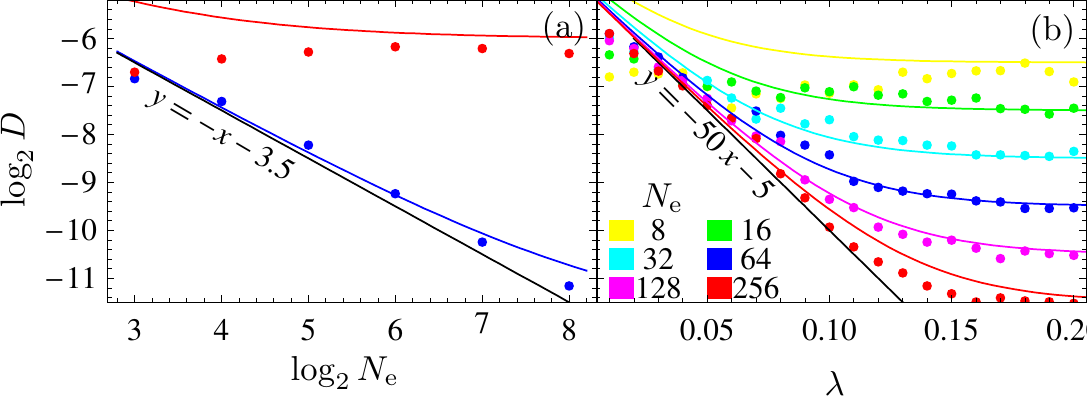}
\caption{We show $D$ [\eref{eq:deferrorwerner}] which measures a
``distance'' in the $CP$ plane between numerical curves for the RMT
  model and the Wigner curve. In (a) its dependence with the size of
  the environment is studied for two cases. For $\lambda=0.02$ (red
  dots) a finite value is approached, whereas for $\lambda=0.14$
  algebraic decay is seen. The black solid line, which is
  proportional to $N_\e^{-1}$, is meant as a guide to the eye.  In
  (b) we plot $D$ as a function of the coupling $\lambda$, for
  various values of $N_\e$.  In the large environment limit, and for
  $\lambda \ll 0.1$, we have a noticeable deviation from the Werner
  curve. In all cases $\Delta=1$.}
\label{fig:transicion}
\end{figure}

To study this situation in more detail, consider a $CP$-curve $C_{\rm
num}(P)$, obtained from our numerical simulations, and define its
``distance'' $D$ to the Werner curve as
\begin{equation}
  \label{eq:deferrorwerner}
  D=\int_{P_{\rm min}}^1\rmd P \left| C_{\rm num}(P)-C_{\rm W}(P)\right|.
\end{equation}
It is sensible to compare $D$ with the unital area $D_u=1/18$. The behavior of
$D$ as a function of the size of the system is shown in
\fref{fig:transicion}(a). For $\lambda=0.14$ (black dots), the error goes to
zero in an algebraic fashion, at least in the range studied.  In fact, from a
comparison with the black solid line we may conclude that the deviation $D$
is inversely proportional to the dimension of $\mcH_\e$.  By contrast, for
$\lambda= 0.02$ (red dots), $D$ tends to a finite value, in line with the
assertion that the numerical results converge to a different curve.  In
\fref{fig:transicion}(b) we plot the error $D$ as a function of $\lambda$,
for different dimensions of $\mcH_\e$.  The results suggest an exponential
decay of $D$ with the coupling strength.  The simplest dependence of the
deviation in agreement with these two observations is 
\begin{equation}\label{eq:depE}
 D_{\Delta=1}=\frac{1}{2^{3.5}N_\rme}+\frac{1}{2^{5+50 \lambda}}.
\end{equation}
We also plot the curves corresponding to this ansatz in
\fref{fig:transicion}. Good agreement is observed for $D \ll D_u$.  Notice the
exponential decay of $D$ with respect to $\lambda$. One can thus, in an
excellent approximation for large $\lambda$, say that for large dimensions
the limiting curve is the Werner curve.  For $\Delta=0$, all studied couplings
numerical convergence to the Werner curve was observed in the large $N_\rme$
limit.

In the presence of TRI the $CP$-curves a similar behavior is observed
whenever $\Delta>0$.  However, in contrast to the GUE case, no saturation of
the deviation $D$ was observed when $\Delta=0$.
In the other configurations considered (the joint and the
separate environment), the behavior is similar. In those cases it is the
largest (of the two) coupling strength which dominates the behavior in the
$CP$ plane (as well, naturally, as in time). 

\begin{figure}
\centering
\includegraphics{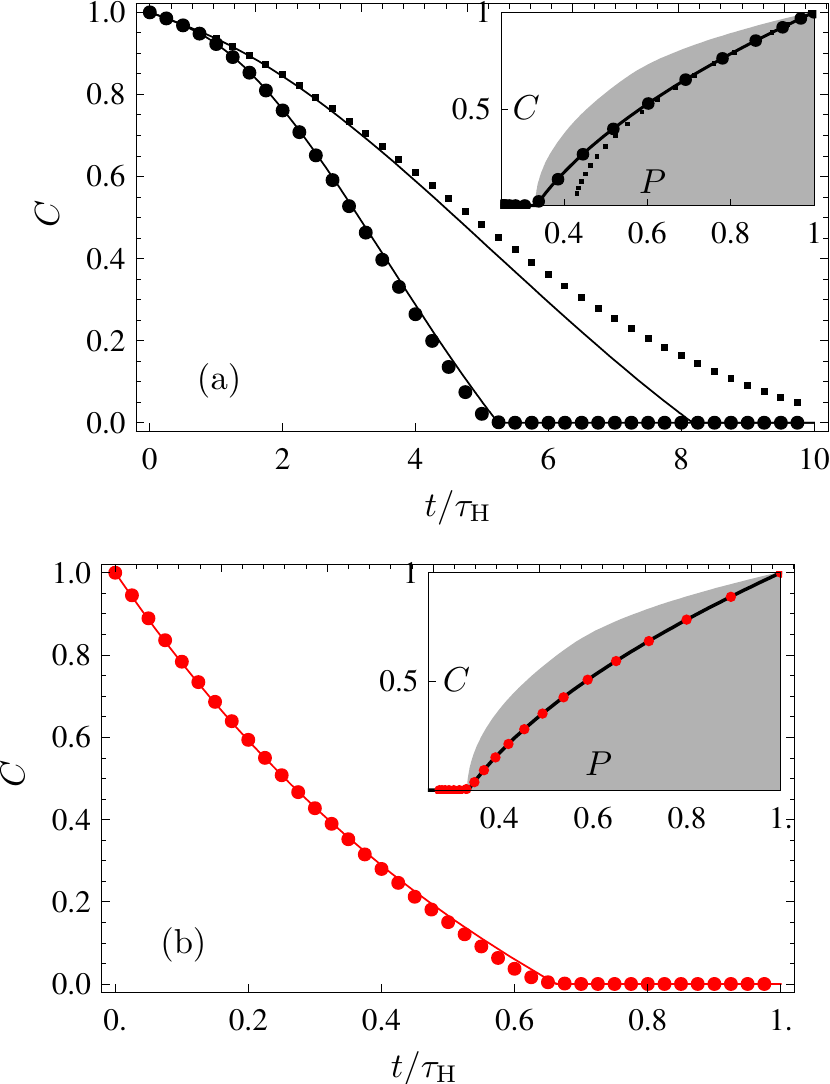}
\caption{We show numerical simulations for the average concurrence as
  a function of time in the joint environment configuration (GUE
  case).  In (a) we consider small couplings
  $\lambda_1=\lambda_2=0.01$ which lead to the Gaussian regime for
  purity decay. The red symbols show the result without internal
  dynamics, whereas the black symbols are
  obtained for fast internal Rabi oscillations
  ($\Delta_1=\Delta_2=10$). The theoretical expectation for
  concurrence decay based on \eref{eq:concurrenceELR} and
  \eref{eq:sepenvfull} is plotted in the corresponding color.  In
  (b) we consider stronger couplings, $\lambda_1=\lambda_2=0.1$, such
  that purity decay becomes essentially exponential (Fermi golden
  rule regime), while the level splitting have been set to
  $\Delta_1=\Delta_2=0.1$ (red symbols). The theoretical expectation
  (red solid line) is based on the same expressions as in (a). The
  insets display the corresponding evolution in the
  $CP$-plane with the same symbols as used in the main graph.  In
  addition, the physically allowed region (gray area) and the Werner
  curve (black solid line) are shown. In all cases $N_\e=1024$}
\label{fig:CPdecay}
\end{figure}

\section{Entanglement decay}\label{sec:concdecay}
Sufficiently close to $P=1$, the above arguments imply a one to one
correspondence between purity and concurrence, which simply reads $C\approx
P$.  This allows to write an approximate expression for the behavior of
concurrence as a function of time
\begin{equation}
  \label{eq:concurrenceLR}
  C_{\rm lr}(t)=P_{\rm LR}(t),
\end{equation}
using the appropriate linear response result for the purity decay. The
corresponding expressions for purity decay have been discussed in detail in
the previous chapters. \Eref{eq:concurrenceLR} has similar limits of validity
as the linear response result for the purity.  As it follows from a linear
response approximation for purity, we call it a linear response expression
for concurrence decay.

In those cases where the deviation from the Werner curve [see
\eref{eq:goodexponwernerCtime}] is small and where the exponentiated linear
response expression \eref{eq:exponentiationone} holds for the average purity,
we can write down a phenomenological formula for concurrence decay, which is
valid over the whole range of the decay
\begin{equation}
  \label{eq:concurrenceELR}
  C_{\rm elr}(t)=C_{\rm W}(P_{\rm ELR}(t)) \; .
\end{equation}
In \fref{fig:CPdecay} we show random matrix simulations for concurrence decay
in the joint environment configuration.  We consider small couplings
$\lambda_1=\lambda_2= 0.01$ which lead to the Gaussian regime for purity
decay, as well as strong couplings $\lambda_1=\lambda_2= 0.1$ which lead to
the Fermi golden rule regime. We find good agreement with the prediction of
\eref{eq:concurrenceELR}, except for the Gaussian regime when we switch-on a
fast internal dynamics in both qubits ($\Delta_1=\Delta_2= 10$) and
consequently $D$ is large. See the insets in \fref{fig:CPdecay}.  Notice how
in both  the heuristic formula and the Monte Carlo simulations, we observe
entanglement sudden death \cite{yu:140404}.

Note that, in the separate environment case, the entanglement of the two
subsystems defined on the spaces $\mcH_1\otimes\mcH_\e$ and
$\mcH_2\otimes\mcH_{\e'}$ is constant in time. For the initial conditions we
use, the entanglement stems entirely from the two qubits. The concurrence of
these two qubits decays because the constant entanglement spreads over all
constituents of the two large subspaces.

\begin{figure}
\centering
\includegraphics{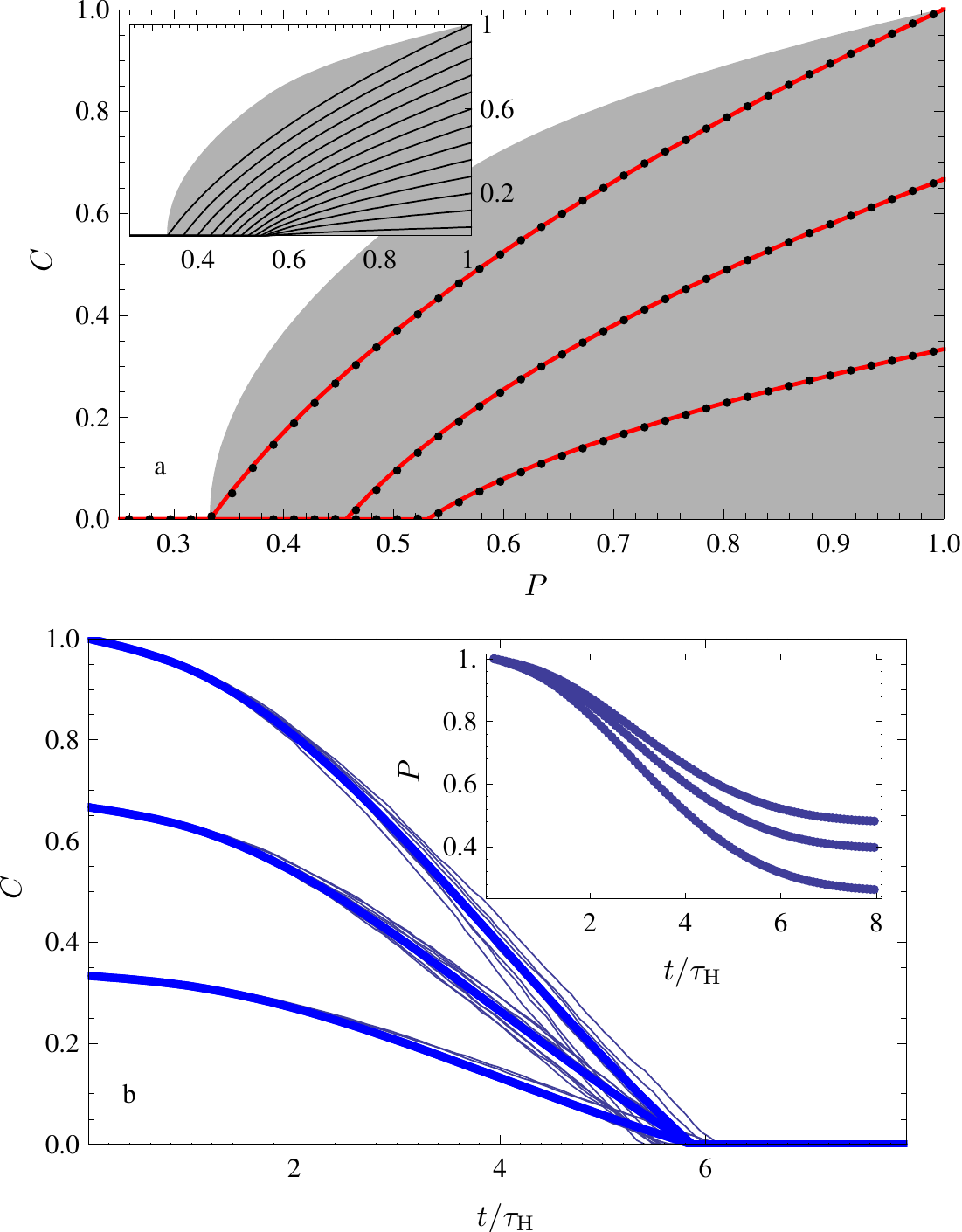} 

\caption{On the top panel the behavior in the $CP$ plane for
  different pure states is shown. We let the initial states have
  $C=1$, $2/3$, or $1/3$. The red curves are the ones described by
  the states when a totally depolarizing channel 
  acts on the first qubit, whereas the black dots
  correspond to Monte Carlo simulations. In the inset we see
  theoretical curves for other initial concurrences. On the lower
  panel in the inset,  we see the exponentiated formula for purity
  decay (blue line) and the Monte Carlo averaged purity (black dots).
  On the main panel we see theoretical result (blue line) and 30
  samples from the numerical ensemble (thin black lines). 
  $\Delta=0$, $\lambda=0.01$, and $\dim \mcH_\e=256$.  }
\label{fig:intermedio_one}
\end{figure}

We now investigated pure initial states of the qubits that are not Bell
states.  Consider the GUE case with no internal Hamiltonian, in the
spectator configuration. As discussed in previous sections, thanks to the
symmetry of the ensemble, we can write the state as
\begin{equation}|\psi_{12}\>=\cos\theta|00\>+\sin\theta|11\>.\end{equation}   
The initial concurrence is $C_0=\sin 2 \theta$; for $\theta =\pi/4$ we obtain
initial Bell states. Applying the totally depolarizing channel to one or both
qubits, results in Werner states. For $\theta \not\equiv \pi/4 \pmod{\pi/2}$,
applying the channel to a single qubit or to both qubits results in different
evolutions in the $CP$ plane. Since we are considering the spectator
configuration it is sensible to apply it to a single qubit. If we parametrize
the depolarizing channel \cite{NC00a} using the Kraus operators
$\{\sqrt{1-\gamma} \openone, \sqrt{\gamma/3}\ \sigma_x, \sqrt{\gamma/3}\
\sigma_y, \sqrt{\gamma/3}\ \sigma_z \}$, we obtain the state
\begin{equation}
\rho=\left(\begin{array}{cccc}
\frac{1}{3} (3-2 \gamma ) \lambda  & 0 & 0 & \frac{1}{3} (3-4 \gamma )
\sqrt{(1-\lambda ) \lambda } \\
 0 & \frac{2}{3} \gamma  (1-\lambda ) & 0 & 0 \\
 0 & 0 & \frac{2 \gamma  \lambda }{3} & 0 \\
 \frac{1}{3} (3-4 \gamma ) \sqrt{(1-\lambda ) \lambda } & 0 & 0 &
\frac{1}{3} (3-2 \gamma ) (1-\lambda )
 \end{array}\right),
\end{equation}
with $C_0=2\sqrt{\lambda(1-\lambda)}$. Concurrence for this state is $\max
\{C_0(1-2\gamma),0\}$ and purity is $1+\left[4\gamma^2(2+C_0)-6\gamma(2+C_0^2)
\right]/9$.  Inverting gamma in terms of purity and the initial concurrence, we
can insert it in the formula for concurrence to obtain the Werner-like
relations
\begin{equation}
    C_{ {\rm W},C_0}=C_0  
        \begin{cases}
	    \frac{3}{2}\sqrt{1-\frac{4(1-P)}{2+C_0^2}}-\frac{1}{2}
	        & \text{if}\  9P>5-C_0^2,\\
            0
	    	& \text{otherwise}
	\end{cases}.
\label{eq:cpae}
\end{equation}
The comparison between numerics and the ansatz is presented in
\fref{fig:intermedio_one}(a). The agreement is more than satisfactory. We also
show in the inset a set of curves for more values of $C_0$. The success of the
description allows to give a heuristic concurrence decay formula for initial
states, not being Bell states along the same lines as in the previous
paragraphs and using \eref{eq:cpae} instead of \eref{eq:goodexponwernerCtime}.
A comparison between our heuristic formula and Monte Carlo simulation
is presented in \fref{fig:intermedio_one}(b).  Note that "sudden
death" of entanglement \cite{yu:140404} persists, though it happens at
different purities. This is also visible for the individual members of the
ensemble. 
\begin{equation}
 C_{ {\rm W},C_0}=\max\left\{ 0,
       \frac{C_0-1}{3}+\frac{1+2C_0}{3} 
              {\rm Re} \frac{-1+\sqrt{1-(1+C_0^2)(3-6P-C_0^2)}}{1+C_0^2}\right\}
\label{eq:oootra}
\end{equation}
is the analogous expression for concurrence in terms of purity, when
both qubits are coupled with similar strength.

\chapter{An example, the KI chain}\label{sec:ki}

\section{Introduction}

In this chapter we continue studying decoherence and entanglement decay using
two qubits as the central system.  We pursue two main objectives. The first
one is to test some of the qualitative and quantitative conclusions we have
derived in the previous chapters, in a deterministic model. The second one is
to explore dynamical regimes that, up to this point, we have ignored:
integrable dynamics and intermediate dynamics (not integrable but with chaos
not fully developed).  We shall  focus on three points.  First we wish  to
study  how dynamics  of the environment, \ie its integrability or chaoticity
influence the behavior of decoherence and internal entanglement.  Second, we
shall analyze to what extent  purity and concurrence may be related.  In
particular we want to test the reach of the relation discovered in
\sref{sec:concurrence}.  Finally, we wish to relate quantitatively the
behavior of purity, in the chaotic regime, with the results obtained with
RMT in the previous chapters.  To implement such a program we need a model
with flexible dynamics, which allows efficient numerics for large Hilbert
spaces. 

The fact that the central system consists of two qubits, makes a spin system
an attractive candidate.  Indeed the kicked Ising spin chain, a model
introduced by Prosen \cite{prosenKI}, was used to study decay of fidelity and
purity \cite{purityfidelity} in echo dynamics for integrable, chaotic (more
precisely mixing~\cite{prosenIntegrableChaos}) as well as for intermediate
cases. Results for purity in echo dynamics can be used for purity decay in
standard forward dynamics if the perturbation is chosen as the coupling
between the central system and the environment, \sref{sec:generalprogram}.
The main advantage of this model is the high computational efficiency that
allows to perform numerical calculations up to twenty
qubits~\cite{pineda:066120} and more on any good workstation. Yet the model
in its original form does not allow for variable Ising interactions, and is
thus not well suited outside of the field of echo dynamics. Experimental
realizations of similar  models have been
proposed~\cite{barjaktarevic:012335} and related studies appeared
in~\cite{lakshminarayan:062334}.
 
In the present chapter we shall generalize this model allowing arbitrary
interactions between the spins, and site dependent kicks (this does not affect
the numerical efficiency of the model, see appendix
\ref{sec:implementationKI}). We then use the generalized model to study the
evolution of purity and concurrence of central system (consisting of two
spins) initially in a Bell state, evolving in environments with different
dynamical properties. As we want concurrence to be affected solely by the
coupling to the environment, we choose non-interacting spins for
the selected pair. Some preliminary results were published 
in \cite{pineda:012305}.  For the environment it is sensible to consider random
states to emulate a bath at fairly high temperature. Using unitary time
evolution of the total system and partial tracing over the environment we can
then calculate concurrence and purity decay of the selected pair, and discuss
their behavior.

To discuss the general model we first recall and discuss a particular case of
the model which, in fact, will help to understand the properties of the
environment (\sref{sec:KIchain}). Then we state the general model and discuss
the particular configurations to be studied (\sref{sec:modelsKI}). Next we
proceed to discuss the behavior in time of both concurrence and purity for
the different configurations and different dynamics, \sref{sec:kievolution}.
Afterwards, in \sref{sec:kiCP}, we discuss the relation between concurrence
and purity.  Finally, we compare purity evolution with the corresponding RMT
model in \sref{sec:KIvsRMT}.

This chapter is largely based on \cite{pineda:012305}; however, we have
learned a lot since its publication. Further research and major rewriting
were thus required to contextualize and discuss the results presented here.
We added new configurations and discuss in detail the physical differences
between the different models. The comparison with RMT is now possible as the
theory is completed. 

\section{The kicked Ising spin chain}\label{sec:KIchain}

We first study the {\em kicked Ising chain} (KIC) \cite{prosenKI}:
A ring of $L$ spin $1/2$ particles which interact with their nearest
neighbors via a homogeneous Ising interaction of dimensionless strength $J$,
and are periodically kicked with a dimensionless homogeneous magnetic field 
$\vec{b}$.The Hamiltonian is thus
\begin{equation}
\label{eq:hamiltonianKI}
H = J \sum_{j=1}^{L} \sigma ^z_j \sigma^z_{j+1} +
\delta_1(t) \sum_{j=1}^{L} 
\vec{b} \cdot \vec{\sigma}_j 
\end{equation}
where $\delta_1(t)=\sum_{n\in \mathbb{Z}} \delta(t-n)$ represents an infinite
train of Dirac delta functions with primitive period one;  $\sigma_j^{x,y,z}$
are the Pauli matrices of particle $j$ and
$\vec{\sigma}_j=(\sigma_j^x,\sigma_j^y,\sigma_j^z)$.  We must also impose
periodic conditions in order to close the ring: $\vec{\sigma}_{L+1} \equiv
\vec{\sigma}_1$.  During the free evolution, \ie between the kicks, the system
evolves with the unitary propagator
\begin{equation} \label{eq:Ising}
  U_\text{Ising}(J) =\exp 
    \left( -\mimath J \sum_{j=1}^{L} \sigma ^z_j \sigma^z_{j+1} \right),
\end{equation}
and the action of the kick is described by the unitary operator
\begin{equation} \label{eq:kick}
  U_\text{kick}(\vec{b})=
    \exp\left(-\mimath \sum_{j=1}^{L} \vec{b} \cdot\vec{\sigma}_j \right).
\end{equation}
The Floquet operator for one period is thus 
\begin{equation} \label{eq:floquet}
	U_\text{KI}=U_\text{Ising}(J)U_\text{kick}(\vec{b}).
\end{equation}

A detailed discussion of the symmetries of the system can be found in
\cite{pp2007}. We summarize the main observations here. Due to the
homogeneity of the kick $\vec b$ and the nearest neighbor interaction $J$,
the system has a rotational symmetry $\vec\sigma_j \to \vec\sigma_{j+1}$.
The dimension of each of the invariant subspaces ($\mcH_k$ with
$k=1,\dots,L$) is close to $2^L/L$ for large $L$.  Each of the invariant
subspaces has an anti-unitary symmetry which, if we choose a base in which
$\vec b \cdot \vec\sigma_j$ is real, is complex conjugation (\ie not
conventional time reversal symmetry).  Since each minimal invariant subspace
has an internal anti-unitary symmetry, it is reasonable to compare with the
GOE or the COE to understand the properties of the system. 

\begin{figure}[t]
\begin{center}\includegraphics{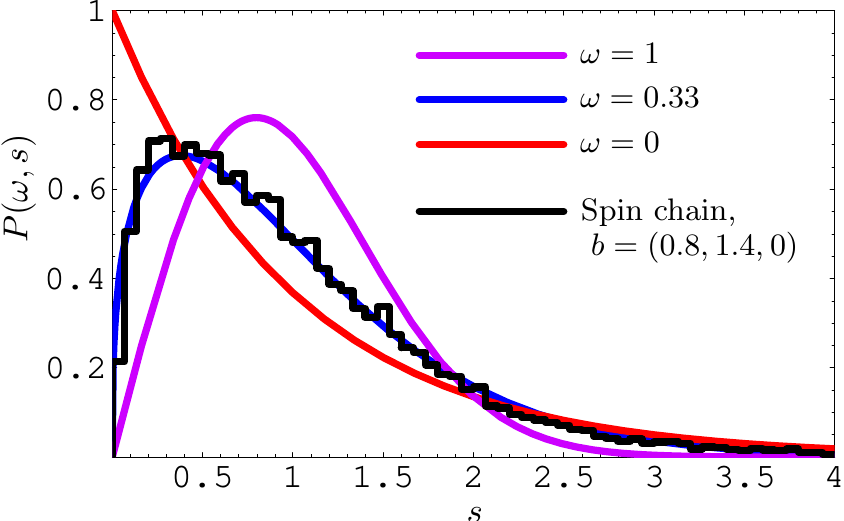} \end{center}
\caption{The nearest-neighbor spacing of the homogeneous spin ring (as a
  black curve) with a magnetic kick $b=(0.8,1.4,0)$. We compare with the
  Brody distribution $P(\omega,s)$. Its limiting values $\omega=1$ and
  $\omega=0$ yield GOE and Poissonian statistics respectively. Intermediate
  values suggest intermediate dynamics. Here we use 15 qubits. }
\label{fig:intermediate}
\end{figure}

As mentioned above the system is very rich regarding its dynamics. Various
parameter combinations yield integrable (meaning analytically solvable)
dynamics. In particular setting the magnetic kick parallel or perpendicular
to the preferred direction of the Ising interaction implies integrability.
However, using $\vec b$ parallel to the Ising interaction yields trivial
integrability (all $\sigma_j^z$ commute with the Hamiltonian), so we shall
use $\vec b$ perpendicular to the $z$ axis. In particular we shall set $\vec
b=(0,1.53,0)$ since for this parameter some oscillations that we shall
observe have a bigger amplitude and period.  In order to obtain chaotic
dynamics one cannot just choose parameters outside the set that yield
integrable dynamics.  One must be ``far enough'' from this set. We verified
that for $J=1$, $b=(1.4,1.4,0)$ (used in this chapter) and $J=0.7$,
$b=(0.9,0.9,0)$ (used in the following one) the system is chaotic (see
\cite{pineda:066120, pp2007}); many statistical tests like the
nearest-neighbor spacing, the form factor, and the skewness, were applied.
Except at extremely large energies, no significant deviations from RMT were
observed.  In order to obtain intermediate dynamics we wanted to be far
enough from both integrability and completely mixing dynamics to have an
appreciable compromise regarding spectral statistics. For $J=1$ and
$b=(0.8,1.4,0)$ we indeed find intermediate spectral statistics. Their
nearest-neighbor spacing is characterized by a Brody parameter
\cite{brodyparameter} of $0.33$, see \fref{fig:intermediate}. The
nearest-neighbor spacing distribution in the Floquet spectrum shows a marked
level repulsion, but it is still quite far from the distribution we expect and
get for the chaotic case.

\section{A generalized kicked Ising model}\label{sec:modelsKI}

The Hamiltonian of the generalized kicked Ising model is
\begin{equation}
\label{eq:fullhamiltonian}
H = \sum_{j>k=1}^{L}J_{j,k} \sigma ^z_j \sigma^z_{k} +
\delta_1(t) \sum_{j=1}^{L} 
\vec{b}_j \cdot \vec{\sigma}_j. 
\end{equation}
The model thus consists of a set of $L$ spin $1/2$ particles coupled to all
other spins by an Ising interaction (first term) and periodically kicked by a
site dependent tilted magnetic field (second term).  

Our model differs from the one stated in \eref{eq:hamiltonianKI} by the fact
that the coupling $J_{j,k}$ is between any pair of particles and has an
arbitrary strength.  This freedom allows to build a variety of models with
different topologies.  Each of them will correspond to a well defined
physical picture, and thus we show how the flexibility of the model can be
exploited to study diverse physical situations in a common framework.  We
also allow site independent kicks to be able to have (or not) internal
dynamics in our central system. 

The central system is composed of two spins, say spins ``1'' and ``2''.  We
are left with $q_\rme=L-2$ spins which are going to be considered as the
environment.  We furthermore diminish the number of parameters by setting
$\vec b_1=\vec b_2=\vec b_\rmc$ and $\vec b_3=\cdots=\vec b_L=\vec b_\rme$.
The fact that we keep the kick in the central system can represent local
operations made by the ``owners'' of each of the qubits, and will not affect
the values of concurrence and purity. We shall furthermore require weak
coupling of the central system to the environment.  Thus $J_{i,1}$ and
$J_{i,2}$ must be much smaller than the typical Ising interaction within the
environment. We set $J_{2,1}=0$ in order to prevent any interaction between
the spins in the central system.  

\begin{figure}
\includegraphics{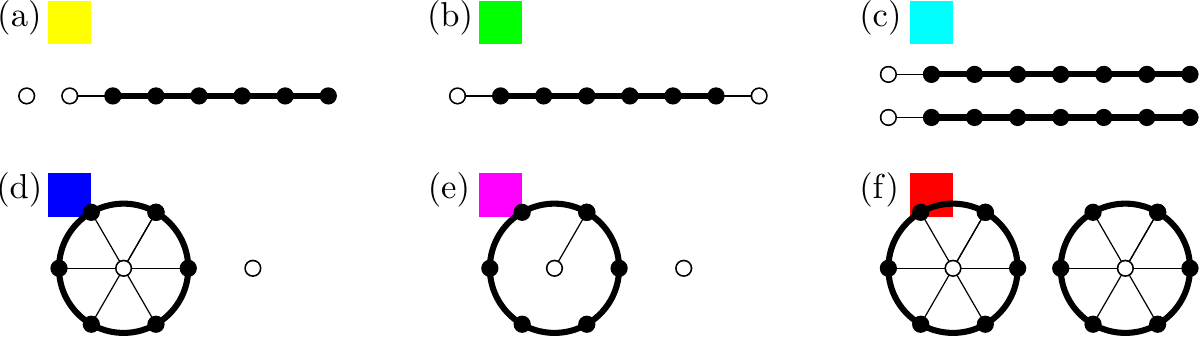}  
\caption{Different  configurations of the coupling of the central system
  (represented as open circles) to the spins that act as environment
  (represented as filled circles). Thick/thin lines represent $J_{\rm
  ce}'$/$J_\rme$ respectively.  The description for each model and its
  features is given in the main text.}  \label{fig:Config}
\end{figure}

We propose in particular six models. They are schematically drawn in
\fref{fig:Config}.  The open circles represent qubits 1 and 2 (the central
system) whereas the filled circles represent the other qubits (environment).
The thin lines represent a weak Ising interaction $J_{\rm ce}'$ of a particle
in the central system with one in the environment. The thick lines represent an
interaction $J_\rme$ among particles in the bath; in all numerical
results presented we fix $J_\rme=1$. Finally we note that the
dimension of the Hilbert space of the environment is $N_\rme=2^{q_\rme}$.
Writing the Hamiltonian for each of the models we propose is now a trivial
task. As an example we write the Hamiltonian of the model (a) in
\fref{fig:Config}: Rewriting the Hamiltonian in terms of central system,
environment and interaction as $H=H_\text{c}+H_\text{e}+H_\text{ce}$ the parts
are given by
\begin{align}
H_\rmc &= \delta_1(t)  \sum_{j=1}^{2} \vec b_\rmc \cdot \vec \sigma_j ,\\ 
H_\text{e} &=J_\rme \sum_{j=3}^{L-1} \sigma ^z_j \sigma^z_{j+1} 
              + \delta_1(t) \sum_{j=3}^{L} \vec b_\rme \cdot \vec \sigma_j \\ 
H_\text{ce} &=J_{\rm ce}' \sigma ^z_{2} \sigma^z_{3}. 
\label{eq:ce}
\end{align}

Each of the models has a characteristic Heisenberg time $\tau_\rmH$. For a
system with no symmetries the Heisenberg time is defined simply as $2\pi/d$,
where $d$ is the mean level spacing. If the system has symmetries, for each
sector we have a Heisenberg time defined again as $2\pi/d$, but $d$ is the
mean level spacing in each sector. We shall stress the Heisenberg time of each
model, as our RMT models show a strong dependence on it [\eg
\eref{eq:spectatorDegenerate}].  We also use a normalized coupling $J_{\rm
ce}$, which gives a model independent coupling strength.  Finally, we want 
to mention the corresponding configuration (spectator, joint environment or
separate environment), as in \sref{sec:twoqubitmodel}.

Please refer to \fref{fig:Config} while reading the descriptions.
\begin{itemize}
\item Model (a) is in the spectator configuration. A single qubit is coupled to
the extreme of an open chain. Due to the breaking of the symmetry of the
coupling, the Heisenberg time of the environment is $\tau_\rmH\approx N_\rme$.
We use the normalized coupling $J_{\rm ce}=J_{\rm ce}'$.
\item The second model is a variation of the previous one. In this the second
qubit is coupled to the other extreme of the chain. It is an example of the
joint environment configuration. Again $\tau_\rmH\approx N_\rme$, however to
have the same effective coupling we must scale $J_{\rm ce}=\sqrt{2} J_{\rm
ce}'$.
\item Model (c) is the last of this family of models in which the environment
is an open chain. In this case $\tau_\rmH\approx \sqrt{N_\rme} \ll N_\rme$. As
we have two coupling spots, we must scale $J_{\rm ce}=\sqrt{2}J_{\rm ce}'$.
This is a good example of  a separate environment Hamiltonian. 
This, and the previous two models, were analyzed in \cite{pineda:012305}.
\item Model (d) is again in the spectator configuration, but here we have a
ring (thus the translational symmetry is not broken) and the coupling is
symmetric. The Heisenberg time is then related to each symmetry sector
separately and is approximately equal to $N_\rme/q_\rme$.  The
coupling must be normalized to $J_{\rm ce}=\sqrt{q_\rme}J_{\rm ce}'$.
\item The next model differs from the previous one only inasmuch that the
coupling is not symmetric. This is enough to mix the invariant subspaces and
thus the Heisenberg time is $\approx N_\rme$. There is a single coupling and
thus it is enough to set $J_{\rm ce}=J_{\rm ce}'$. 
\item Model (f) is another example of a separate environment configuration. 
It has the smallest
Heisenberg time, since the coupling does not break the symmetry of the ring.
Thus, it is approximately $\sqrt{N_\rme}/q_\rme$. The coupling is normalized to
$J_{\rm ce}=\sqrt{q_\rme}J_{\rm ce}'$.
\end{itemize}
It is pertinent to mention, that the degree of chaoticity was obtained for
cyclic chains, while the chains representing some of our environments are
open.  Yet,  for large number of spins, this is irrelevant.

The non-unitary evolution of the central system alone is calculated along the
same lines as for the RMT models. We perform a unitary evolution of the whole
system  (yielding state $|\psi (t)\>$)  and then we partial trace over the
environment. That is
\begin{equation}
\rho(t)=\tr_\e |\psi (t)\> \< \psi (t)|. 
\end{equation}
We shall consider an initial condition
\begin{equation} \label{eq:InitialCondition}
|\psi(t=0) \> = |\psi_\text{Bell}\> \otimes |\psi_\text{env}\>,\
\end{equation}
\textit{i.e.} a general Bell state [any state such that
$C(|\psi_\text{Bell}\>\<\psi_\text{Bell}|)=1$] not entangled with the
environment, which is in a pure random state [for models (a), (b), (d), and (e)]
or in a tensor product of pure random states in each section of the environment
[for models (c) and (f)].

\section{The evolution of concurrence and purity}\label{sec:kievolution}

We now present the results of our numerical calculations of both concurrence
and purity of the selected pair of spins as a function of time.  We first
concentrate on the short time behavior.

\begin{figure}
\includegraphics{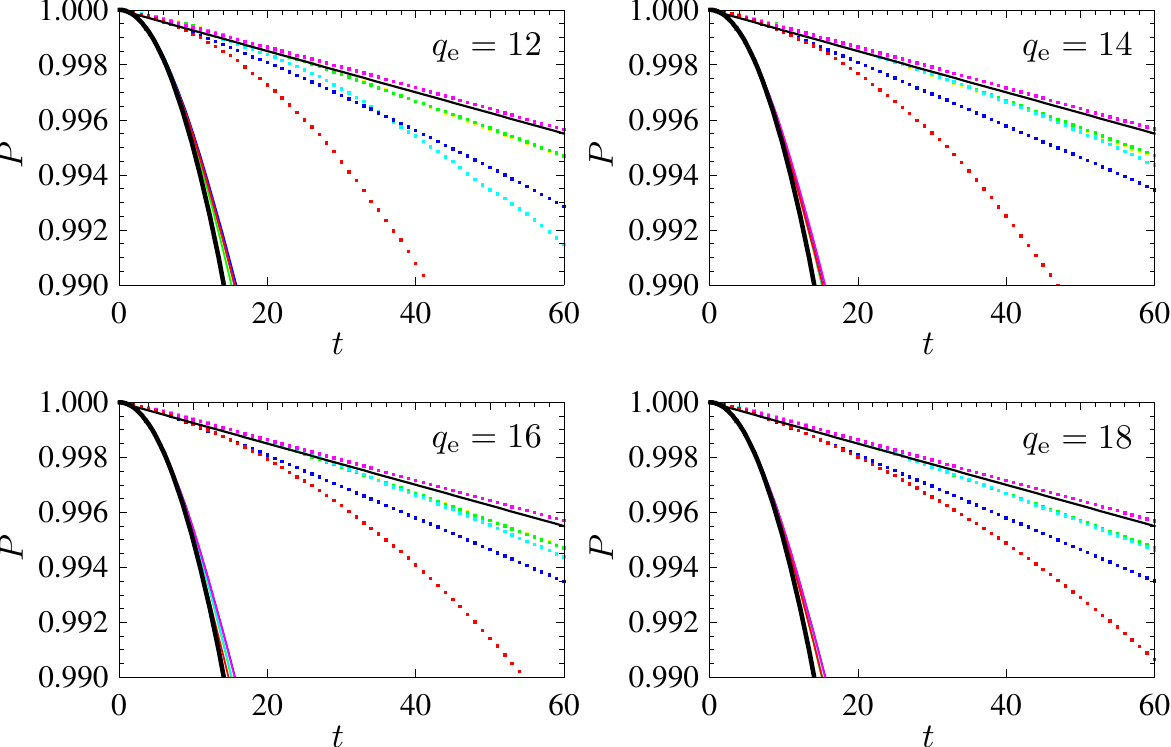}
\caption{These figures show the evolution of purity for a single initial
  condition per configuration, for chaotic (points) and integrable (thin
  lines) dynamics.  All different configurations from \fref{fig:Config} are
  tested (we keep the corresponding color coding). In the chaotic case,
  purity shows a linear dependence on time  whereas in the integrable one, it
  exhibits a quadratic decay. We also plot the line $1- 3 J_c^2 t$ and the
  parabola $1-2J_c^2 t^2$ with thick black curves, as guides. We vary the
  size of the environment to understand the scaling with $q_\rme$. The
  corresponding plots for concurrence are indistinguishable from the ones
  presented here. We have set $J_\textrm{c}=0.005$ and $\vec b_\rmc=\vec
  b_\rme$. }
  \label{fig:timeevolution}
\end{figure}

Fig.~\ref{fig:timeevolution} shows the  time evolution of purity of the
selected pair of spins for the initial state consisting of a Bell pair in the
central system and the appropriate random state in the environment.  We choose
the environment to be chaotic in one case and integrable in the other, using
the corresponding parameters mentioned in \sref{sec:KIchain}.  The chaotic
environment leads to linear decay for some time interval, while the integrable
environment leads to a quadratic decay. The interval in which the linear decay
can be observed depends on two things: the configuration and the dimension of
the environment. Indeed, for configuration (f) we have the shortest linear
decay but we also have the shortest Heisenberg time. The reader is encouraged
to check table \ref{tab:heis} to understand better our numerical results.
Recall that in all our derived formulae, we predicted linear decay before
Heisenberg time. The situation for the integrable case is different.  We
observed a marked quadratic decay, independent of the configuration and the
size of the environment. We believe that one should be able to explain the
results (for the integrable situation) with exact analytic calculations,
however the aim of this chapter is not to explain the details, but to
illustrate the qualitative differences between having a chaotic and regular
environment.  

\begin{table}[h]\label{tab:heis}
\begin{center}
\begin{tabular}[h]{|l|c|c|c|c|c|c|}\hline
 & (a) {\hueoneosix $\blacksquare$} & (b) {\hueoneothree $\blacksquare$}
 & (c) {\hueoneotwo $\blacksquare$} & (d) {\huetwoothree $\blacksquare$}
 & (e) {\huefiveosix $\blacksquare$}& (f) {\huezero $\blacksquare$}\\\hline\hline
 $q_\rme=12$& 4096   & 4096   & 64   & 341.3  & 4096  & 10.7 \\ \hline 
 $q_\rme=14$& 16348  & 16348  & 128  & 1170.3 & 16348 & 18.3 \\ \hline
 $q_\rme=16$& 65536  & 65536  & 256  & 4096   & 65536 & 32   \\ \hline
 $q_\rme=18$& 262144 & 262144 & 512  & 14563  & 262144& 56.9 \\ \hline
\end{tabular}
\end{center}
\caption{Numerical values of the Heisenberg times for the different configurations
of the environment, and different sizes of it.}
\end{table}

\begin{figure}[t]
\includegraphics{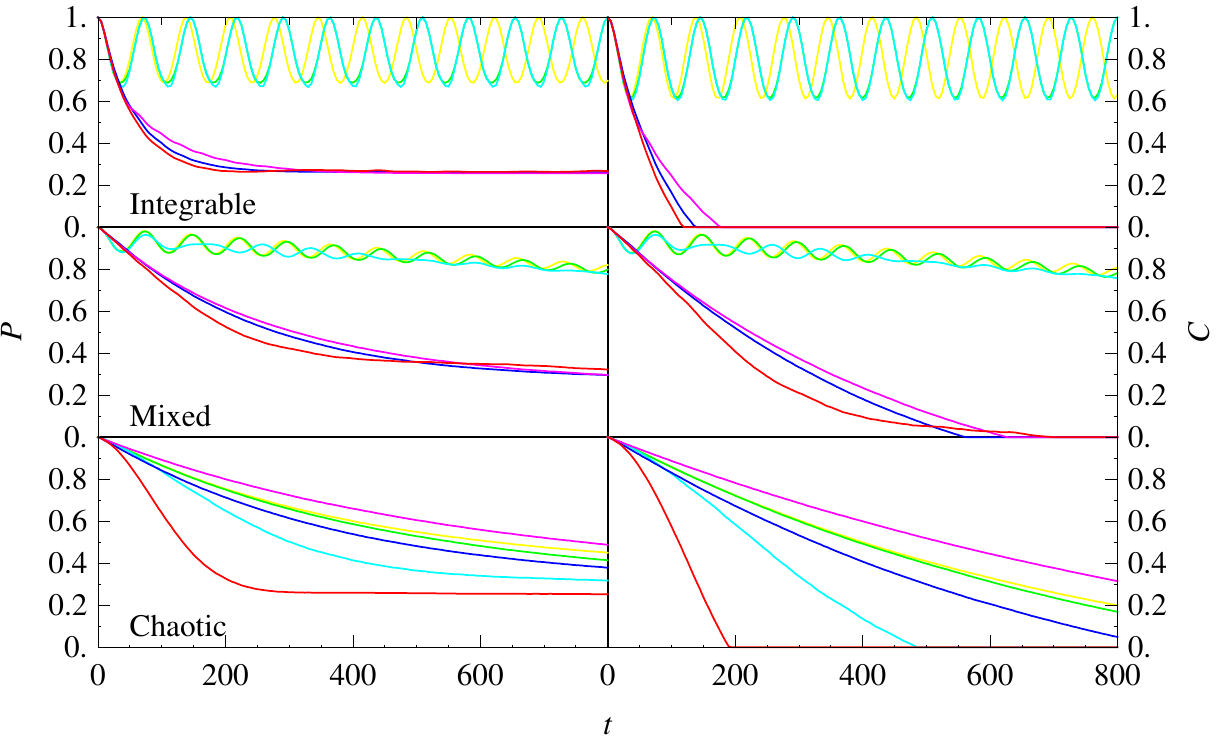}
\caption{Long-time behavior of purity (left) and concurrence (right) for an
  arbitrary initial condition in integrable (top), non-ergodic and
  non-integrable (middle), and fully chaotic (bottom) regimes.  For the
  integrable regimes we observe recurrences for the open environment
  configurations and fast, possibly Gaussian, decay in the others. For the
  intermediate case we observe in some instances the superposition of
  oscillations together with the exponential decay.  For the chaotic case we
  observe exponential decay together with a marked dependence on the kind of
  environment.  We have set $q_\rme=16$, $J_\textrm{c}=0.02$, and $\vec
  b_\rmc=\vec b_\rme$.}
\label{fig:longCP}
\end{figure}

We now proceed to look at the long-time evolution of concurrence and purity
again for all configurations.  In Fig.~\ref{fig:longCP} we show both
concurrence and purity for integrable, intermediate and chaotic situations.

For the integrable case we see that for some configurations [(a), (b), and
(c)] after the initial quadratic decay there is a full revival of both
concurrence and purity. However the autocorrelation function of the full
system (not shown) drops fast and reaches negligible values even after the
first few kicks. It is also worthwhile to note that the oscillations of
purity imply increase in some cross-correlation
function~\cite{prosen:062108}.  For the other configurations there is no such
revival, and the decay seems Gaussian; this is in agreement with the
quadratic decay observed for large purities.  Notice that in all the
configurations in which there is a revival the coupling is to one end of a
chain.  Notice how concurrence drops to exactly 0 after a finite time. 

For the chaotic case we observe a monotonic decay of both purity and
concurrence.  The speed at which they decay, though initially the same, is
quite different for the different configurations. This can be explained with
the huge differences in the Heisenberg times of the environment.  

For the mixed case, we observe a compromise among the behavior in the
integrable case and the chaotic one.  Interestingly in all cases we observed
that concurrence drops identically to zero after some time, in contrast with
purity which does no reach its minimum.

\section{A relation between concurrence and purity}\label{sec:kiCP}

The rather similar pictures emerging for concurrence and purity suggest that
a simple relation between the two might emerge. Indeed, from the results of
the previous chapter this could be expected for the chaotic case.
Unfortunately we are not able to access such a big range of values of the
coupling as in the previous section; too strong coupling leads to total
concurrence decay after a single period (remember that we are dealing with
Floquet operators). Too weak couplings imply too long runs as we have to
integrate one step at a time.  On the other hand, we are going to be able to
test the relation not only on chaotic situations but also in other dynamical
regimes such as integrable environments and intermediate environments. 

\begin{figure}
\begin{center}\includegraphics{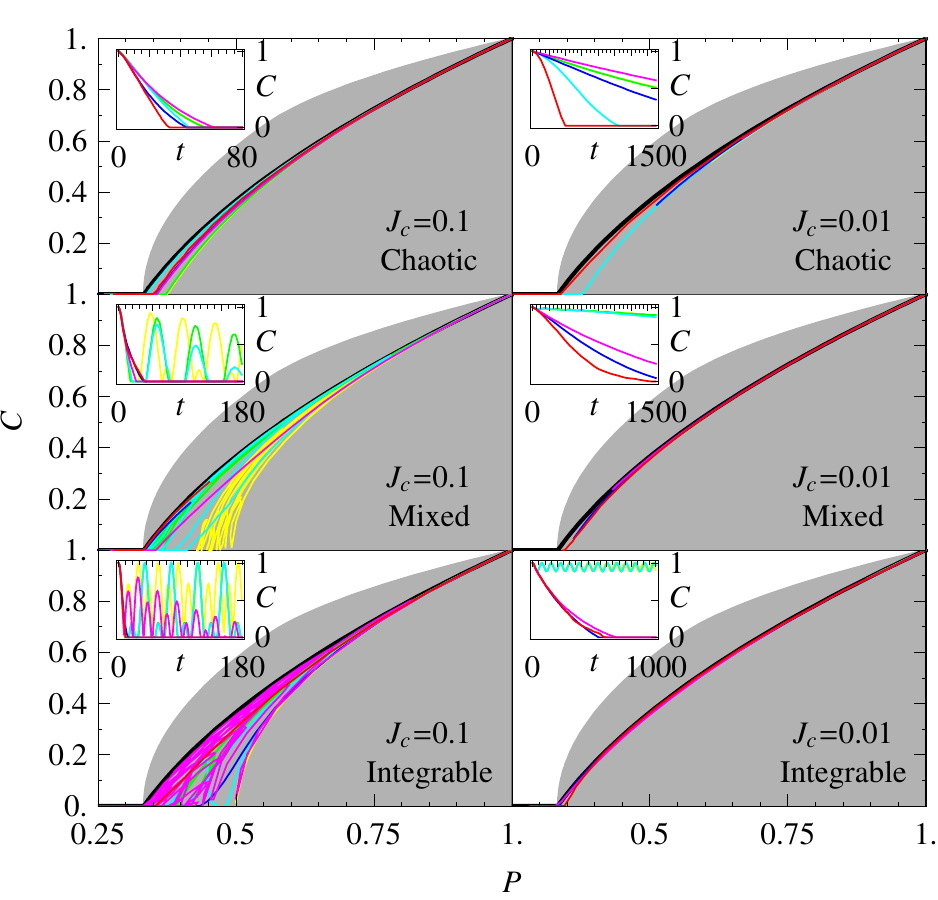}\end{center}
\caption{The evolution of a random initial condition, in the $(C, P)$ plane,
  for different dynamical regimes: integrable, intermediate, and chaotic, for
  two different coupling strengths. All configurations are tested, the color
  indicates the configuration, as in \fref{fig:timeevolution}. In the
  thermodynamical limit all curves seem to fall in the unital region.  For
  smaller couplings there is a tendency to stick to the Werner curve.  For
  stronger couplings, in the integrable and intermediate regime, they seem to
  fill the unital area. We set here $q_\rme=16$, and $\vec b_c=\vec b_\rme$.
  The gray region and the black thick line have the same meaning as in
  \fref{fig:exampleacuuWerner_one}.}
\label{fig:varydynamics}
\end{figure}

We plot in \fref{fig:varydynamics} concurrence against purity in all the
dynamical regimes studied so far. We do so for two strengths of the
perturbation, namely $J_\rmc=0.1$ and $0.01$.  A wide variety of behaviors
are observed. One thing that stands out is that all curves lie in the unital
region. Indeed for  $J=0.1$, in the integrable and intermediate regime we
observe a filling of the unital region [recall it from
\fref{fig:exampleacuuWerner_one}].  This is not a finite dimension effect: we
verified that the behavior remains for increasing dimension of the
environment in both cases.  For small dimensions, and in configuration (b)
one observes that the curves do not lie in the unital region and explore a
larger area. This can be explained: one has a common environment which, in
fact, has memory (due to its small dimension).  An example of this behavior
can be found in figure 5 of \cite{pineda:012305}. 

For the other cases, where there is no filling of the unital region, one can
suspect self averaging.  This  can be tested calculating the standard
deviation of purity $\sigma_P$ , one time step before concurrence is zero.
This is equivalent to examining the width of the coloured curves in
\fref{fig:exampleacuuWerner_one}. We numerically observed that for some set
of parameters, $\sigma_P \propto 1/N_\rme$ whereas for other it approaches a
finite value. Though a tendency to accumulate towards the Werner curve is
clear from the pictures presented, in some cases it is not more than a
tendency.  Indeed, even in the chaotic case and when having self averaging,
the limiting curve not always crosses the line $C=0$ at $P=1/3$ (like the
Werner curve), but a slighter bigger value. Again our numerics show that this
is not a finite size effect. However, all the limiting curves are similar to
the Werner curve, and that still can be used for approximations, as long as
we recall it is only an approximation.  Though all six configurations are
physically quite different and the  individual behavior of purity and
concurrence is quite affected, the relation between the two is entirely
robust.

Cases (c) and (f) lead to an instructive lesson: Here we have two uncoupled
environments, and we start with a pure state  in each of these.  The
purity of the uncoupled subsystems  will remain unchanged,  but the purity and
concurrence of the  initial Bell pair will decay.  Thus one might consider
seeing a paradox, but this is not the case; the entanglement of the pair
is simply spread over  all the  system  with  time.  

Up to this point our results are in full qualitative agreement with the ones
found in previous chapters. This leads us to make a tighter comparison with 
the RMT models developed before. 

\section{Comparison with RMT predicted behavior}\label{sec:KIvsRMT}

Our KI model has some parameter ranges in which we can consider the system to
be ``quantum chaotic''. A quantitative comparison of purity decay for the
dynamical system and the RMT model is sound. Statistical tests and analytical
arguments, indicate that one can consider the homogeneous KI as a typical
member of the GOE. Thus we shall compare with the results obtained in
\sref{sec:goeone} and \sref{sec:goetwo}.  We first list some differences
between the two models.

$\bullet$ The most serious difference is the structure of the coupling.
Relations of the form eqs. (\ref{eq:GOEconmute}) and (\ref{eq:GUEconmute})
used to define the coupling in the RMT model, generically yield non-separable
operators, whereas the coupling for the KI model is indeed separable and of
the form $\sigma_z \otimes V_\rme$. Dynamics involving separable operators
have special properties. We must not be surprised that this particular
structure of the coupling affects the behavior. 
 
$\bullet$ The theory developed in earlier chapters is for an ensemble of
systems. Here we are dealing with a single system. Self averaging takes care
of this issue. Numerical tests indicate that as the size of the environment
increases, the behavior for a typical Hamiltonian approaches rapidly the
behavior of the ensemble average (see \fref{fig:decayGOE}).  We can safely
ignore this problem as long as we go to high enough dimensions.  However, for
very detailed discussions this must be taken into consideration (as in
\cite{pineda:066120}).

$\bullet$ We are using environments with either exact rotational symmetries
[(d), (e), and (f)] or with some approximate translational symmetry [(a),
(b), and (c)].  But we have not developed a theory for chaotic environments
with discrete unitary symmetries. 

One might be tempted to assume that the overall result is an average of the
results in the different sectors, as happens with fidelity $f$:  Given an
echo operator $M$ with a discrete unitary symmetry, and a set of states
$\{|\psi_s\>\}$ living in the different symmetry sectors, $$ f=(\sum_s
\alpha^*_s \<\psi_s|) M (\sum_{s'} \alpha_{s'} |\psi_{s'}\>)= \sum_s
|\alpha_s|^2 \<\psi_s|M|\psi_s\>.$$  For purity the situation is different, 
\begin{equation}
P\left(\tr_1 \sum_{s,s'} \alpha_s \alpha_{s'} |\psi_s\> \<\psi_{s'} |\right) 
  \ne \sum_s |\alpha_s|^2 P(\tr_1|\psi_s\> \<\psi_s |).
\end{equation}
Let $\alpha_1=\alpha_2=1/\sqrt{2}$, $|\psi_{s=1}\>=|00\>$, and
$|\psi_{s=2}\>=|11\>$.  In this case purity of the reduced system is minimal
($1/2$) whereas the purity of the reduced state of each of the components is
maximal ($1$). 

However, a quantity that is affected immediately when superposing spectra is
the form factor. The form factor of a superposition of many, say GOE, spectra
approaches that of a Poissonian one.  Thus, a simplified 
way to incorporate this effect is to replace in our formulae the GOE form
factor $b_2^{(1)}(t)$ by the one of a random spectra $b_2(t)=0$.

$\bullet$ During the first chapters we assumed time independent Hamiltonians.
The KI Hamiltonian (\ref{eq:fullhamiltonian}) depends explicitly on time.
However, due to its periodic nature, one can construct its corresponding
Floquet operator. The circular ensembles \cite{dyson} are used to model such
operators \cite{haakebook}. Circular and Gaussian ensembles share all the
properties regarding its eigenvectors.  The eigenphases of the former and the
eigenvalues of the later share most properties but a precise mathematical
relation among them is cumbersome to state.  We can naively consider the
circular ensembles as an exponentiated (and approximated) version of the
Gaussian ensembles.  We thus do not need to modify our theory to take into
account this issue.

$\bullet$ With \eref{eq:GOEconmute} we are assuming certain normalization of
the coupling.  One can adjust the coupling constant $\lambda$ so any $V$ has
the given normalization [see \eref{eq:hlambda}].  In our case, we can expect
that for models (a) and (e) the decay, at least for short times, does not
depend on the number of qubits. This can be achieved substituting $\lambda
\to \alpha J_{\rm ce}/\sqrt{\tau_\rmH}$ with $\alpha$ an $N_\rme$ independent
number.  The leading term (with respect to time) is now independent of
$\tau_\rmH$ and $N_\rme$.

\begin{figure}[t]
\begin{center} \includegraphics[width=.85\textwidth]{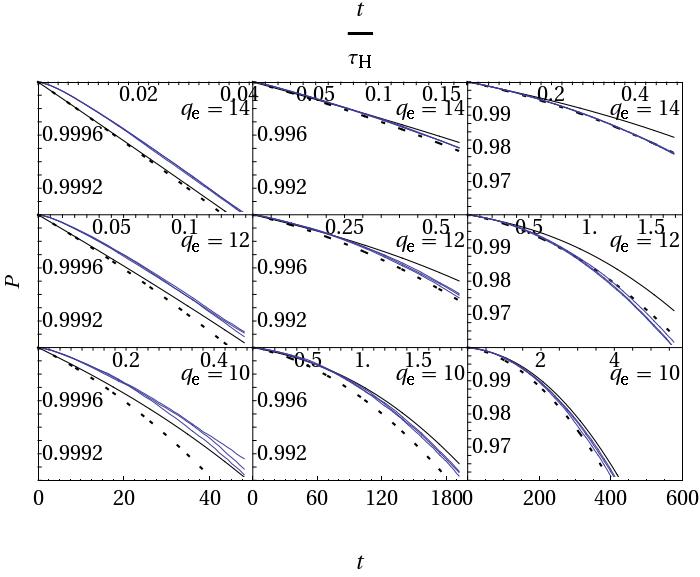} \end{center}
\caption{We plot, with blue thin lines, the evolution of purity for 10
  initial condition in model (d). We vary the number of qubits vertically and
  the time ranges horizontally.  The formula derived from our RMT model
  \eref{eq:RMTKI} is plotted as a thin black line.  The same formula, but
  assuming no correlations within the spectra, is plotted as a dashed line.
  We use $J_{\rm ce}'=0.0005$ to be in the linear response regime for all
  plots. See text for a complete description.}
\label{fig:comparisonKIRMT}
\end{figure}

For simplicity we eliminate the internal Hamiltonian in the dynamical model
($H_\rmc=0$) to use the RMT formulae with $\Delta=0$. We use configuration
(d) for the comparison to eliminate possible border effects.  Its Heisenberg
time is moderately small and thus we can access easily time regimes
comparable with the Heisenberg time. The effective size of the environment is
still large ($2^{q_\rme}$) [compared with separate environment configurations
($\sqrt{2^{q_\rme}}$)] so the large dimension limit is achieved faster. 

Our adapted RMT formula (using Bell states as initial conditions) reads
\begin{equation} \label{eq:RMTKI} 
P(t)=1-\alpha \left( \frac{J_{\rm ce}'}{\sqrt{q_\rme}} \right)^2 
   \left\{ 
      3 t \tau_H + \frac{4}{\tau_\rmH} t^2-\frac{3}{\tau_\rmH} B_2^{(1)}(t) 
   \right\}.
\end{equation}

The numerical results are summarized in \fref{fig:comparisonKIRMT}. There we
find 9 plots for the evolution of purity in the KI chain. Each blue line
corresponds to a different realization of the initial condition
\eref{eq:InitialCondition}.  We include formula (\ref{eq:RMTKI}) with and
without the correlation term $\frac{3}{\tau_\rmH} B_2^{(1)}(t)$ as solid and
dashed black lines respectively.  We increase the size of the environment
(from bottom to top)to appreciate how the large dimension limit is
approached.  We also vary the plot range to examine different time
regimes of the problem. In the plots, two time scales are available.  The
lower scale is the number of time steps (number of applications of the
Floquet operator) while the upper one is time in units of the Heisenberg
time. 

We first encounter the Zeno regime, in which the discreteness of the spectrum
induces deviations of the RMT formulae; for a finite spectrum, what we take
as a $\delta$ function in \eref{eq:thecorrelation}, has a finite width and
height, causing an initial quadratic decay.  Indeed we can see on the left
most plots that the deviation lasts a few kicks. The effect is more visible
for $q_\rme=12$ and 14.  The next regime is the linear one, sometimes called
Fermi golden rule regime.  We observe linear decay as predicted by RMT
whenever $1\ll t \ll \tau_\rmH$. We use this regime to obtain the fitting
parameter $\alpha=0.21$.  Next we observe a crossover in which deviations
from the linear behavior are evident (starting at $t\approx \tau_\rmH/4$).
For $q_\rme=10$ this regime overlaps slightly with the Zeno regime. After
the crossover we find the Gaussian regime, in which a quadratic decay of
purity is manifest.  The clearest picture of this behavior is seen for
$q_\rme=10$ where the Heisenberg time is the smallest.

As we increase the size of the environment we observe better agreement
between the dynamical model and the RMT formula with no correlation.  For
$q_\rme=14$ the agreement is virtually perfect for all time regimes beyond
the Zeno regime.  Comparing the behavior for different environments, we can
observe the dramatic effect of the Heisenberg time on decoherence: The denser
the spectrum, the later we arrive to the Gaussian regime. Environments with
tighter spectra cause less damage to the central system. Other configurations
also show qualitative agreement with the RMT formula developed before.



\begin{figure}[t]
\begin{center} \includegraphics{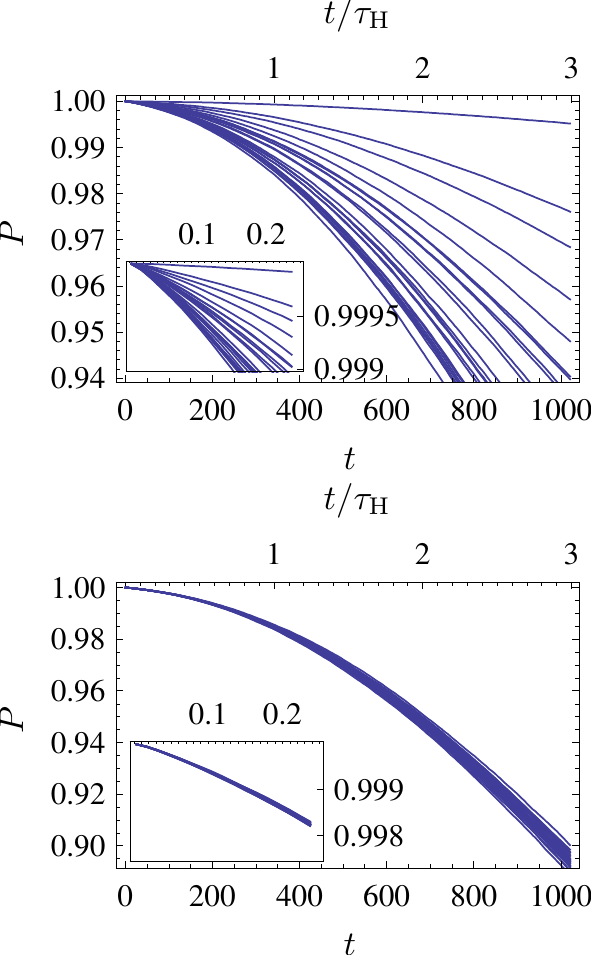} \end{center}
\caption{Evolution of 25 random initial conditions for configuration (d).  In
  the top panel we prepare separable initial conditions, whereas in the
  lower panel entangled ones.  We observe no self-averaging both for long
  times and short times (in the insets). This might be due to the separable
  structure of the coupling. $J_{\rm ce}=0.0005$ and $q_\rme=12$.}
\label{fig:noselfaveragingKIRMT}
\end{figure}

The results of \sref{sec:goeone} predict that for arbitrary separable
conditions, on a TRI system (or with any other anti-unitary symmetry), we
have a finite spreading of purity, even in the large dimension limit. This
effect becomes important after the Heisenberg time. We argued why we expect
self averaging (with respect to the initial conditions) for Bell states and
indeed observed it in the large dimension limit. 

We tested these two findings in our KI models. As expected, we found no self
averaging for separable ones. However, we observed the same effect before and
after the Heisenberg time. This anomaly can be due to the separable structure
of the coupling, which implies a preferred base within the central system.
From that point of view, this irregularity is analogous to the effect of an
internal Hamiltonian.  Interestingly, we again observe that for entangled
states we do have self averaging. This suggests that the ``transport'' of
symmetries via entanglement is a general property that is not limited to the
topics discussed during this thesis.

\chapter{Quantum memories}\label{sec:nqubit}

\newcommand{\jvec}{\vec j} 
\begin{quote} {\em Divide et impera.}
\end{quote}

Physical devices capable of storing faithfully a quantum state, \ie
quantum memories, are crucial for any quantum information task.
While different types of systems have been considered, most of the
effort is concentrated in manipulating qubits as they are essential
for most quantum information tasks \cite{NC00a}.  For a general
quantum memory (QM) it is necessary to record, store, and retrieve an
arbitrary state. For quantum communication and quantum computation
initialization in the base state, single qubit manipulations, a two
qubit gate (e.g. control-not), and single qubit measurements are
sufficient.  For arbitrary quantum memories recording and retrieving
the state is still an experimental challenge but initializing a qubit
register and individual measurement is mostly mastered.  Faithful
realization of two qubit gates together with better isolation from
the environment are the remaining obstacles for achieving fully
operational quantum technology.  However the huge effort done in the
community has born some fruit \cite{Julsgaard2004, Chou2005,
Chaneliere2005, Eisaman2005, cyclicQM, 2004quant.ph..4115P}.

Understanding decoherence of one, two, and $n$ qubits has been of
interest for some time. In \cite{Fedichkin2004} the authors calculate
decoherence of a QM for a spin-boson model using a measure designed for
their purposes.  In the limit of slight decoherence they obtain various
results including additivity.  Experimental results are also available
\cite{krojanski:090501, krojanski:062319}.  In the previous chapters we
have studied one and two qubit decoherence using both a random matrix
formulation and a specific dynamical model.  One would wish to
encompass some of this progress in a general picture.

In the present chapter we address this problem using a standard measure
of decoherence, namely purity.  We obtain analytic expressions for the
decoherence of a QM during the storage time.  Specifically we discuss a
QM composed of a set of individual qubits interacting with some
environment. The expressions given are based on previous knowledge of
the decoherence of a single qubit entangled with some non-interacting
spectator. Their validity is limited to small decoherence, \ie large
purity of the QM.  Note that the latter is not a significant restriction,
within the context of quantum information processing,
due to the high fidelity requirements of quantum error correction
codes.  We again assume that the entire system is subject to unitary
time-evolution, and that decoherence comes about by entanglement
between the central system (CS) and the environment.  Spurious
interactions inside the central system are neglected.  

A further and critical assumption is the independence of the coupling
of different qubits with the environment {\it in the interaction
picture}.  This is justified if the couplings are already independent
in the Schr\"odinger picture or if we have rapidly decaying
correlations due to mixing properties of the environment
\cite{purityfidelity}.  Physically, the first would be more likely if we
talk about qubits realized in different systems or degrees of freedom,
while the second seems plausible for many typical environments.

The central result is a decomposition of the decoherence of the full
QM, coupled to a single or several environments into a sum of terms.
Each of these describes the decoherence of a single qubit in a
``spectator configuration'' \eref{sec:spectatorintro}.
Recall, the CS consists of two non-interacting parts,
one (the qubit) interacting with the environment and the other (the
rest of the QM) not. This configuration is non-trivial if the two parts
of the CS are entangled. Apart from the above assumptions, this result
does not depend on any particular property of the environment or
coupling. Thus, it can be applied to a variety of models.

The general relation is obtained in linear response approximation and
leads to explicit analytic results if the spectral correlations of the
environment are known, see \sref{sec:nlinear}.  We test successfully the
results in our random matrix model (\sref{sec:nrmt}).  Finally, we
perform numerical simulations for four qubits, interacting with the
kicked Ising spin chain as an environment, in \sref{sec:nki}.

\section{The calculation}\label{sec:nlinear}

We now outline the calculation for $n$ qubits. Since it just represents
a variation in the indices of some calculations presented in appendix
\ref{sec:calculationSpec}, we skip some details and go through it
pretty fast. 

The central system (our QM) is composed of $n$ qubits. Thus, its
Hilbert space is $\mcH_\rmqm=\bigotimes_{i=1}^{n}\mcH_i$, where
$\mcH_i$ are the Hilbert spaces of the qubits.  The Hilbert space of
the environment is again denoted by $\mcH_\rme$. The Hamiltonian reads
\begin{equation}\label{eq:fullham}
H= H_\rmqm + H_\rme + \lambda V,\qquad 
\lambda V=\sum_{i=1}^{n} \lambda_i V^{(i)}.
\end{equation}
Here, $H_\rmqm=\sum_{i=1}^{n} H_i$, where $H_i$ acts on $\mcH_i$,
whereas $H_\rme$ describes the dynamics of the environment, and
$\lambda V$ the coupling of the qubits to the environment. The strength
of the coupling of qubit $i$ is controlled by the parameter
$\lambda_i$, while $\lambda=\max \{|\lambda_i|\}$. The (possibly
time-dependent) Hamiltonian gives rise to the unitary evolution
operator $U_\lambda (t)$.

We consider two different settings. In the first one $V^{(i)}$ acts on
the space $\mcH_i\otimes \mcH_\rme$, \ie all qubits interact with a
single environment called {\it joint environment}.  In the second one
each qubit interacts with a {\it separate environment}. Thus the
environment is split into $n$ parts, $\mcH_\rme = \bigotimes_{i=1}^{n}
\mcH_{\rme,i}$ and $V^{(i)}$, in \eref{eq:fullham}, acts only on
$\mcH_i\otimes \mcH_{\rme,i}$.  The first case would be typical for a
quantum computer, where all qubits are close to each other, while the
second would apply to a non-local quantum network.  Both configurations
are defined in perfect analogy with the ones, with corresponding names,
considered in \sref{sec:twoqubitmodel}.

Again we choose the initial state to be the product of a pure state of
the central system (the QM) and a pure state of the environment
\begin{equation}
|\psi_0\>=|\psi_\rmqm \> |\psi_\rme \>,
\quad |\psi_\rmqm \>\in \mcH_\rmqm,\, |\psi_\rme \>\in \mcH_\rme.
\label{eq:initialstate}
\end{equation}
In the separate environment configuration we furthermore assume that
$|\psi_\rme \>=\bigotimes_i |\psi_{\rme,i }\>$ with $|\psi_{\rme,i} \>
\in \mcH_{\rme,i}$, which corresponds to the absence of quantum
correlations among the different environments.

Purity evolves as $P(t)=\tr \rho_\rmqm ^2(t)$, with $\rho_\rmqm(t)=
\tr_\rme U_\lambda (t) |\psi_0\> \<\psi_0| U^\dagger_\lambda (t)$. To
calculate purity (or any other quantity that depends solely on the
Schmidt coefficients) we replace the forward time evolution $U_\lambda
(t)$ by the echo operator $M(t)=U_0 (t)\, U_\lambda (-t)$ where $U_0
(t) $ gives the evolution without coupling, recall
\sref{sec:generalprogram}. Using Born expansion for the echo operator
and defining $\lambda \tilde V_t = U_0 (t) \lambda V U_0^\dagger (t)$
(the coupling at time $t$ in the interaction picture)  purity is then
given by
\begin{equation}
\label{eq:purityA}
	P(t)=1-2\lambda^2 \into {\rm Re}\,
           A(\tau,\tau')+\Or\left(\lambda^4\right), 
\end{equation}
with 
\begin{multline}
 A(\tau,\tau')= 
  p[\tilde V_\tau \tilde V_{\tau'} \varrho_0\otimes\varrho_0] - 
    p[\tilde V_{\tau'} \varrho_0 \tilde V_\tau \otimes\varrho_0]    \\
  +p[\tilde V_\tau \varrho_0\otimes \tilde V_{\tau'} \varrho_0] 
    - p[\tilde V_{\tau'} \varrho_0\otimes \varrho_0 \tilde V_\tau] , 
\end{multline}
and $\varrho_0=|\psi_0 \>\<\psi_0|$ [recall  that $p [ \rho_1 \otimes
\rho_2]= \tr( \tr_\rme  \rho_1 \tr_\rme \rho_2)$].  Note that the
linear terms vanish identically after tracing out the environment.
Considering the form of the coupling in \eref{eq:fullham} we can
decompose $\lambda^2 A(\tau,\tau')=\sum_{i,j} \lambda_i \lambda_j
A^{(i,j)}(\tau,\tau')$ with
\begin{multline}\label{eq:AIJdecom} 
A^{(i,j)}(\tau,\tau')=
	p[\tilde V^{(i)}_\tau \tilde V^{(j)}_{\tau'} \varrho_0
		\otimes\varrho_0]
	- p[\tilde V^{(i)}_{\tau'} \varrho_0 \tilde V^{(j)}_\tau 
		\otimes\varrho_i]\\
	+ p[\tilde V^{(i)}_\tau \varrho_0
		\otimes \tilde V^{(j)}_{\tau'} \varrho_0]
	- p[\tilde V^{(i)}_{\tau'} \varrho_0
		\otimes \varrho_0 \tilde V^{(j)}_\tau] . 
\end{multline}
$\tilde V^{(i)}_t= U_0^\dagger(t)\, V^{(i)}  U_0(t)$ in analogy with
$\tilde V_t$.  Equation~(\ref{eq:purityA}) is given as a double sum in
the indices of the qubits.  In a diagonal approximation ($i=j$), $P(t)$
is expressed in terms of the purities $P^{(i)}_{\rm sp}(t)$ which
correspond to the purity decay of the CS in a spectator configuration
where only qubit $i$ is interacting with the environment. Purity then
reads as
\begin{eqnarray}\label{eq:spectatorpurity} 
  P(t)&=& 1- \sum_{i=1}^n\left( 1- P^{(i)}_{\rm sp}(t) \right),\\ 
  P^{(i)}_{\rm sp}(t)&=& 
  	1- 2 \lambda_i^2\into A^{(i,i)} (\tau,\tau') 
		+ \Or\left(\lambda_i^4\right).\nonumber 
\end{eqnarray}
This is our central result of this chapter. The diagonal approximation
is justified in two situations: First if the couplings $V^{(i)}$ of the
individual qubits are independent from the outset, as would be typical
for the separate environment configuration or for the random matrix
model of decoherence. Second, if the couplings in the interaction
picture become independent due to mixing properties of the environment,
as would be typical for a ``quantum chaotic'' environment.

\section{A RMT example}\label{sec:nrmt} 

To illustrate the case of independent couplings mentioned above, random
matrix theory provides a handy example. Such models were discussed in
previous chapters and in \cite{pinedaRMTshort, pinedalong,
1464-4266-4-4-325, otroRMT}. 

In chapter \ref{sec:twoqubit} purity decay was computed in
linear response approximation for two qubits, one of them being the
spectator. For the sake of simplicity we choose the joint environment
configuration, no internal dynamics for the qubits, and $H_\rme$ and
$V^{(1)}$ as typical members of the Gaussian unitary ensemble. Purity
is then given by
\begin{eqnarray}
P^{(1)}_{\rm sp}(t) &=& 1-\lambda_1^2 (2-p_1) f(t); \label{eq:gue}\\
f(t) &=& t\, \max \{t,\tau_\rmH \} 
   + \frac{2}{3\tau_\rmH}\, \min \{ t,\tau_\rmH\}^3.
\end{eqnarray}
Here, $\tau_\rmH$ is the Heisenberg time of the environment and $p_1$ is
the {\it initial} purity of the first qubit alone, which measures its
entanglement with the rest of the QM, see
\eref{eq:spectatorDegenerate}. As we required  a priori independence of
the couplings, we can now insert \eref{eq:gue} in
\eref{eq:spectatorpurity} to obtain the simple expression
\begin{equation}\label{eq:sepgen}
	P(t)= 1-f(t) \sum_{i=1}^{n}\lambda_i^2 (2-p_i),
\end{equation}
where $p_i$ is the initial purity of qubit $i$. In the presence of
internal dynamics the spectator result is also known and can be
inserted. As an example, we apply the above equation to an initial GHZ
state ($(|0\cdots0\>+ |1\cdots1\>)/\sqrt{2}$). Then all $p_i=1/2$ and
we obtain $P(t)=1-(3/2)f(t)\sum_i \lambda_i^2$. For a W state
$(|10\ldots0\>+|01\ldots0\>+\cdots+|00\ldots1\>)/\sqrt{n}$, purity for
each qubit is $p_i=(n^2-2n+2)/n^2 \approx 1$, in the large $n$ limit,
and purity decays as $P(t)=1-f(t)\sum_i \lambda_i^2$.  

\begin{figure}
\begin{center} \includegraphics{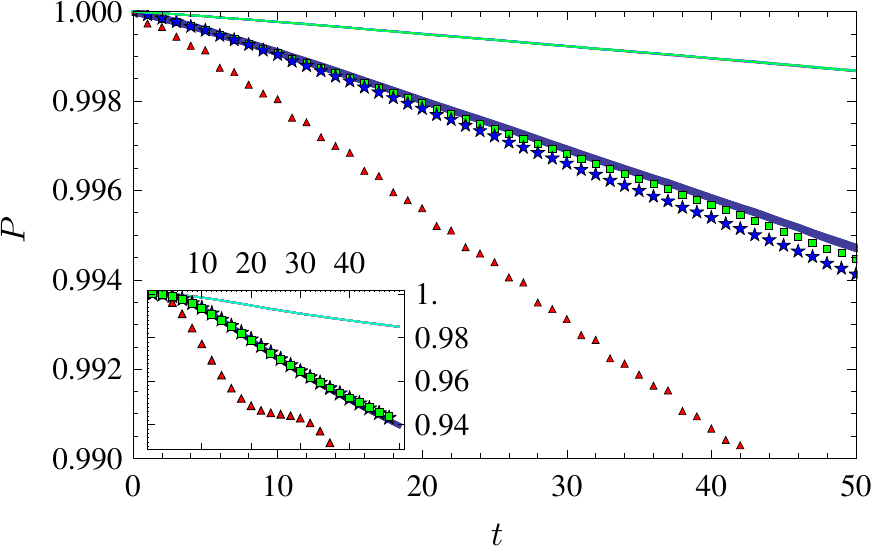} \end{center}
\caption{ Purity decay for a GHZ state is shown for the environment
  described by \eref{eq:hamkie}. The couplings are: all qubits to one
  spin (red triangles), nearby spins (blue stars), and maximally
  separated spins (green squares). The theoretical result (thick line)
  is calculated in terms of the $P^{(i)}_{\rm (sp)}(t)$ (thin lines).
  The main figure corresponds to the case of a mixing environment and
  the inset for an integrable one.  }
\label{fig:kichaos}
\end{figure}

\section{A dynamical model}\label{sec:nki}

We now use the homogeneous kicked spin (\sref{sec:KIchain}) as an environment
for $n$ qubits.  We consider its chaotic and integrable
regimes. We shall thus use a particular realization of
\eref{eq:fullhamiltonian} in the spirit of chapter \ref{sec:ki}.

The main assumption in \eref{eq:spectatorpurity} is the fast decay of
correlations for couplings of different qubits to the environment. For the
random matrix model discussed above this is trivially fulfilled.  Yet,
integrable environments are commonly used~\cite{caldeiraleggett} and one may
wonder whether \eref{eq:spectatorpurity} can hold in such a context. We shall
therefore study a dynamical model where a few qubits are coupled to an
environment represented by a kicked Ising spin chain using identical coupling
operators for all qubits.  In this model, the variation of the angle of the
external kicking field allows the transition from a ``quantum chaotic'' to an
integrable Hamiltonian for the environment~\cite{prosenKI, pp2007}(a).  For
convenience we separate it in the environment and the central system.  The
Hamiltonian of the environment, namely the homogeneous kicked Ising spin
chain, recall, is given by
\begin{equation}
H^{(\rme)}= \sum_{i=0}^{L-1}  \sigma_x^{(\rme,i)} \sigma_x^{(\rme,i+1)}
 +\boldsymbol{\delta}(t) \sum_{i=0}^{L-1} b \cdot \sigma^{(\rme,i)} 
\label{eq:hamkie}
\end{equation}
where $\boldsymbol{\delta}(t)=\sum_{n\in \mathbb{Z}} \delta(t-n)$ (\ie
time is measured in units of the kick period), $L$ the number of spins
in the environment, $\sigma^{(\rme,i)}= (\sigma_x^{(\rme,i)},
\sigma_y^{(\rme,i)}, \sigma_z^{(\rme,i)})$ the Pauli matrices of spin
$i$, and $b$ the dimensionless magnetic field with which the chain is
kicked ($\hbar=1$). We close the ring requiring $\sigma^{(\rme,L)}
\equiv \sigma^{(\rme,0)}$. The Hamiltonian of the QM is
\begin{equation}
  H^{(\rmqm)}=\boldsymbol{\delta}(t) \sum_i  b \cdot
\sigma^{(\rmqm,i)},
\end{equation}
where $\sigma^{(\rmqm,i)}$ is defined similarly as
for the environment. The coupling is given by 
\begin{equation}
\lambda V=\lambda \sum_i \sigma_x^{(\rmqm,i)} \sigma_x^{(\rme,j_i)}. 
\end{equation}
The $j_i$'s define the positions where the qubits of the QM are coupled
to the spin chain.  Equivalently we could  use kicked Ising couplings
and a time-independent field \cite{dodo}.

To implement a chaotic environment, we use here $b=(0.9,0.9,0)$.   We
use a ring consisting of $L=12$ spins for the environment and $4$
additional spins for the qubits of the QM. The coupling strength is
$\lambda=0.005$. In Fig.~\ref{fig:kichaos}, we study purity decay when
all four qubits are coupled to the same spin, to neighboring spins, and
to maximally separated spins, $\vec j=(0,3,6,9)$. The initial state is
the product of a GHZ state in the QM and a random pure state in the
environment. We compare the results with \eref{eq:spectatorpurity}
(thick line) obtained from simulations of the spectator configuration
(thin solid line). Coupling the spectator to different positions in the
chain yields near identical results for $P(t)$, so we can see only one
line.  The figure demonstrates the validity of
\eref{eq:spectatorpurity} for well separated and hence independent
couplings.  For coupling to neighboring spins decay is slightly faster
while the sum rule does not hold if all qubits are coupled to the same
spin.

\begin{figure}
\begin{center} \includegraphics{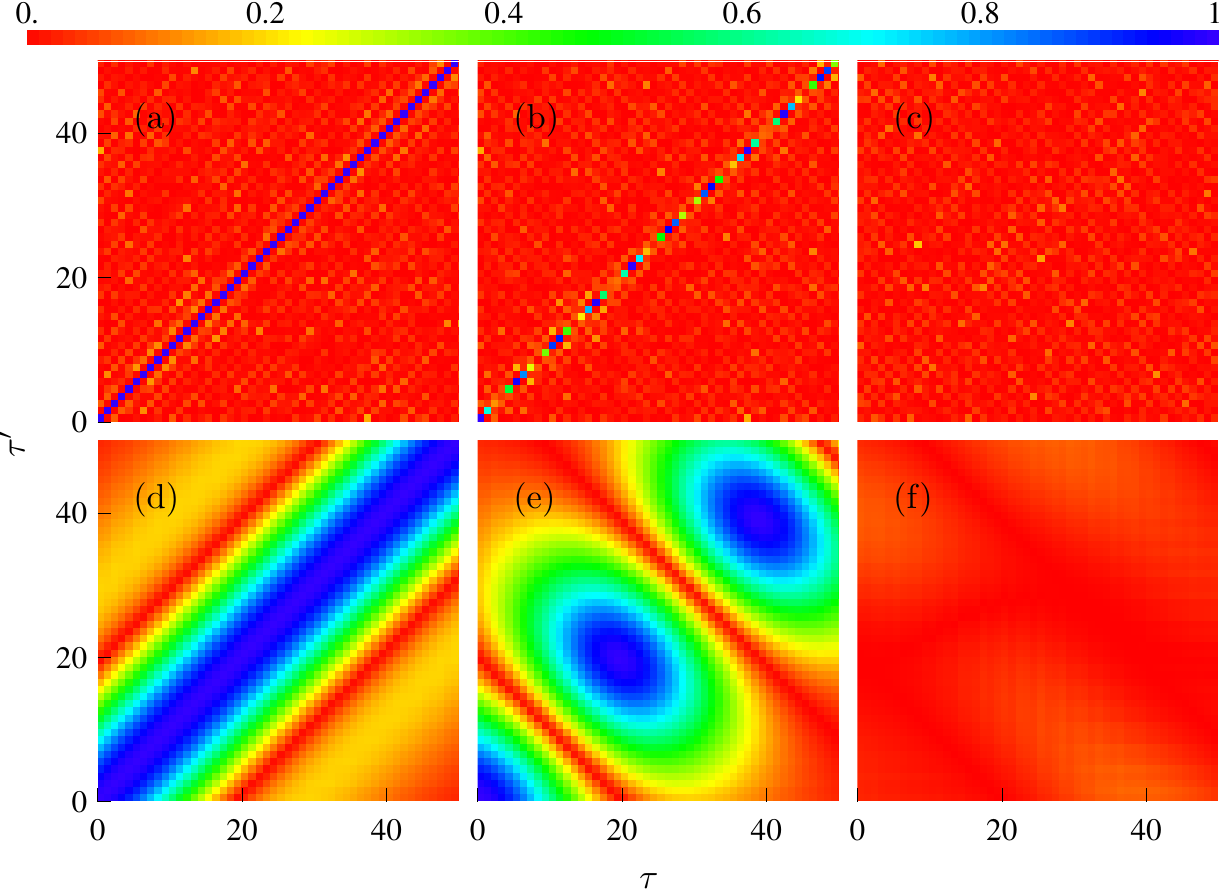} \end{center}
\caption{ The color-coded absolute value of the real part of
  $\<\psi_0|\tilde V^{(i)}_\tau \tilde V^{(j)}_{\tau'} |\psi_0\>$ given
  in \eref{eq:domterm} as a function of $\tau$ and $\tau'$ for the
  chaotic [(a) (b) (c)] and the integrable [(d) (e) (f)] spin chain
  environment ($L=8$).  (a),(d) show the autocorrelation function
  ($i=j$); (b), (e) and (c) (f) the cross correlation function for
  couplings to the same spin or opposite spins in the ring
  respectively.  }
\label{fig:kicorrel}
\end{figure}

A similar calculation for integrable environments  with $b=(0,1.53,0)$
yields  faster purity decay than in the mixing case as expected from
general considerations in \cite{reflosch}.  Nevertheless the sum rule
is again well fulfilled except if we couple all qubits to the same
spin.  This leads us to check the behavior of the correlation function
\begin{equation}\label{eq:domterm}
p[\tilde V^{(i)}_\tau \tilde V^{(j)}_{\tau'}
\varrho_0\otimes\varrho_0] = \<\psi_0|\tilde V^{(i)}_\tau \tilde
V^{(j)}_{\tau'} |\psi_0\>. 
\end{equation}
This is the simplest and usually largest term in $A^{(i,j)}$,
\eref{eq:AIJdecom}. Figure \ref{fig:kicorrel} shows this quantity for chaotic
(\ie mixing) and integrable environments in the first and second row,
respectively.  The first column shows the autocorrelation function ($i=j$).
The second and third columns give the cross correlation function ($i \ne j$)
when the qubits are coupled to the same or opposite spins, respectively.  In
the latter case, correlations are always small thus showing that also in
integrable situations our condition can be met.  Non-vanishing correlations,
as in \fref{fig:kicorrel}(b) and \fref{fig:kicorrel}(e) lead to deviations
from the sum rule. The pronounced structure for the integrable environment,
\fref{fig:kicorrel}(e), may be associated to the oscillations of purity decay
(inset of \fref{fig:kichaos}).

Another example where the conditions of our derivation are not met is
the following. A Bose-Einstein condensate in which the $n$ atoms have
two immiscible internal states \cite{dalvitBEC} could be interpreted as
a QM.  The symmetry of the wave function reduces the dimension of the
Hilbert space and causes high correlations among the couplings to the
environment.  It is not surprising then that decoherence scales
differently (as $n^2$).

\chapter{Conclusions}

We now proceed to formulate the conclusions of this work. We start by
particular conclusions for each chapter, and afterwards to the general
conclusions and perspectives. 


In chapter \ref{sec:onequbit} we analyzed the decoherence (measured by purity
during all the thesis) of a single qubit immersed in an environment. For this
purpose we introduced a model with unitary dynamics in the qubit plus
environment space.  Both the environment Hamiltonian and the interaction were
chosen as random matrices from the classical ensembles.  We solved purity
decay in terms of correlation functions of the environment. 

We obtained linear and quadratic decay before and after the Heisenberg time
respectively. In general, purity of a superposition of eigenstates decays
faster than an eigenstate. The presence of an internal Hamiltonian in the
qubit was observed to contribute to stability. For the GOE case, different
initial conditions in the Bloch sphere yield different behaviors of
purity and von Neumann entropy, even in the large environment limit.

Monte Carlo simulations showed that exponentiating the analytical
formulae, as explained in appendix \ref{sec:exponentiation}, allowed
extending the analytical result outside the scope of linear response
formalism.


In chapter \ref{sec:twoqubit} we used the spectator configuration (introduced
in the introduction of the thesis), in which only a part (one qubit) of the
central system (two qubits) is coupled to an environment.  This constituted
the first of three models analyzed.  We solve it  using
extensively the symmetries of the ensembles. We found some elements in common
with the single qubit case: the same qualitative dependence on time (linear
and quadratic before and after the Heisenberg time, respectively; a dependence
on an internal Hamiltonian, namely having a big energy splitting in the qubit
increases stability; finally, a reasonable agreement with a heuristic
exponentiation.  With the second qubit we added a new element, entanglement.
We observed that entanglement allowed taking the invariance properties of one
qubit into the other.  We also observed that entanglement induces a faster
decoherence. For the TRI breaking case we found that, due to entanglement,
self averaging is expected when using Bell states.

The other two models (joint environment and separate environment), which
possess a very different physical interpretation, were reduced to the
spectator configuration using the independence of the couplings. Their
behavior is similar, up to a scaling of the coupling constant.


In chapter \ref{sec:concurrence} we studied, for two qubits coupled to an
environment, the relation between purity and concurrence as both evolve in
time. Both the environment and the coupling are chosen from the classical
ensembles. In order to analyze their relation we introduce the $CP$ plane. We
found that in the large dimension limit of the environment, concurrence and
purity are restricted to a small region defined using unital channels. In
fact, in the large dimension limit, for all parameters studied, the channels
induced by the random Hamiltonian approach rapidly the unitality condition.
For large purities the allowed unital region is almost one dimensional, which
allows to write a one to one relation for concurrence and purity.  In some
cases the curve described by the system in the $CP$ plane coincides with the
Werner curve.  The deviation from it, for a particular level splitting in the
coupled qubit, is given in an empirical formula which depends on the coupling
strength and the size of the environment. This deviation goes rapidly to zero
as the size of the environment and the coupling increase. Assuming that
purity and concurrence are related via the Werner curve, it is possible to
use our knowledge from the previous chapters to give a formula for
concurrence decay.


In chapter \ref{sec:ki} we studied the same problem as in previous chapters
(one and two qubit decoherence), but using particular dynamical models. The
models consists of $n$ spins interacting pairwise with individually
adjustable strengths, and kicked with a magnetic field.  The dynamical system
is flexible enough to allow different configurations and symmetries of the
environment and coupling.  Moreover, varying the parameters we can have
qualitatively different kinds of dynamics, namely chaotic, intermediate 
and integrable.  In the chaotic regime of the
environment we obtained quantitative agreement with the RMT formulae obtained
earlier for purity decay.  For the integrable case we observed quadratic
decay, sometimes with strong revivals (when the coupling was to an open end
of a chain).  The intermediate case displays oscillations superimposed to a
linear decay. In the chaotic case we observe that the effective Heisenberg
time (determined, among other things, by the symmetry of both the environment
and the coupling) plays an important roll in decoherence.

Purity and concurrence behave in a qualitatively similar way for
all configurations.  The relation between the two, for sufficiently large
environments, follows an analytic expression we get for Werner states quite
closely, although the specific dynamics does not produce Werner states.


Finally, in chapter \ref{sec:nqubit}
we calculated generic decoherence  of a $n$-qubit quantum memory
as represented by purity decay. For large purities it is globally given
in linear response approximation by a sum rule in terms of the purities
of the individual qubits entangled with the remaining register as
spectator. This sum rule depends crucially on the absence or rapid decay
of correlations between the couplings of different qubits in the
interaction picture. We prove that this is fulfilled by sufficiently
chaotic environments or couplings but, using a spin chain model as an
environment, we find that even if the latter is integrable correlations
can be absent and the sum rule holds.  While exceptions can be
constructed we have a very general tool to reduce the decoherence of a
QM of $n$ qubits to the problem of a single qubit entangled with the
rest of the QM. Furthermore the result is equally valid if we perform
local operations on the qubits.


Some possible  extensions or generalizations of the  thesis are now
outlined. 
\begin{itemize}
\item Several measures for entanglement of multi-qubit systems exist. 
Depending on the particular problem to be treated, one or the other 
might be useful. Preliminary results \cite{steffanprivate2007} indicate 
that for multi-qubit systems a relation, like the one found for
two qubits, might be used to obtain formulae for entanglement 
decay.
\item In order to deepen our knowledge of the behavior of integrable
and intermediate systems, one should study particular models.
\item As many interactions are indeed separable, a RMT formulation
for separable interactions of the qubit(s) with the environment
might proof useful to understand the similarities and differences with
this ``more realistic'' coupling. 
\item The sum rule studied in chapter \ref{sec:nqubit} also applies
if the register is split into arbitrary sets of qubits. In fact, one and
two qubit gates are known to be universal for quantum computation
\cite{NC00a}, so we lay the foundation for the computation of
decoherence during the execution of a general algorithm.  This
extension only requires the knowledge of the decoherence of a pair
suffering the gate operation while entangled with the rest of the
register. As each gate is different, we will have to take a step by
step approach, which at each step will only involve one and two qubit
decoherence. The linear response approximation will be sufficient due
to the high fidelity requirements of quantum computation.
\end{itemize}

Summarizing, we have developed a tool to study decoherence of qubits based on
the spectator configuration and on random matrix theory.  We have compared
the results found with a kicked Ising model.  We almost always observed
linear/quadratic decay of purity before/after the Heisenberg time of the
environment and self averaging both with respect to the Hamiltonian and to
the initial conditions. The exceptions are: (i) In the presence of an
internal Hamiltonian, the central system eigenstates decay linearly even
after Heisenberg time. (ii) If we have a time reversal invariant Hamiltonian,
no self averaging with respect to the initial conditions is observed after
the Heisenberg time. (iii) In the KI model, and presumably due to the
separable structure of the Hamiltonian, no self averaging, before the
Heisenberg time, was observed.  Entanglement enhances decoherence and may
transport symmetry properties from one qubit to the other.  When studying two
qubits we also studied their internal entanglement, as quantified by
concurrence. A one to one correspondence was often found, characterized by
the Werner curve. Thus, formulae for entanglement decay are also available.
For a larger number of qubits a sum rule can be applied in a wide range of
circumstances to express purity decay of a group of particles in terms of
purity decay of each particle in a spectator configuration. In all cases,
heuristic exponentiation leads to good agreement beyond linear response
scope.

\appendix \noappendicestocpagenum \addappheadtotoc
\renewcommand{\chaptername}{Appendix}

\chapter{One-qubit purity decay for pure and mixed states}
\label{sec:calculationSpec}
Here, we consider the one-qubit decoherence in a more general context.
One-qubit decoherence as described in \sref{sec:onequbit} deals with a
separable initial state $\varrho_0= \rho_1 \otimes |\psi_{\rm e}\> \<\psi_{\rm
e}|$, $\rho_1= |\psi_1\> \<\psi_1|$. Thus $\varrho_0$ is a pure state of a
single qubit coupled to an environment in the pure state $|\psi_{\rm e}\>$. We
consider here arbitrary, finite dimensions of $\mcH_1$ and $\mcH_{\rm e}$, and
we allow $\rho_1$ to be mixed. The latter poses some problems on the physical
interpretation of purity as a measure for decoherence. However, these can be
avoided by taking the point of view of the environment, as it becomes entangled
with the central system.  That means we consider the purity of the state of the
environment after tracing out the central-system's degrees of freedom; see
\eref{eq:bothpuritiesequal} and \eref{eq:ahora}.  If we consider the
entanglement with a spectator being responsible for the mixedness of $\rho_1$,
we can use the following results to describe purity decay in the spectator
model.

We will derive explicit expressions for the average purity $\< P(t)\>$ as a
function of time, where the average is only over the random coupling.  These
expressions involve a few elemental spectral correlation functions, whose
properties will be discussed in \ref{aB}.  As detailed in
\sref{sec:generalprogram}, we work with the echo operator $M(t)$ in the linear
response approximation \eref{eq:bornexpansion}. Note that $I(t)$ is Hermitian,
while, due to time-order inversion, $J(t)$ is not.

This section is slightly boring, but since this is a Ph.D. thesis, the details
must be incorporated. My thanks an admiration to the devoted reader who 
reached this section.

\section{General calculation}
Let $\varrho_1$ and $\varrho_2$ be two operators acting on the Hilbert
space $\mcH'= \mcH_1\otimes\mcH_{\rm e}$. We define the purity form $p[\,
\cdot\, ]$ as a function of pairs of such operators,
\begin{equation}\label{aA:defP}
  p[\varrho_1\otimes\varrho_2]= \tr_\e[\tr_1(\varrho_1)\,
\tr_1(\varrho_2)]\; .
\end{equation}
Since any linear operator acting on $\mcH' \otimes \mcH'$ can be expanded in
terms of separable operators of the form $\varrho_1\otimes\varrho_2$,
linearity implicitly defines the purity form for arbitrary linear operators.
For arbitrary operators $A,B$ acting both on $\mcH'$, the purity form
has the
following property:
\begin{equation}\label{B:pfprop}
  p[A \otimes B] = p[A^\dagger \otimes B^\dagger]^* = p[B \otimes A] =
p[B^\dagger \otimes A^\dagger]^* \; .
\end{equation}
The purity defined in \eref{eq:defpurity} can be expressed in terms of
the purity form as
\begin{equation}
P(\tr_\e \varrho)= p[\varrho\otimes\varrho] \; .
\end{equation}
Note that on the RHS of this equation, we first take the trace over the central
system, \ie we consider the purity of the state of the environment after
tracing out the central degrees of freedom. With this little twist we can
describe the one-qubit decoherence and the spectator model at the same time.

To average purity, we use \eref{eq:bornexpansion} to compute
$\varrho^M(t)\otimes \varrho^M(t)$ in linear response approximation
\begin{equation}
\< P(t)\>= \<\, p[\varrho^M(t)\otimes\varrho^M(t)]\, \>
         = p[\< \varrho^M(t)\otimes\varrho^M(t)\>] \; .
\end{equation}
Keeping terms only up to second order in $\lambda$:
\begin{align}
\varrho^M(t)&\otimes \varrho^M(t) = \varrho_0 \otimes \varrho_0 \\
&- \mimath\lambda \left[I\varrho_0\otimes\varrho_0 
     - \varrho_0 I^\dagger\otimes\varrho_0 + \varrho_0\otimes I\varrho_0 
     - \varrho_0 \otimes \varrho_0 I \right]^\dagger  \nonumber\\
& - \lambda^2 \left[ J\varrho_0 \otimes \varrho_0 
     + \varrho_0 J^\dagger\otimes \varrho_0 
     + \varrho_0 \otimes J\varrho_0 
     + \varrho_0 \otimes \varrho_0 J \right]^\dagger \nonumber\\
& + \lambda^2  \big[ I \varrho_0 I^\dagger \otimes \varrho_0 
     - I \varrho_0 \otimes I\,\varrho_0 
     + I \varrho_0 \otimes \varrho_0 I^\dagger  \nonumber \\
&\quad\quad\quad\quad +\varrho_0 I^\dagger \otimes I \varrho_0 
     - \varrho_0 I^\dagger \otimes \varrho_0 I^\dagger 
     + \varrho_0 \otimes I\, \varrho_0 I^\dagger \big].
 \nonumber
\end{align}
In the next step, we perform the ensemble average over the coupling. 
The odd terms in $\lambda$ vanish as purity (even without averaging) must be
real.  For the remaining terms, we use the properties in \eref{B:pfprop} and
the fact that $I(t)$ is Hermitian, to obtain
\begin{equation}\label{eq:pajai}
\< P(t)\> = P(0) -\lambda^2 (A_J - A_I)
\end{equation}
with $P(0)=p[\varrho_0\otimes\varrho_0]$ and
\begin{align}
A_J &= 4\, {\rm Re} p[\< J\>  \varrho_0\otimes\varrho_0]\nonumber\\
A_I &= 2\left( p[\< I \varrho_0 I\> \otimes \varrho_0]
   - {\rm Re} p[\< I \varrho_0 \otimes I \varrho_0\>]
   + p[\< I \varrho_0\otimes\varrho_0 I\>] \right) .
\end{align}
In order to pull out the same double integral of $A_J$ and $A_I$, we use the
time ordering symbol $\mathcal{T}$. It allows to write for the average
purity
as a function of time
\begin{equation}
\< P(t)\>= P(0) -2\lambda^2\int_0^t\rmd\tau\int_0^t\rmd\tau'\;
   {\rm Re}\, A_{\rm JI}\label{B:pulrdef}\\
\end{equation}
with
\begin{multline}\label{B:AJIdef}
A_{\rm JI}=
   p[\mathcal{T}\<\tilde V \tilde V'\> \varrho_0\otimes\varrho_0]
   - p[\<\tilde V' \varrho_0 \tilde V\> \otimes\varrho_0] \\
   + p[\<\tilde V \varrho_0\otimes \tilde V' \varrho_0\>]
   - p[\<\tilde V' \varrho_0\otimes \varrho_0 \tilde V\>] \; ,
\end{multline}
where $\tilde V$ and $\tilde V'$ are short forms for the coupling matrices
$\tilde V(\tau)$ and $\tilde V(\tau')$, respectively. The arguments $\tau$ and
$\tau'$ of the coupling matrices are interchanged in the second and fourth term
of $A_{\rm JI}$. This does not change the value of the integral of course, but
it facilitates to handle common terms in the following considerations.

The integrand $A_{\rm JI}$ in \eref{B:AJIdef} consists of four terms.
These terms will be considered, one after the other. We will always first
average the argument of the purity form. Only then we perform the partial
traces over subsystem $\mcH_1$, and therefore the final trace over the
environment.
The averaging is only over the coupling, and it is done by applying two
simple rules, as described below.

For the coupling $V_{1,{\rm e}}$ in the product eigenbasis of $H_0$, we
use either random Gaussian unitary (GUE) or orthogonal (GOE) matrices.
Their
statistical properties are completely characterized by the following second
moments (for notational ease we ignore the subscript $_{1,{\rm e}}$ for a
moment)
\begin{equation}
\< V_{ij} V_{kl}\> = \delta_{il}\, \delta_{jk}
   + \chi_{\rm GOE}\; \delta_{ik}\, \delta_{jl} \; ,
\label{B:GEdef}
\end{equation}
where $\chi_{\rm GOE} = 1$ if $V$ is taken from the GOE, while
$\chi_{\rm GOE} = 0$ if $V$ is taken from the GUE. In the interaction
picture
we then find:
\begin{equation}
\< \tilde V_{ij}\, \tilde V_{kl}\> = \big (\, \delta_{il}\,
   \delta_{jk}\; \rme^{-\mimath (E_j-E_i)\, (\tau-\tau')} + \chi_{\rm GOE}\;
   \delta_{ik}\, \delta_{jl}\; \rme^{-\mimath (E_j-E_i)\,
(\tau+\tau')}\, \big )
   \; .
   \label{B:tVextdef}
\end{equation}

\section{The GUE case}
\label{SU}
We now procede to calculate one by one the contributions.

\paragraph{\boldmath $p[\mathcal{T} \< \tilde V \tilde V'\>
\varrho_0\otimes\varrho_0](t)$:}
We first compute the average
\begin{align}
\mathcal{T}\< \tilde V \tilde V' \varrho_0 \otimes \varrho_0\> &=
   \sum_{ij kl mn} |ij\>\; \mathcal{T}\< \tilde V(\tau)_{ij,kl}
   \tilde V(\tau')_{kl,mn}\>  \< mn| \varrho_0\otimes\varrho_0
\nonumber\\
&= \sum_{ij kl} |ij\>  \rme^{-\mimath (E_{kl}-E_{ij}) |\tau-\tau'|}
   \< ij| \varrho_0 \otimes \varrho_0  ,
\end{align}
where we have used the averaging rule, \eref{B:tVextdef}, as well as
the fact that the time-ordering operator $\mathcal{T}$ requests to
exchange $\tau$ with $\tau'$ whenever $\tau < \tau'$.  The indices
$i$, $k$ and $m$ denote basis states in $\mcH_1$, while the indices
$j$, $l$ and $n$ denote basis states in $\mcH_\rme$. We can rewrite
that expression in a more compact form by employing the diagonal
matrices $\bm{C}_x(\tau)$, defined in \eref{B:CCxdef}:
\begin{multline}
\mathcal{T}\< \tilde V \tilde V' \varrho_0 \otimes \varrho_0\> =
  \left(\bm{C}_1(|\tau-\tau'|) \otimes \bm{C}_{\rm e}(|\tau-\tau'|) \right) \\
\times   \left(\rho_1 \otimes |\psi_{\rm e}\>\<\psi_{\rm e}|\right)
   \otimes
   \left(\rho_1 \otimes |\psi_{\rm e}\> \<\psi_{\rm e}|\right).
\end{multline}
Now we apply the partial traces over subsystem $\mcH_1$
\begin{equation}
\tr_1\big (\, \< \tilde V\, \tilde V'\>\, \varrho_0\, \big )\,
\tr_1\, \varrho_0 = C_1(|\tau-\tau'|)\; \bm{C}_{\rm e}(|\tau-\tau'|)\,
   |\psi_{\rm e}\>\, \<\psi_{\rm e}| \; .
\end{equation}
The final trace over the environment yields
\begin{equation}
p[\mathcal{T}\< \tilde V\, \tilde V'\> \varrho_0\otimes\varrho_0](t)=
   C_1(|\tau-\tau'|)\; C_{\rm e}(|\tau-\tau'|) \; .
\end{equation}

\paragraph{\boldmath $p[\< \tilde V' \varrho_0 \tilde V\> \otimes
\varrho_0](t)$:}
We first compute the average
\begin{align}
\<\tilde V' \varrho_0 \tilde V\> \otimes \varrho_0 &=
   \sum_{ij kl mn pq} |ij\> \left\< \tilde V'_{ij,kl}
   \< kl|\varrho_0|mn\> \tilde V_{mn,pq}\right\>  \< pq| \otimes
\varrho_0
\nonumber\\
&= \sum_{ij kl} |ij\>\rme^{-\mimath (E_{ij}-E_{kl}) (\tau-\tau')}
   \< kl|\varrho_0| kl\> \< ij| \otimes \varrho_0 ,
\end{align}
where we have used \eref{B:tVextdef}.
We apply the partial traces over subsystem $\mcH_1$
\begin{align}
\tr_1\left( \<\tilde V' \varrho_0 \tilde V\> \right) \tr_1\varrho_0 
 &= \left( \sum_{ik} \rme^{-\mimath (E_i-E_k) (\tau-\tau')} 
       \< k|\rho_1|k\> \right) \\
&\qquad \times
   \sum_{jl} |j\> \rme^{-\mimath (E_j-E_l) (\tau-\tau')}
   \< l|\psi_{\rm e}\> \< \psi_{\rm e}|l \> \< j|\psi_{\rm e}\>
   \< \psi_{\rm e}| \nonumber\\
&= C_1(\tau-\tau')
   \sum_{jl} |j\>\; \rme^{-\mimath (E_j-E_l)\, (\tau-\tau')} \nonumber\\
 & \qquad \times
   \< l|\psi_{\rm e}\>\, \< \psi_{\rm e}|l \>\, \< j|\psi_{\rm e}\>\,
   \< \psi_{\rm e}| \nonumber .
\end{align}
The final trace over the environment yields
\begin{align}
p[\<\tilde V' \varrho_0 \tilde V\> \otimes \varrho_0]
 &= C_1(\tau-\tau') \sum_{jl} \<\psi_{\rm e}|l\> \rme^{\mimath E_l (\tau-\tau')}
   \< l|\psi_{\rm e}\>\<\psi_{\rm e}|j\> 
   \rme^{-\mimath E_j (\tau-\tau')} \< j|\psi_{\rm e}\> \nonumber\\
&= C_1(\tau-\tau') S_{\rm e}(\tau-\tau')  ,
\end{align}
where $S_x(\tau)$ is defined in \eref{B:Sxdef}.

\paragraph{\boldmath $p[\< \tilde V \varrho_0\otimes \tilde V'
\varrho_0\>](t)$:}
We first compute the average
\begin{align}
   \<\tilde V\, \varrho_0\otimes \tilde V'\, \varrho_0\>& =
   \sum_{ij\, kl\, mn\, pq} |ij\>\; \left\< \tilde V_{ij,kl}\; \< kl|\,
   \varrho_0 \otimes |mn\>\; \tilde V_{mn,pq}\right\>  \< pq|\,
   \varrho_0 \nonumber\\
&= \sum_{ij\, kl} |ij\>\; \rme^{-\mimath (E_{kl}-E_{ij})\, (\tau-\tau')}\;
   \< kl|\; \varrho_0 \otimes |kl\>\, \< ij|\; \varrho_0 \; .
\end{align}
Then we apply the partial traces over subsystem $\mcH_1$
\begin{multline}
\big \<\, \tr_1\big (\, \tilde V\, \varrho_0\, \big )\,
   \tr_1\big (\, \tilde V'\, \varrho_0\, \big )\, \big \> =
\big (\, {\textstyle\sum_{ik}} \rme^{-\mimath (E_k-E_i)\, (\tau-\tau')}\,
   \< k|\rho_1|i\>\, \< i|\rho_1|k\> \, \big ) \nonumber\\
\times \sum_{jl} |j\>\; \rme^{-\mimath (E_l-E_j)\,
(\tau-\tau')}\;
   \< l|\psi_{\rm e}\> \, \<\psi_{\rm e}|l\> \, \< j|\psi_{\rm e}\> \,
   \<\psi_{\rm e}| \; .
\end{multline}
The final trace over the environment yields
\begin{equation}
p[\< \tilde V\varrho_0\otimes\tilde V'\varrho_0\>]=
   S_1(\tau-\tau')\; S_{\rm e}(\tau-\tau') \; .
\end{equation}

\paragraph{\boldmath $p[\< \tilde V'\varrho_0\otimes\varrho_0\tilde V\>](t)$:}
We first compute the average
\begin{align}
\< \tilde V'\, \varrho_0\otimes\varrho_0\, \tilde V\> &=
   \sum_{ij\, kl\, mn\, pq} |ij\>\; \left\< \tilde V'_{ij,kl}\; \< kl|\,
   \varrho_0 \otimes \varrho_0\, |mn\>\; \tilde V_{mn,pq}\right\> \< pq|
\nonumber\\
&= \sum_{ij\, kl} |ij\>\;
   \rme^{-\mimath (E_{ij}-E_{kl})\, (\tau-\tau')}\; \< kl|\, \varrho_0
\otimes \varrho_0\, |kl\>\; \< ij|
\end{align}
We apply the partial traces over subsystem $\mcH_1$
\begin{multline}
 \big \<\, \tr_1\big (\, \tilde V\, \varrho_0\, \big )\,  
    \tr_1\big (\, \varrho_0\, \tilde V'\, \big )\, \big \> =
   \big (\, {\textstyle\sum_{ik}} \rme^{-\mimath (E_i-E_k)\, (\tau-\tau')}\,
   \< k|\rho_1|i\>\, \< i|\rho_1|k\> \, \big ) \nonumber\\
\times \sum_{jl} |j\>\; \rme^{-\mimath (E_j-E_l)\,
(\tau-\tau')}\;
   \< l|\psi_{\rm e}\> \, \<\psi_{\rm e}|l\> \, \< j| \; .
\end{multline}
The final trace over the environment yields
\begin{equation}
p[\< \tilde V\varrho_0\otimes\varrho_0 \tilde V'\>]=
   S_1(\tau-\tau')\; C_{\rm e}(\tau-\tau') \; .
\end{equation}

\section{The GOE case}
\label{sec:appGOE}  
In the GOE case, the average over the coupling yields, besides the
previously
considered GUE-term an additional one; see \eref{B:tVextdef}. In the
following we will redo the calculation of the previous subsection, but
consider only that additional term. As a reminder, we add the subscript $_2$
to the brackets which denote the ensemble average.

\paragraph{\boldmath $p[\mathcal{T}\< \tilde V \tilde V'\>_2
\varrho_0\otimes\varrho_0](t)$:}
We first compute the average
\begin{align}
  \mathcal{T}\< \tilde V\, \tilde V'\>_2 \, \varrho_0 \otimes \varrho_0 &=
  \sum_{ij\, kl\, mn} |ij\>\; \mathcal{T}\< \tilde V(\tau)_{ij,kl}\;
  \tilde V(\tau')_{kl,mn}\>_2\; \< mn|\;
\varrho_0\otimes\varrho_0\nonumber\\
& = \sum_{ij} |ij\> \; \< ij|\; \varrho_0 \otimes \varrho_0\\
& = \varrho_0 \otimes \varrho_0\; ,
\end{align}
where we have used the averaging rule, \eref{B:tVextdef}. Here the
time-ordering operator has no effect.
Now we apply the partial traces over subsystem $\mcH_1$
\begin{equation}
\tr_1\big (\, \< \tilde V\, \tilde V'\>_2 \, \varrho_0\, \big )\,
\tr_1\, \varrho_0 = |\psi_{\rm e}\>\, \<\psi_{\rm e}| \; .
\end{equation}
The final trace over the environment yields
\begin{equation}
p[\< \tilde V\, \tilde V'\>_2 \varrho_0\otimes\varrho_0](t)= 1 \; .
\end{equation}

\paragraph{\boldmath $p[\< \tilde V' \varrho_0 \tilde V\>_2 \otimes
\varrho_0](t)$:}
We first compute the average
\begin{align}
\<\tilde V'\, \varrho_0\, \tilde V\>_2 \otimes \varrho_0 &=
   \sum_{ij\, kl\, mn\, pq} |ij\>\; \left\< \tilde V'_{ij,kl}\;
   \< kl|\varrho_0|mn\>\; \tilde V_{mn,pq}\right\>_2\; \< pq|
\otimes\varrho_0
\nonumber\\
&= \sum_{ij\, kl} |ij\>\;
   \rme^{-\mimath (E_{ij}-E_{kl})\, (\tau+\tau')}\;
   \< kl|\varrho_0| ij\>\; \< kl| \otimes \varrho_0 \; ,
\end{align}
where we have used \eref{B:tVextdef}.
We apply the partial traces over subsystem $\mcH_1$
\begin{equation}
\tr_1\big (\, \<\tilde V'\, \varrho_0\, \tilde V\>_2\, \big )\;
\tr_1\, \varrho_0 =
   \sum_{jl} |j\>\; \rme^{-\mimath (E_j-E_l)\, (\tau+\tau')}\;
   \< l|\psi_{\rm e}\>\, \< \psi_{\rm e}|j \>\, \< l|\psi_{\rm e}\>\,
   \< \psi_{\rm e}| \; .
\end{equation}
The final trace over the environment yields
\begin{align}
p[\<\tilde V' \varrho_0 \tilde V\> \otimes \varrho_0]&=
   \sum_{jl} \< l|\psi_{\rm e}\> \;\rme^{\mimath E_l\, (\tau+\tau')}\;
   \< l|\psi_{\rm e}\> \; \<\psi_{\rm e}|j\> \;
   \rme^{-\mimath E_j\, (\tau+\tau')}\; \<\psi_{\rm e}|j\> \nonumber\\
&= S'_{\rm e}(-\tau-\tau') \; ,
\end{align}
where $S'_x(\tau)$ is defined in \eref{B:Spxdef}.

\paragraph{\boldmath $p[\< \tilde V \varrho_0\otimes \tilde V'
\varrho_0\>_2](t)$:}
We first compute the average
\begin{align}
   \<\tilde V\, \varrho_0\otimes \tilde V'\, \varrho_0\>_2 &=
   \sum_{ij\, kl\, mn\, pq} |ij\>\; \left\< \tilde V_{ij,kl}\; \< kl|\,
   \varrho_0 \otimes |mn\>\; \tilde V_{mn,pq}\right\>_2  \< pq|\,
   \varrho_0 \nonumber\\
&= \sum_{ij\, kl} |ij\>\; \rme^{-\mimath (E_{kl}-E_{ij})\, (\tau+\tau')}\;
   \< kl|\; \varrho_0 \otimes |ij\>\, \< kl|\; \varrho_0 \; .
\end{align}
Then we apply the partial traces over subsystem $\mcH_1$
\begin{multline}
 \big \< \tr_1\big ( \tilde V \varrho_0 \big )
   \tr_1\big (\tilde V' \varrho_0 \big ) \big \>_2 =
\big ( {\textstyle\sum_{ik}} \rme^{-\mimath (E_k-E_i) (\tau+\tau')}
   \< k|\rho_1|i\>\< k|\rho_1|i\> \big ) \\
\times \sum_{jl} |j\> \rme^{-\mimath (E_l-E_j)
(\tau+\tau')}
   \< l|\psi_{\rm e}\>  \<\psi_{\rm e}|j\>  \< l|\psi_{\rm e}\> 
   \<\psi_{\rm e}|  .
\end{multline}
The final trace over the environment yields
\begin{equation}
p[\< \tilde V\varrho_0\otimes\tilde V'\varrho_0\>_2]=
   S'_1(\tau+\tau')\; S'_{\rm e}(\tau+\tau') \; .
\end{equation}

\paragraph{\boldmath$p[\< \tilde V'\varrho_0\otimes\varrho_0\tilde
V\>_2](t)$:}
We first compute the average
\begin{align}
\< \tilde V'\, \varrho_0\otimes\varrho_0\, \tilde V\>_2 & =
   \sum_{ij\, kl\, mn\, pq} |ij\>\; \left\< \tilde V'_{ij,kl}\; \< kl|\,
   \varrho_0 \otimes \varrho_0\, |mn\>\; \tilde V_{mn,pq}\right\>_2 \< pq|
\nonumber\\
& = \sum_{ij\, kl} |ij\>\;
   \rme^{-\mimath (E_{ij}-E_{kl})\, (\tau+\tau')}\; \< kl|\, \varrho_0
\otimes
   \varrho_0\, |ij\>\; \< kl| \; .
\end{align}
We apply the partial traces over subsystem $\mcH_1$
\begin{multline}
\big \<\, \tr_1\big (\, \tilde V\, \varrho_0\, \big )\,  
    \tr_1\big (\, \varrho_0\, \tilde V'\, \big )\, \big \>_2 =
   \big (\, {\textstyle\sum_{ik}} \rme^{-\mimath (E_i-E_k)\, (\tau+\tau')}\,
   \< k|\rho_1|i\>\, \< k|\rho_1|i\> \, \big ) \nonumber\\
\times \sum_{jl} |j\>\; \rme^{-\mimath (E_j-E_l)\,
(\tau+\tau')}\;
   \< l|\psi_{\rm e}\> \, \<\psi_{\rm e}|j\> \, \< l| \; .
\end{multline}
The final trace over the environment yields
\begin{equation}
p[\< \tilde V\varrho_0\otimes\varrho_0 \tilde V'\>_2]=
   S'_1(-\tau-\tau')\; .
\end{equation}

\section{The general solution}
If the coupling matrix is taken either from the Gaussian
unitary (GUE) or orthogonal (GOE) ensemble, we find
\begin{multline}
A_{\rm JI}= \big [\, C_1(|\tau-\tau'|) - S_1(\tau-\tau')\, \big ]\,
   \big [\, C_{\rm e}(|\tau-\tau'|) - S_{\rm e}(\tau-\tau')\, \big ] \\
+ \chi_{\rm GOE} \big [ 1 - S'_{\rm e}(-\tau-\tau') + S'_1(\tau+\tau')
   S'_{\rm e}(\tau+\tau') - S'_1(-\tau-\tau') \big ] .
\label{B:pulrres}
\end{multline}
These expressions are derived from the previous two sections, for the GUE
case in \ref{SU}, for the GOE case in \ref{sec:appGOE}. The correlation
functions $C_x, S_x, C'_x$ and $S'_x$ with $x\in \{1,{\rm e}\}$ are defined and
discussed in \ref{aB}.

If the dimension of the Hilbert space of the environment goes
to infinity, the corresponding correlation functions simplify, as discussed
in~\ref{aB}. We are then left with
\begin{equation}\label{B:pulrires}
  A_{\rm JI}= \big [\, C_1(|\tau-\tau'|) - S_1(\tau-\tau')\, \big ]\;
     \bar C(|\tau-\tau'|) + \chi_{\rm GOE}\big [\, 1 -
S'_1(-\tau-\tau')\, \big ] \; .
\end{equation}

\section{Some particular correlation functions} \label{aB}

Averaging over the perturbation (\ie the coupling) leads to expressions
which may involve the diagonal matrix
\begin{equation}
  \bm{C}_x(\tau)= \sum_{ik} |i\>\; \rme^{-\mimath (E_k-E_i)\, \tau}\; \< i|
  \qquad x\in 1,{\rm e}\; .
  \label{B:CCxdef}
\end{equation}
Here, $x$ denotes one of the two subsystems, either the qubit or the
environment. Evidently, the energies $E_k$ are the eigenvalues of the
corresponding Hamiltonian. In the derivations in this appendix, the expectation
value of $\bm{C}_x(\tau)$ with respect to the initial state
$\varrho_0=\rho_1\otimes\rho_\e$ are of particular importance. These are
denoted by
\begin{equation}\label{B:Cxdef}
  C_x(\tau)= \tr\big (\, \bm{C}_x(\tau)\; \rho_x\, \big )
  	\qquad x\in 1,{\rm e} \; .
\end{equation}
In this work, $\rho_{\rm e}= |\psi_\e\> \<\psi_\e|$ is always a pure state.

We will also encounter another type of correlation function, which may be
defined as follows
\begin{equation}\label{B:Sxdef}
  S_x(\tau)= \tr_x\big [\, \rme^{-\mimath H_x\, \tau}\, \rho_x\,
     \rme^{\mimath H_x\, \tau}\, \rho_x\, \big ] \; .
\end{equation}
If the initial state $\rho_x$ is pure, this quantity becomes the return
probability.  At $\tau=0$ it gives the purity of $\rho_x$. In the case of GOE
averages, we also encounter the correlation function
\begin{equation}\label{B:Spxdef}
  S'_x(\tau)= \tr_x\big [\, \rme^{-\mimath H_x\, \tau}\, \rho_x\,
     \rme^{\mimath H_x\, \tau}\, \rho_x^T, \big ] \; .
\end{equation}

In the main part of this paper, we focus on the limit, where the
dimension of
the environment(s) becomes infinite. In that case it makes sense to perform
an additional spectral average over $H_\e$ and/or $H_{\e'}$. In the case of
$\bm{C}_\e(t)$ this yields
\begin{equation}\label{B:Cbardef}
\< \bm{C}_{\rm e}(\tau)\> = \bar C(\tau)\; \openone_{\rm e}
  \bar C(\tau)= \< C_\e(\tau)\> = 1+ \delta(\tau/\tau_H)
      - b_2^{(\beta)}(\tau/\tau_H) \; ,
\end{equation}
where $b_2^{(\beta)}(t)$ is the two-point spectral form factor with time
measured in units of the Heisenberg time $\tau_H$ for the corresponding
spectral ensemble. In the limit of large dimension $N_\e=\dim(\mcH_\e)$
we find
for the other correlation functions:
\begin{align}
S_\e(\tau)&= \tr\left(
   \rme^{-\mimath H_\e \tau} |\psi_\e\> \<\psi_\e|
   \rme^{\mimath H_\e\tau} |\psi_\e\> \<\psi_\e| \right) \\
S'_\e(\tau)&= \tr\left(
   \rme^{-\mimath H_\e \tau} |\psi_\e\> \<\psi_\e|
   \rme^{\mimath H_\e\tau} |\psi_\e^*\> \<\psi_\e^*| \right) \; .
\end{align}
For random pure states, as considered in the present work, both correlation
functions are at most of order one at $\tau=0$. For $\tau>0$ they drop very
quickly and soon become of order
$N_\e^{-1}$. This happens on the same time scale,
where $C_\e(\tau)$ drops from values of the order $N_\e$ to values of
the order
one. In that sense we consider these correlation functions to contribute
only
$\Or(N_\e^{-1})$ corrections to the result given in
\eref{B:pulrires}.

A single qubit is
a two level system. The most general pure initial state is given by
$|a\>= |1\>\, a_1 + |2\>\, a_2$, with $|a_1|^2 + |a_2|^2 = 1$. The most
general
mixed state is given by
$\rho_1= \lambda_1\, |a\>\, \<a| + \lambda_2\, |b\>\, \<b|$, with
$\lambda_1, \lambda_2 \ge 0$, real, $\lambda_1+\lambda_2=1$, and
$|a\>,\, |b\>$ arbitrary pure states with $\< a|b\> =0$. We will now
investigate the behaviour of the different correlation functions
${\rm Re} C_1(\tau), S_1(\tau)$, and $S_1'(\tau)$ as it depends on the
initial state and the Hamiltonian $H_1= |1\> E_1 \< 1| + |2\> E_2 \< 2|$.

\begin{itemize}
\item[(a)] Assume $\rho_1= |a\>\, \< a|$. Then
\begin{equation}
C_1(\tau)= \sum_{ik} |a_i|^2\, \rme^{-\mimath (E_k-E_i)\, \tau}
 = 1 + \cos\Delta\tau +\mimath (|a_2|^2 - |a_1|^2)\, \sin\Delta\tau \; ,
\end{equation}
holds, so
\begin{equation}
{\rm Re}\, C_1(\tau)= 1 + \cos\Delta\tau \qquad \Delta= E_2-E_1 \; .
\end{equation}

\item[(b)] For $\rho_1= \lambda_1\, |a\>\, \<a| + \lambda_2\, |b\>\,
\<b|$ we
still have
\begin{equation}\label{aB:ReC1}
  {\rm Re}\, C_1(\tau)= 1 + \cos\Delta\tau \; .
\end{equation}
\item[(c)] Assume $\rho_1= |a\>\, \< a|$. Then we find for
$S_1(\tau)= s(|a\>; \tau)$
\begin{equation}
S_1(\tau)= \sum_{ik} |a_i|^2 |a_k|^2\, \rme^{-\mimath (E_i-E_k)\, \tau}
 = |a_1|^4 + |a_2|^4 + 2\, |a_1|^2 |a_2|^2\, \cos\Delta\tau \; .
\end{equation}
This expression only depends on the absolute values squared of the
coefficients
$a_1$ and $a_2$. Therefore we may parametrize them without loss of
generality
as $a_1= \cos\phi$ and $a_2= \sin\phi$. We then find
\begin{equation}\label{aB:S1pure}
S_1(\tau)= g_\phi + (1- g_\phi)\, \cos\Delta\tau\qquad
g_\phi= |a_1|^4 + |a_2|^4= 1- \frac{1}{2}\; \sin^2\, 2\phi \; .
\end{equation}

\item[(d)] For a general mixed state
$\rho_1= \lambda_1\, |a\>\, \<a| + \lambda_2\, |b\>\, \<b|$ we find
\begin{align}
S_1(\tau) &= \lambda_1^2\, s(|a\>; \tau) + \lambda_2^2\, s(|b\>; \tau) +
   2\, \lambda_1\lambda_2\, \sum_{ik} a_i a_k^* b_k b_i^*\, \cos
(E_i-E_k)\tau
\nonumber\\
&= \lambda_1^2\, s(|a\>; \tau) + \lambda_2^2\, s(|b\>; \tau) + \nonumber \\
&\quad \qquad   4\, \lambda_1\lambda_2\, |a_1|^2 |b_1|^2 \big (\, 1 -
\cos\Delta\tau\, \big ),
\end{align}
where we have used that $\< a|b\> = a_1 b_1^* + a_2 b_2^* = 0$.
Note that the coefficients $a_1, a_2, b_1, b_2$ may be arranged into a
square
unitary matrix, and must therefore be of the following general form
\begin{equation}\label{aB:abunit}
\left(\!\!\!\begin{array}{cc} a_1 & b_1\\ a_2 & b_2
\end{array}\!\!\!\right)=
\rme^{\mimath\vartheta}\left(\!\!\!\begin{array}{cc}
   \rme^{\mimath\xi}\, \cos\phi & \rme^{\mimath\chi}\, \sin\phi\\
   -\, \rme^{-\mimath\chi}\, \sin\phi & \rme^{-\mimath\xi}\, \cos\phi
\end{array}\!\!\!\right) \; .
\end{equation}
This shows that
$s(|a\>; \tau)= s(|b\>; \tau) = g_\phi + (1- g_\phi)\, \cos\Delta\tau$,
and that
\begin{multline}
S_1(\tau)= (\lambda_1^2 + \lambda_2^2) \big [\,
   g_\phi + (1-g_\phi)\, \cos\Delta\tau\, \big ] \\
   + 2\, \lambda_1\lambda_2\,
   (1-g_\phi)\, \big (\, 1 - \cos\Delta\tau\, \big ) \; .
\end{multline}
Since $\lambda_1+\lambda_2= 1$ it is convenient to set $\lambda_1=
\cos^2\theta$
and $\lambda_2= \sin^2\theta$ such that we may write
$\lambda_1^2 + \lambda_2^2 =g_\theta$ and obtain
\begin{equation}\label{aB:S1mixed}
S_1(\tau)= 1- g_\theta- g_\phi + 2\, g_\theta g_\phi + (2g_\theta-1)
(1-g_\phi)\,
   \cos\Delta\tau \; .
\end{equation}

\item[(e)] Assume again that $\rho_1= |a\>\, \<a|$. Then we find for
$S'_1(\tau)= s'(|a\>; \tau)$
\begin{eqnarray}
S'_1(\tau)= |a_1|^4 + |a_2|^4 + a_1^2 (a_2^*)^2\,
\rme^{\mimath\Delta\tau} +
   a_2^2 (a_1^*)^2\, \rme^{-\mimath\Delta\tau} \nonumber\\
 = g_\phi + 2\, |a_1|^2 |a_2|^2\, \cos(\Delta\tau+2\eta)\nonumber\\
 = g_\phi + (1-g_\phi)\, \cos(\Delta\tau+2\eta)\qquad
\eta= {\rm arg}(a_1) - {\rm arg}(a_2) \; .
\label{aB:S1ppure}\end{eqnarray}
However, as discussed in \sref{sec:goeone}, a natural symmetry around
the $y$ axis (in the Bloch sphere picture) appears for $\Delta=0$. In
that case, using $\gamma$ as defined
in that section, one can prove that
\begin{equation}
    1-S'_1(\tau)\big|_{\Delta=0}=(1-g_\phi)[1- \cos(2\eta)]=\sin^2\gamma
    \label{aB:S1ppurenD}
\end{equation}
using elementary geometric and trigonometric considerations.

\item[(f)] For a general mixed state
$\rho_1= \lambda_1\, |a\>\, \<a| + \lambda_2\, |b\>\, \<b|$ we find
\begin{equation}
S'_1(\tau)= \lambda_1^2\, s'(|a\>; \tau) + \lambda_2^2\, s'(|b\>; \tau)
 + 2\, \lambda_1\lambda_2\, {\rm Re}\, \sum_{ik} a_i a_k^* b_k^* b_i\,
   \rme^{-\mimath (E_i-E_k)\, \tau} \; .
\end{equation}
It follows from \eref{aB:abunit} that
${\rm arg}(b_1) - {\rm arg}(b_2) = {\rm arg}(a_1) - {\rm arg}(a_2) -
\pi$, such
that the equality $s'(|a\>; \tau)= s'(|b\>; \tau)$ holds, just as in the
case of
$S_1(\tau)$. Therefore we may write
\begin{align}
S'_1(\tau) & = g_\theta \big [\, g_\phi + (1-g_\phi)\, \cos(\Delta\tau +2\eta)\,
   \big ] \\
 &  \qquad + 2\, \lambda_1\lambda_2\, (1-g_\phi)\, {\rm Re}\big (\, 1 +
   \rme^{\mimath (\Delta\tau +2\eta -\pi)}\, \big ) \nonumber\\
  & = 1- g_\theta- g_\phi + 2\, g_\theta g_\phi + (2 g_\theta -1)(1-g_\phi)\,
   \cos(\Delta\tau +2\eta) \; .
\label{aB:S1pmixed}\end{align}
Note that the only difference to $S(\tau)$ in case (d) is the additional
phase
$2\eta$ in the argument of the cosine function. Using the same angle
$\gamma$
defined with \eref{eq:psi12goe}, in complete analogy with
\eref{eq:initialOneGOE},
and using \eref{aB:S1ppurenD} we obtain
\begin{equation}
    1-S'_1(\tau)\big|_{\Delta=0}=1-g_\theta+(2g_\theta-1)\sin^2\gamma.
    \label{aB:S1pmixednD}
\end{equation}

\end{itemize}

\chapter{Implementing the evolution of the KI models}
\label{sec:implementationKI}

In this appendix we describe how to implement numerically Hamiltonians
describing a set of qubits with pairwise variable Ising interaction, and
kicked with a periodic, and site dependent, magnetic field.  We have used
this systems in chapters \ref{sec:ki} and \ref{sec:nqubit}, the reason being
(i) the flexibility of the dynamical model to describe various physical
situations, (ii) it has various dynamical regimes, and (iii) it is has an
efficient implementation making it useful for numerical studies.

Recall \eref{eq:fullhamiltonian}, the Hamiltonian of the system is 
\begin{equation}
H = \sum_{j>k=1}^{L}J_{j,k} \sigma ^z_j \sigma^z_{k} +
\delta_1(t) \sum_{j=1}^{L} \vec{b}_j \cdot \vec{\sigma}_j. 
\end{equation}
which gives rise to the Floquet operator
$U_\text{KI}=U_\text{Ising}U_\text{kick}$ with 
\begin{equation} 
  U_\text{Ising}=\exp 
    \left( -\mimath \sum_{j>k =1}^{L} J_{j,k} \sigma ^z_j \sigma^z_k \right),\quad
  U_\text{kick}=
    \exp\left(-\mimath \sum_{j=1}^{L} \vec{b}_j \cdot\vec{\sigma}_j \right).
\end{equation}
However notice that
\begin{equation} 
  U_\text{Ising} = \prod_{j>k=1}^{L}\exp \left(-\mimath J_{j,k}\sigma ^z_j \sigma^z_k\right)
  ,\quad 
  U_\text{kick}= \prod_{j=1}^{L}\exp \left(-\mimath \vec{b}_j \cdot\vec{\sigma}_j\right)
\end{equation}
Thus the evolution operator is decomposed trivially into single and two qubit
operators that can be efficiently implemented in a computer. Notice that we do
not need to explicitly diagonalize or even store the complete evolution
operator thus allowing the implementation of twice as much spins than with 
brute force diagonalization.

The state  can be stored in a complex array with indices ranging from
$0$ to $2^L-1$. The state (expressed in the computational basis)
\begin{equation}
|\psi\>=\sum_{i_{L-1},i_{L-2},\cdots,i_0=0}^{1}
\alpha_{i_{L-1} i_{L-2} \cdots i_0} |i_{L-1} i_{L-2} \cdots i_0\>,
\end{equation}
can be  stored in a complex array of dimension $2^L$, if in the position
$\sum_j i_j 2^j$ we store the value $\alpha_{i_{L-1} i_{L-2}\cdots i_0}$. 
A Fortran 90 routine implementing the Ising interaction with the state stored
in this form is the following.
\begin{verbatim}
subroutine Action_Ising_Interaction(psi,l,m,J)
complex(kind(1d0)),intent(inout)  :: psi(0:)
real(kind(1d0)),intent(in)        :: J
integer,intent(in)                :: l,m
integer                           :: i
do i=0,size(psi)-1
 If(btest(i,l).eqv.btest(i,m)) then
  psi(i)=exp(-(0d0,1d0)*J)*psi(i)
 else
  psi(i)=exp((0d0,1d0)*J)*psi(i)
 end If
end do
end subroutine Action_Ising_Interaction
\end{verbatim}

A routine to implement the magnetic kick is now presented. 
\begin{verbatim}
subroutine Application_Of_Kick(psi,b,l)
complex(kind(1d0)),intent(inout)  :: psi(0:)
real(kind(1d0)),intent(in)        :: b(:)
complex(kind(1d0))                :: Matrix_For_Kick(2,2)
complex(kind(1d0))                :: z(2)
integer,intent(in)                :: l
integer                           :: j_left,j_right,left_digit
integer                           :: UpperNumber,LowerNumber
if(sum(b**2).eq.0d0) return
Matrix_For_Kick=Matrix_For_Magnetic_Kick(b)
do j_left=0,size(psi)/2**(1+l)-1
 left_digit=ishft(j_left,l+1)
 do j_right=0,2**l-1
  LowerNumber=left_digit+j_right
  UpperNumber=IBset(LowerNumber,l)
  z(1)=psi(LowerNumber); z(2)=psi(UpperNumber)
  z=MatMul(Matrix_For_Kick,z)
  psi(LowerNumber)=z(1); psi(UpperNumber)=z(2)
 end do
end do
end subroutine Application_Of_Kick
\end{verbatim}																															
The previous routine uses the following function.
\begin{verbatim}
function Matrix_For_Magnetic_Kick(b)
complex(kind(1d0))         :: Matrix_For_Magnetic_Kick(2,2)
real(kind(1d0)),intent(in) :: b(3)
real(kind(1d0))            :: n(3),phi,cosphis2,sinphis2
phi=norma(b);    n=b/phi
Matrix_For_Magnetic_Kick(1,1)=&
   cos(phi)-(0d0,1d0)*n(1)*sin(phi)
Matrix_For_Magnetic_Kick(2,2)=&
   conjg(Matrix_For_Magnetic_Kick(1,1))
Matrix_For_Magnetic_Kick(1,2)=&
   -(n(3)+(0d0,1d0)*n(2))*sin(phi)
Matrix_For_Magnetic_Kick(2,1)=&
   -conjg(Matrix_For_Magnetic_Kick(1,2))
end function Matrix_For_Magnetic_Kick
\end{verbatim}

\chapter{The exponentiation}\label{sec:exponentiation}

The linear response formulae for purity derived throughout the thesis are
expected to work for high purities or, equivalently, when the first two terms
in the series \eref{eq:bornexpansion} approximate the echo operator well; it is
of obvious interest to extend them to cover a larger range. The exact solution
for  fidelity decay has, in some cases, been possible with some effort, using
super-symmetry techniques. This has been partly due to the simple structure of
fidelity, but trying to use this approach for a more complicated object such as
purity seems to be out of reach for the time being.  Exponentiating the
formulae obtained from the linear response formalism has proved to be in good
agreement with the exact super-symmetric and numerical results for  fidelity,
if the perturbation is not too big. The exponentiation of the linear  response
formulae  for  purity  can  be  compared  with  Monte Carlo simulations in
order to prove its  validity.  We wish to explain the details required to
implement this procedure in this appendix.

Given a linear response formula $P_{\rm LR}(t)$ [for which $P(0)=1$], and an
expected asymptotic value for infinite time $P_\infty$, the exponentiation
reads
\begin{equation}
  \label{eq:ELRextension}
  P_{\rm ELR}(t)=P_\infty+(1-P_\infty)\exp\left[- \frac{1-P_{\rm
LR}(t)}{1-P_\infty}\right].
\end{equation}
This particular form guaranties that 
\begin{equation} P_{\rm ELR}(t) \approx P_{\rm LR}(t)\end{equation}
for short times, and that 
\begin{equation} \lim_{t \to \infty}P_{\rm ELR}(t)=P_\infty.\end{equation}

The particular value of $P_\infty$ will depend on the physical situation; in
our case it will depend on the configuration and on the initial condition. We
now turn to an estimation of this quantity for the spectator configuration,
under some very general assumptions.

Let us write the initial condition in the qubits as
\begin{equation}
 |\psi_{12}(\theta)\>=\cos\theta|\tilde 0_1 \tilde0_2\> +
                         \sin\theta|\tilde 1_1 \tilde 1_2\>
\end{equation}
for some rotated qubits $|\tilde 0_i\>$, $|\tilde 1_i\>$.  We assume that for
long enough times, decoherence is described for the uncoupled qubit by the
following channel:
\begin{equation}
  \mathcal{E}_\textrm{td}[\rho]
    =\frac{\openone}{2}
    =\frac{1}{4}\left(\sigma_x \rho \sigma_x+\sigma_y \rho \sigma_y
                           +\sigma_z \rho \sigma_z+ \rho \right).
\end{equation}
This  is the  depolarizing channel  with full strength~\cite{NC00a}. This
channel is unital. We now evaluate $\mathcal{E}_\textrm{td}\otimes \openone
[|\psi_{12}(\theta)\>\<\psi_{12}(\theta)|]$.  Note that
$\mathcal{E}_\textrm{td}\otimes \openone [|00\>\<00|] =  \openone \otimes
|0\>\<0|/2$ and $\mathcal{E}_\textrm{td}\otimes \openone [|11\>\<11|] =
\openone \otimes |1\>\<1|/2$. For the cross terms we obtain:
\begin{equation}
\mathcal{E}_\textrm{td}\otimes \openone[|00\>\<11||]
  =\frac{|10\>\<01| - |10\>\<01| -|00\>\<11|+|00\>\<11|}{4}=0,
\end{equation}
and also $\mathcal{E}_\textrm{td}\otimes \openone[|11\>\<00||]=0$.  Using the
linearity of the channel and the results found, we see that 
\begin{equation}
  \mathcal{E}_\textrm{td}\otimes \openone
   [|\psi_{12}(\theta)\>\<\psi_{12}(\theta)|]=
   \frac{1}{2}
         \begin{pmatrix}
	    \cos^2\theta&0&0&0\\
	    0&\cos^2\theta&0&0\\
	    0&0&\sin^2\theta&0\\
	    0&0&0&\sin^2\theta
	\end{pmatrix}.
\end{equation}
In the end, we estimate the asymptotic value of purity as
\begin{equation}
  \label{eq:estpurity}
  P_\infty =\frac{\cos^4\theta + \sin^4\theta}{2}=\frac{g_\theta}{2},
\end{equation}
where $g_\theta$ was defined in \eref{eq:defg}.

For both the joint  and the separate environment configuration we use
the value
\begin{equation}
P_\infty
   =P\left(\mathcal E_{\rm d}\otimes\mathcal E_{\rm d}
[|\psi_{12}(\theta)\>\<\psi_{12}(\theta)|]\right) =\frac{1}{4}.
\end{equation}

Good   agreement   is  found   with   Monte Carlo   simulations  for  moderate
and strong couplings. Obtaining the value of $P_\infty$ for weak coupling
should be possible using perturbation theory on the eigenvectors. This
calculation is still missing.

\chapter{Entanglement} \label{sec:entappendix}

Here we want to discuss or comment on some technical aspects that do not
need to be in the introduction. We prefer to include them in an appendix to
ease the reading of the main part of the text. 

Recall from \eref{eq:separablecondition} that a pure state of a bipartite
system is separable if it is the tensor product of states in the individual
systems.  The generalization for multipartite systems is straightforward.
Let us have $n$ particles, each in a Hilbert space $\mcH_i$, $i=1,\cdots,n$.
A state $|\psi\>$ is said to be separable if it can be written as
\begin{equation}
	|\psi\>=\bigotimes_{i=1}^n |\psi_i\>,\quad(|\psi_i\>\in \mcH_i),
\label{eq:multipartiteentanglement}
\end{equation}
compare with \eref{eq:separablecondition}. Otherwise it is called entangled.
A mixed, bipartite system in state $\rho$ is said to be separable if it can
be written as a convex combination of separable states. That is, if 
\begin{equation}
	\rho=\sum_i p_i \rho_i^{(A)} \otimes \rho_i^{(B)} 
\label{eq:entangledmixed}
\end{equation}
where $p_i > 0$ and $\rho_i^{(A)}$ ($\rho_i^{(B)}$) are density matrices of
system $A$ ($B$). Though this definition is simpler from a conceptual point
of view, often a more comfortable formulation in terms of pure states if
obtained expressing the density matrices in terms of projectors. Thus a
bipartite mixed state $\rho$ is said to be separable if it can be written as
\begin{equation}
	\rho=\sum_i p_i 
	|\psi_i^{(A)}\> \<\psi_i^{(A)}| \otimes 
	     |\psi_i^{(B)}\> \<\psi_i^{(B)}| 
\label{eq:entangledmixedalt}
\end{equation}
with $|\psi_A\> \in \mcH_A$, $|\psi_B\> \in \mcH_B$ and $p_i$ real and
positive.  For the multipartite case one should require, instead of
\eref{eq:entangledmixed}, that 
\begin{equation}
 \rho=\sum_i p_i \otimes_{k=1}^n \rho_i^{(k)}
\end{equation}
with $\rho_i^{(k)}$ being a density matrix of space $\mcH_k$ and the $p_i$s
probabilities (\ie still a convex combination of separable states).

\section{Quantifying two qubit entanglement}\label{sec:quantifyent} 

Here we want to present the entanglement of formation, and its relation to 
concurrence.

Let us first define the entanglement of formation.  The entanglement of
formation $E_F$ measures the number of Bell pairs needed to create an
ensemble of pure states representing the state to be
studied~\cite{firstconcurrence}. We define $E_F$ by stages.  The
entanglement of formation for a pure state is the von Neumann  entropy of
the reduced density matrix. Given $|\psi\>$ with Schmidt coefficients
$\lambda_1$ and $\lambda_2=\sqrt{1-\lambda_1^2}$, $E_F(|\psi\>
\<\psi|)=-\lambda_1 \log_2 \lambda_1 - \lambda_2 \log_2
\lambda_2$\footnote{Notice the close relation with the classical entropy.}.
For mixed states we have
\begin{equation}
 E_F(\rho)=\min_{\{p_i,|\psi_i\>\}} \sum_i p_i E_F(|\psi_i\> \<\psi_i|) 
 \quad \text{such that} \quad
 \rho =  \sum_i p_i |\psi_i\> \<\psi_i|
\label{eq:entanglementodformation}
\end{equation}
and the $p_i$s are positive. This represents the minimum amount of 
entanglement needed to create an ensemble that reproduces $\rho$. 
All the requirements discussed above are fulfilled. 

Concurrence $C$ is defined via the entanglement formation:
\begin{equation}
 E_F(\rho)=h\left(\frac{1+\sqrt{1-C(\rho)^2}}{2} \right);
 \quad h(x)=-x\log_2(x)-(1-x)\log_2(1-x).
\label{eq:defconcent}
\end{equation}
This weired definition is inspired in the fact that a simple and close
formula for $C(\rho)$ was obtained. 

\section{Some generalizations}

Two possible generalizations for the entanglement measures would be
desirable: (i) Allow qudits instead of qubits (\ie having more that 2
levels). Furthermore one could like to quantify the entanglement in systems
with infinite or a continuous number of levels. (ii) Allow the Hilbert space
of the system to have a more complicated partition, Namely let
$\mcH=\otimes_{k=1}^n \mcH_k$ with $n$ being bigger than 2; for example if
one wants to examine the entanglement between three particles, $n=3$. The
cases in which the state representing the system is pure are again much
simpler that when the state is mixed. 

For both cases the concept of entanglement is easy to define.  A relevant
review of entanglement measures for qudits is \cite{plenio-2007-7}.   Some
measures of entanglement, for the multipartite case, are discussed in
\cite{andrereview}.  As the reader can find, this generalizations are not
simple, unique or easy to calculate.

\chapter{RMT: various aspects} \label{sec:RMTtech}

In this appendix we wish to discuss some technicalities about 
the random matrix ensembles used in this thesis. It includes a 
small discussion about the ensembles with/without time reversal 
symmetry, the construction of the ensembles, and concepts 
as the Heisenberg time, level density, and form factor.

We start by discussing the difference between the unitary ensembles (GUE and
CUE) and the orthogonal ensembles (GOE and COE).  We first describe the GOE.
This is an ensemble of Hermitian matrices with the special property of being
real.  As we know, any Hamiltonian describing a system with an anti-unitary
symmetry (as time reversal) can be represented as a real matrix {\it without
diagonalizing it}. The COE corresponds to the unitary version of this
ensemble, for each $H$ in the GOE, a corresponding $\exp (\imath H)$ is found
in the COE.   On the other hand, when the physical system has no underlying
anti-unitary symmetry, it must be described by arbitrary Hermitian matrices.
The GUE is the appropriate ensemble as it has no anti-unitary symmetry. The
set corresponding to this ensemble is the set of all Hermitian matrices. The
CUE is the unitary version of this ensemble. The most common example of this
distinction involves a particle in a chaotic billiard. With/without a magnetic
field the level statistics resemble those of the GUE/GOE.  Numerical
\cite{GOEGUEnum} and experimental \cite{GOEGUEexp, KramersSeligman}
confirmation has been reported. 

We now turn our attention to the construction of the ensembles used
in this thesis. The neatest way to perform that construction is using
the arguments of least information (see \cite{balian, dyson}
or,  for a more modern view, \cite{mello-Les-Houches} section 2).
We wish to recall that an ensemble is a set and a probability
distribution, which can be regarded as a measure. The set shall
be no problem to define, but the measure is more delicate. 

We start by defining the entropy of a distribution of matrices (objects)
$A$.  Let $P(A)$ be a probability distribution on a set matrices $A$ (or of
anything). The information of the ensemble is measured via the information
entropy \cite{Shannon1948}:
\begin{equation}
     I[P(A)]=- \int  P(A) \log P(A)\rmd A
\label{eq:shannonentropy}
\end{equation}
Notice that we need an {\em a prior} measure $\rmd A$. This measure, in all
the cases considered here, is an invariant (under a suitable operation)
measure. If there is a group structure underling (like for the GOE and GUE
with matrix addition and the CUE with matrix multiplication)
this measure coincides with the Haar measure. Otherwise (for the COE), 
it is slightly
more complicated. For details see
\cite{cartanRMT, mehta}.  The information entropy, for a given $P(A)$, can
be interpreted in  a number of ways \cite{Shannon1948}.  For example, it
measures the amount of information in a given member of the ensemble. For
bigger entropy, the information in each member is smaller (\ie its members
are closer to the anonymity).  

Let us start by constructing the CUE, which stands for circular unitary
ensemble. This is an ensemble of unitary matrices of given dimension. 
The
distribution that we use for the ensemble is simply the homogeneous one,
$P(A)={\rm constant}$.  Intuitively it is clear that, for this set, it is
the ensemble with greatest information entropy, see \eref{eq:shannonentropy}.
We can do an analogous treatment for orthogonal matrices to obtain the
circular orthogonal ensemble (COE), using a uniform ``probability'' for all
orthogonal matrices.  

For Hermitian operators, as compared to the unitary ones, the situation is
more complicated.  Think of Hermitian $1\times1$ operators (real numbers).
Its Haar measure is simply the usual Lebesgue  measure.  The whole set has an
unbounded measure hence it cannot be normalized; working naively with
$P(A)\propto 1$ results in diverging integrals. Moreover, for the real line
(and using the usual measure) there does not exists  such a thing as a
uniform distribution.  Similar problems result for larger matrices due to
the non-compactness of the set under consideration.  However one can create
a least information ensemble if one is willing to postulate an additional
condition.  We shall require that the average of $\tr H^2$ (the
``strength'' of the operator) is some fixed number.  Its particular value
is irrelevant and will only impose some normalization condition in the
ensemble.  Notice that this condition is basis independent.  One can solve
the problem of maximizing the entropy constrained to the condition above,
using Lagrange multipliers.  The solution is
\begin{equation}
 P(H)\propto e^{-\lambda \tr H^2}
\label{eq:NONO}
\end{equation}
with $\lambda$ a real (ensemble dependent) number.  

In all the cases (COE, CUE, GOE, and GUE) we can see that the distribution
is invariant under unitary (for the CUE and GUE) or orthogonal (for the COE
and GOE) transformations, \eg for the GUE, $P(H)=P(U^{-1} H U)$ for any
unitary operator $U$, as the only condition we imposed is basis
independent.  The invariance properties of the ensemble will be used
heavily to simplify the problem under consideration and group together
equivalent cases during the thesis. 

A practical way to realize the Gaussian ensembles (G*E) is to choose the
matrix elements as Gaussian random variables with a given standard deviation
$\sigma$.  Consider the independent elements of a Hermitian matrix $H$. We
have that $\tr H^2=\sum_i H_{ii}^2+ 2 \sum_{i<j} H_{ij}H_{ji}$ and then the
distribution \eref{eq:NONO} can be factorized as
\begin{equation}
 P(H)\propto \prod_{i<j=1}^{N} e^{-2\lambda H_{ij}H_{ji}}
                  \prod_{i=1}^{N} e^{-\lambda H_{ii}^2}.
\label{eq:NONOgoe}
\end{equation}
For the GOE (of fixed dimension $N$) we choose: the diagonal elements as real
variables with $\sigma^2=1$ (forget about the global, unimportant
normalization constant); for the elements above the diagonal,  real variables
with $\sigma^2=2$; and the others (below the diagonal) are chosen to make the
matrix symmetric.  The matrix elements average  like
\begin{equation}
 \overline{V_{ij}}=0,\, \overline{V_{ij} V_{kl}}=
           \delta_{il} \delta_{jk}+\delta_{ik} \delta_{jl}.
\label{eq:GOEconmute}
\end{equation}
Of course the matrix  elements can be multiplied by a convenient (possibly $N$
dependent) factor to achieve certain particular condition.  Under similar
considerations one can show that a typical GUE element  is obtained as follows.
Choose the diagonal elements from a real Gaussian ensemble with $\sigma=1$. The
upper triangle of the matrix are complex numbers; the real and imaginary parts
are chosen independently from a real Gaussian ensemble with
$\sigma=1/\sqrt{2}$.  The lower elements are chosen to make the matrix
Hermitian.  Its matrix elements average as
\begin{equation}
 \overline{V_{ij}}=0,\, \overline{V_{ij} V_{kl}}= \delta_{il} \delta_{jk}.
\label{eq:GUEconmute}
\end{equation}

\begin{figure}[t]
\begin{center}
\includegraphics{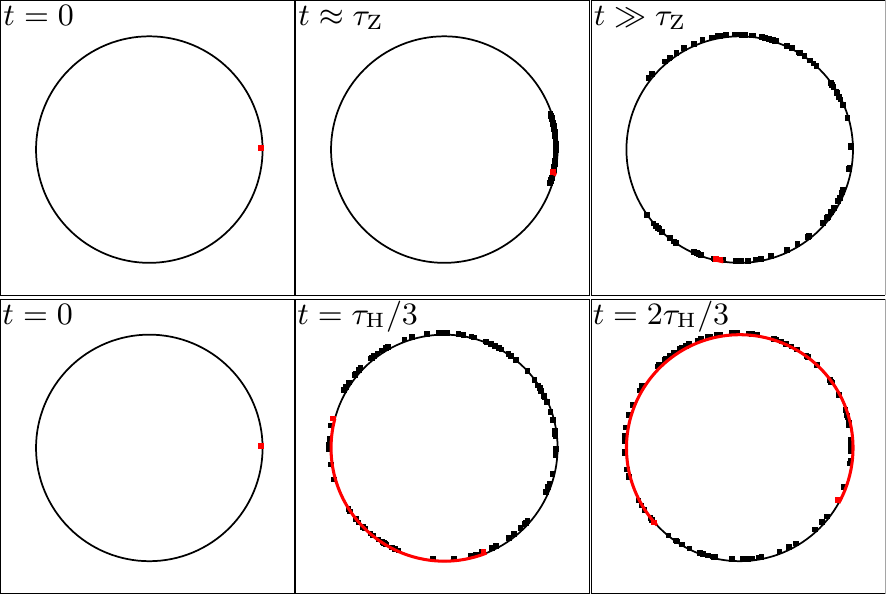}
\end{center}
\caption{Visualization of the Zeno time (above) and the Heisenberg time
  (below).  Eigenvalues $E$ are plotted in the unit circle as dots as time
  passes, using $\exp( -\imath E t/\hbar)$. An arc between 2 particular
  adjacent eigenvalues is shown in red. The Zeno time $\tau_\rmZ$ is close to
  the first time chosen all eigenvalues cover the unit circle.  Heisenberg
  time $\tau_\rmH$ is the {\it average} time when adjacent eigenvalues cross
  in the unit circle. 100 Poissonian eigenvalues were used for these plots.}
\label{fig:times}
\end{figure}

\begin{figure}[t]
\begin{center}
\includegraphics{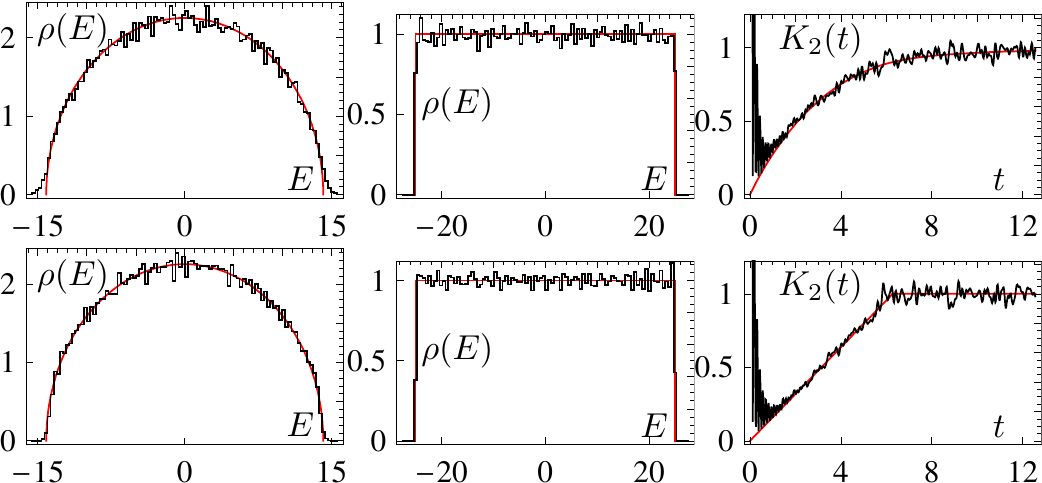}
\end{center}
\caption{We see here the level density of an ensemble of 800 elements, of
  the GOE (top) and GUE (bottom) with $N=50$. We show, from left to right,
  the level density \eref{eq:density}, the unfolded level density (see
  \eref{eq:unfold}) and the form factor $K_2$ Eqs.  \ref{eq:b2defGOE} and
  \ref{eq:b2defGUE}.}
\label{fig:examplesRMT}
\end{figure}

At this point we want to introduce the concept of level density, sometimes
also called density of states. This is the number of levels per unit interval.
The level density is a particular feature of each system, even of chaotic
ones. Random matrices have their own level density. The circular ensembles
have a constant level density (within the interval $[0,2\pi )$) whereas the
Gaussian ensembles have an ellipsoidal level density (often referred to as
circular). In the large dimension limit it is given by 
\begin{equation}
 \rho(E)=\frac{\sqrt{N}}{\pi}\sqrt{1-\frac{E^2}{4N}}.
\label{eq:density}
\end{equation}
for the normalization previously proposed, see \fref{fig:examplesRMT}. Other
concept closely related to the level density is the Heisenberg time
$\tau_\rmH$. This is the typical time in which two neighboring levels 
cross in the unit circle (\ie when $(E_{i+1}-E_i) t/\hbar \approx 2\pi$). Of
course if the level density is not uniform, the Heisenberg time is also not
uniform.  Sometimes it is convenient to have a uniform Heisenberg time.
This amounts to have constant density of states throughout the spectrum. If
one uses the circular ensembles this is automatically granted. Otherwise, the
constant density can be achieved using the unfolding procedure. Normally it
requires adjusting an average density but, since for the GUE and GOE we know
the average density, we can calculate its cumulative function
\cite{guhr98random}, and use that to unfold the spectrum. Let $E_c$ be the
smallest expected energy (eigenvalue) of your spectrum. The unfolded energy
$\phi_i$ corresponding to the original eigenvalue $E_i$ is
\begin{equation}
  \phi_i=\frac N\pi \left(\sin^{-1}e_i +e_i\sqrt{1-e_i^2} \right),\,
  e_i=\frac{E_i}{E_c}.
\label{eq:unfold}
\end{equation}
Another important time scale is the Zeno time $\tau_\rmZ$. 
It is the time the eigenvalues are far from the origin, but are still much
smaller than $2\pi$.
Both concepts are sketched in 
\fref{fig:times}.

An important characteristic of the RMT spectra are its correlations.  We can
use the form factor
\begin{equation}
        K_2(t):=\frac{1}{N} 
          \left| \sum_{j=1}^{N} e^{\imath t E_j}\right| ^2
\label{eq:formfactor}
\end{equation}
to quantify this correlation.
In the large dimension limit we have that
\begin{equation}
        \overline{K_2(t)}=K_2^{(\beta)}(t)=
          1+\delta\left( \frac{t}{\tau_\rmH} \right)
           - b_2^{(\beta)}\left( \frac{t}{\tau_\rmH} \right).
\label{eq:formfactorRMT}
\end{equation}
The index $\beta$ characterizes the ensemble, $\beta=1$ for the GOE (and COE)
and $\beta=2$ for the GUE (and CUE).
The functions $b_2^{(\beta)}$ are given by
\begin{equation}
b_2^{(1)}(t)=\begin{cases}
1-2|t|+|t| \ln (2|t|+1)         & \text{if\ }|t|\le 1,\\
-1+|t|\ln \frac{2|t|+1}{2|t|-1} & \text{if\ }|t|>1
\end{cases}
\label{eq:b2defGOE}
\end{equation}
and
\begin{equation}
b_2^{(2)}(t)=\begin{cases}
1-|t|& \text{if\ }|t|\le 1,\\
0& \text{if\ }|t|>1
\end{cases},
\label{eq:b2defGUE}
\end{equation}
see \fref{fig:examplesRMT}.

A characteristic features of the GOE and GUE spectra is the hole in the
correlation function (for a random spectra, $K_2(t)= 1+\delta(
t/\tau_\rmH)$). This hole has been shown to induce a certain stability in
fidelity decay. In this work it will also display some enhancement of
stability though not as spectacular as for fidelity decay. 

\chapter{On the numerics of RMT and quantum information}
\label{sec:implementationRMTQI}

We now describe some tricks to implement some aspects of RMT. We
also include some of the routines useful for quantum information
simulation. All routines are for Fortran 90. 
$\mathtt{pi}=\pi$ and $\mathtt{I} =\imath$.

\section{Random matrices}
To build the random matrices, as noted in Appendix \ref{sec:RMTtech}, one can
use random Gaussian variables. Most low level languages provide
homogeneous distributions in the unit interval $[0,1)$. A transformation from
this distribution (characterizing variables $u$ and $v$),
to a complex Gaussian one (characterizing 
variable $z$) of width $\sigma$ and centered
at $x_0$  is provided by
\begin{equation}
z= \sigma \rme^{2\pi \imath v} \sqrt{-2 \log u }+x_0
\label{eq:transformhomogeneoustogaussian}
\end{equation}
Its implementation is straightforward, however if $u=0$ (which occurs
numerically with very small, but finite probability) the routine could crash.
If instead of selecting $u \in [0,1)$ we select $1-u \in [0,1)$, the problem
is fixed. The routine is called {\tt RandomGaussian}. Its first argument
is $\sigma$ and the second one $x_0$. 

In view of the considerations of the  Appendix \ref{sec:RMTtech},
to build a GUE matrix, one has to choose the diagonal elements
real and with $\sigma^2 = 1 $ whereas for the off diagonal elements 
$\sigma^2 = 1/2 $. The routine to build the GUE matrix is:
\begin{verbatim}
  subroutine generate_GUE_matrix(matrix)
    implicit none
    integer                         :: nsize,j1,j2
    complex(kind(1d0)), intent(out) :: matrix(:,:)
    nsize=size(matrix(1,:))
    do j1=1,nsize
       matrix(j1,j1)=real(RandomGaussian(1d0,0d0))
       do j2=j1+1,nsize
          matrix(j1,j2)=RandomGaussian(1/sqrt(2d0),0d0)
          matrix(j2,j1)=conjg(matrix(j1,j2))
       end do
    end do
  end subroutine generate_GUE_matrix  
\end{verbatim}
A routine to generate a GOE matrix is
\begin{verbatim}
  subroutine generate_GOE_matrix(matrix)
    implicit none
    integer                         :: nsize,j1,j2
    real(kind(1d0)), intent(out)    :: matrix(:,:)
    nsize=size(matrix(1,:))
    do j1=1,nsize
       matrix(j1,j1)=&
           real(RandomGaussian(sqrt(2d0),0d0),kind(1d0))
       do j2=j1+1,nsize
          matrix(j1,j2)=real(RandomGaussian(),kind(1d0))
          matrix(j2,j1)=matrix(j1,j2)
       end do
    end do
  end subroutine generate_GOE_matrix  
\end{verbatim}
The normalization is such that \eref{eq:GOEconmute} holds.  
To obtain the spectrum of such matrices one must diagonalize numerically a
dense matrix.  The {\tt LAPACK90} package provides this routines conveniently
packed for Fortran 90. 

Most RMT results refer to the center of the spectrum,
where the density of states is constant. To have better agreement of the
numerics with the theory one can unfold the spectrum, \ie force an
approximate constant level density. Given a density $\rho(E)$, the function
$\eta(e)=\int_{\min\{E\}}^{e} \rmd E \rho(E)$ maps $E$ to an ``unfolded''
energy $e$. Since 
\begin{equation}
 \int_{-1}^e \frac{2 N}{\pi} \sqrt{1-E^2} =
  \frac {N}{\pi}\left( e\sqrt{1-e^2}+\pi-\sin^{-2} e \right)
\label{eq:unfoldRMT}
\end{equation}
(which would assume an spectrum with ranging roughly from $\pm 1$), the
following routine gives an unfolded spectrum, ranging from 0 to $N$.
\begin{verbatim}
  function approxcumulative(Energy,size_spectrum)
    real(kind(1d0))                :: approxcumulative
    real(kind(1d0)), intent(in)    :: energy
    integer, intent(in)            :: size_spectrum
    real(kind(1d0))                :: rescaled_energy
    rescaled_energy=Energy*sqrt(pi)/size_spectrum
    if (rescaled_energy <= -1D0) approxcumulative=0d0
    if (rescaled_energy >= 1D0) approxcumulative=pi
    if ((-1d0<rescaled_energy).and.(rescaled_energy<1d0)) &
         approxcumulative=(asin(rescaled_energy)+ &
	    rescaled_energy*sqrt(1-rescaled_energy**2)+pi/2)
    approxcumulative=(approxcumulative-pi/2)*size_spectrum/pi
  end function approxcumulative
\end{verbatim}
We assume in this function that the spectrum ranges from $\pm N/\sqrt \pi$.
The unfolding procedure can also work with the same routine presented above
as the level density is again a semicircle.

\section{Quantum information}
During the numerical implementation of the ideas exposed in the thesis, 
we have to do operations that involve tensor product of states [as in
\eref{eq:initialonequbit}], like 
\begin{equation}
 |\psi \> = |\psi_A \> \otimes |\psi_B\>
\label{eq:tensorstate}
\end{equation}
with $|\psi_A \> \in \mcH_A$ and $|\psi_B \> \in \mcH_B$. We shall assume
that both $\mcH_A$ and $\mcH_B$ are spaces of one or more qubits. This
allows to take advantage of the bitwise operations available in all programing
languages.

In order to explain the implementation of tensor products its better to use
an example. We restrict our selfs to tensor products of qubit spaces.
Consider an space composed of $L$ qubits, numbered from 0 to $L-1$: $\mcH=
\mcH_{L-1} \otimes \cdots \otimes \mcH_0$. Now consider the two subspaces
$\mcH_A= \mcH_{a_{A-1}} \otimes \cdots  \otimes \mcH_{a_0}$ of dimension
$2^A$ and $\mcH_B= \mcH_{b_{B-1}} \otimes \cdots  \otimes \mcH_{b_0}$ of
dimension $2^B$ such that $\mcH= \mcH_A \otimes \mcH_B$. Assume that $L=7$,
$\mcH_A$ is the subspace of qubits 0, 3 and  5 and thus $\mcH_B$ is the
subspace of qubits 1, 2, 4, 6.  One can label unambiguously $\mcH_A$ with a
number $n_A = \sum_i 2^{a_i}$. In our case, $n_a=2^0+2^3+2^5=41$. Analogously
one can label $\mcH_B$ with $n_b=2^1+2^2+2^4+2^6=86$ (notice that
$n_a+n_b=2^L -1$ and thus are not independent). Using tensor notation, one
can associate an single index $\mu$ of $\mcH$ with two indices, one in
$\mcH_A$ and one in $\mcH_B$. If one uses the notation explained in Appendix
\ref{sec:implementationKI} one must combine the binary bits of index $\mu$
with the bits of the numbers representing each space (say $i_A$ in $\mcH_A$
and $i_B$ in  $\mcH_B$) . As an example assume that $\mu=55$. One obtains
$n_a=5$ and $n_b=7$, as shows the following table.
\begin{center}
\begin{tabular}{|c|ccccccc|}\hline
qubit number &   6&5&4&3&2&1&0 \\ \hline 
$\mu=55$     &   0&1&1&0&1&1&1 \\ \hline
$n_a=41$     &   0&1&0&1&0&0&1 \\
$i_A= 5$     &    &1& &0& & &1 \\ \hline
$n_b=86$     &   1&0&1&0&1&1&0 \\
$i_B= 7$     &   0& &1& &1&1&  \\ \hline      
\end{tabular}
\end{center}
A routine implementing the procedure is
\begin{verbatim}
  subroutine exdig(number_in,L,n1out,n2out,nwhich)
    implicit none
    integer, intent(in)  :: L,nwhich,number_in
    integer, intent(out) :: n1out,n2out
    integer              :: j,numin, L1,L2
    n1out=0
    n2out=0
    numin=number_in
    L1=bits_on_one(nwhich)
    L2=L-L1
    do j=0,L-1
       if (btest(nwhich,j)) then
          if (btest(numin,0)) n1out=ibset(n1out,0)
          n1out=ishftc(n1out,-1,L1)
       else
          if (btest(numin,0)) n2out=ibset(n2out,0)
          n2out=ishftc(n2out,-1,L2)
       endif
       numin=ishftc(numin,-1,L)
    end do
    return
  end subroutine exdig
\end{verbatim}
Here $\mathtt{number\_in}=\mu$, $\mathtt{L}=L$, $\mathtt{n1out}=i_A$ ,
$\mathtt{n2out}=i_B$, and $\mathtt{nwhich}=n_a$. 
A routine to calculate tensor products of real matrices, that takes advantage
of the previous routine is the following.
\begin{verbatim}
  subroutine TensorProduct_real_routine(H_a,H_b,nwhich,H_out)
    integer, intent(in)            :: nwhich
    real(kind(1d0)), intent(in) :: H_a(0:,0:),H_b(0:,0:)
    real(kind(1d0)), intent(out):: H_out(0:,0:)
    integer                        :: total_size, j1, j2, &
                        j1a, j1b, j2a, j2b, qubits
    total_size=size(H_a,1)*size(H_b,1)
    qubits=IntegerLogBase2(total_size)
    do j1=0,total_size-1
       do j2=0,total_size-1
          call exdig(j1,qubits,j1a,j1b,nwhich)
          call exdig(j2,qubits,j2a,j2b,nwhich)
          H_out(j1,j2)=H_a(j1a,j2a)*H_b(j1b,j2b)
       enddo
    enddo
\end{verbatim}
The function {\tt IntegerLogBase2} calculates the integer logarithm in base
2, of its argument (it must be a power of 2).  Again $\mathtt{nwhich}=n_a$,
$\mathtt{H\_A}$ ($\mathtt{H\_B}$) is the operator acting in  $\mcH_A$
($\mcH_A$).

The last routine that we wish to document is one for performing partial traces. 
This routine calculates the partial trace over $\mcH_B$ of a pure state
$\rho=\tr_B |\psi\> \< \psi |$. The
input parameters are the state $|\psi\>$ ({\tt statin}) and $n_a$ ({\tt nwhich}).
The output parameter {\tt rho} is $\rho$. 
\begin{verbatim}
  subroutine safe_PartialTrace(statin,rho,nwhich)
    implicit none
    complex(kind(1d0))          ::  statin(:)
    complex(kind(1d0))          ::  rho(0:,0:)
    integer, intent(in)         ::  nwhich
    integer                     ::  &
                        err(3), n_final,j1,j2,qubits
    complex(kind(1d0)), pointer ::  &
                         staout(:,:), st1(:), st2(:)
    qubits=IntegerLogBase2(size(statin))
    n_final=bits_on_one(nwhich)
    allocate (staout(0:2**(qubits-n_final)-1,0:2**n_final-1),&
          stat=err(2))
    allocate (st1(0:2**(qubits-n_final)-1),&
          st2(0:2**(qubits-n_final)-1), stat=err(3))
    if (any(err /= 0)) then
       print*,"error en PartialTrace"
       stop
    endif
    call unmsta(statin,staout,nwhich,qubits)
    do j1=0,2**n_final-1
       do j2=0,2**n_final-1
          rho(j1,j2)=dot_product(staout(:,j2),staout(:,j1))
       enddo
    enddo
    deallocate(staout, st1, st2, stat=err(1))
    return
    contains
      subroutine unmsta(statin,staout,nwhich,qubits)
        implicit none
        integer, intent(in)             :: nwhich,qubits
        complex(kind(1d0)), intent(in)  :: statin(:)
        complex(kind(1d0)), intent(out) :: staout(:,:)
        integer                         :: j, ncol, nrow
        staout=0d0
        do j=0,size( statin)-1
          call exdig(j,qubits,ncol,nrow,nwhich)
          staout(nrow+1,ncol+1)=statin(j+1)
        enddo
        return
      end subroutine unmsta
  end subroutine safe_PartialTrace
\end{verbatim}
The function {\tt bits\_on\_one} counts the number of bits of its argument,
set to $1$. The subroutine {\tt unmsta} reorganizes the input state as a
matrix, to be able to express the partial trace as a dot product of the
vectors composing this matrix.  All the programs, and routines are available
upon explicit request to the author.

\chapter{Two minor technicalities}
\section{The Form Factor} \label{sec:FormFactor}

We define the spectral form factor as
\begin{equation}
  K(t)=\frac{1}{N}\sum_{i=1}^N e^{\imath t E_i}.
\end{equation}
Some authors  rescale the  energy (or,  equivalently the  time). The factor in
front is used to obtain  an asymptotic average value of one  for most spectra.
This quantity  has been studied extensively in the context of quantum
chaos~\cite{guhr98random, JPA-volRMT}

We want to evaluate a double integral of the two-level form factor for the GUE
case:
\begin{equation} \label{eq:doubleb}
  B_2^{\beta=2}(t)
    :=\int_0^t \d \tau \int_0^\tau \d \tau' 
        b_2^{\beta=2}\left(\frac{\tau'-\tau}{\tau_H}\right)
     =\int_0^t \d \tau \int_0^\tau \d \tau' 
        b_2^2\left(\frac{\tau'}{\tau_H}\right)
\end{equation}
Let  us suppose  first that  $0<t<\tau_H$, and thus replace $b_2^2$  by a
linear expression:
\begin{align}
  \label{eq:doublebsec}
  B_2^2(t)&=\int_0^t \d \tau \int_0^\tau \d \tau' 
             b_2^{\beta=2}\left(\frac{\tau'-\tau}{\tau_H}\right)\\
        &=\int_0^t \d \tau \int_0^\tau \d \tau' 
             \left(1-\frac{\tau'}{\tau_H}\right)\\
        &=\frac{t^2}{2}-\frac{t^3}{6\tau_H}.
\end{align}
If $t>\tau_H$ the expression must be strictly linear since 
$b_2^2(t \geq \tau_H)=0$, and with slope given by 
$\int_0^{\tau_H}b_2^2(t/\tau_H)=\tau_H/2$. Furthermore, since 
$B_2^2(t)$ must be continuous, we also have the $y$ intersect of the 
linear function. This leave us with the result
\begin{equation}
  \label{eq:finalBtwotwo}
  B_2^2(t)=\begin{cases} 
             \frac{t^2}{2}-\frac{t^3}{6\tau_H} &
                   \text{if $0\leq t<\tau_H$}\\[.2cm]
             \frac{t\tau_H}{2}-\frac{\tau_H^2}{6}&\text{if $t\geq\tau_H$}
           \end{cases}.
\end{equation}

\section{A small proof of the Born expansion}
\label{sec:bornproof}

Consider the Hamiltonian $H_\lambda=H_0+\lambda W$.  The
definition of the state ket in the interaction picture is~\cite{sakurai}
\begin{equation}
  \label{eq:defstateinteract}
  |\psi(t)\>_I = U_0^\dagger (t) U_\lambda (t)|\psi(0)\>
\end{equation}
where $U_\lambda (t)=\exp(\imath t  H_\lambda)$. Consider the equation of
motion of the ket in the interaction picture~\cite{sakurai}:
\begin{equation}
  \label{eq:motioninteract}
  \imath \frac{\d |\psi(t)\>_I}{\d t}= \lambda \tilde{W} |\psi(t)\>_I
\end{equation}
with $\tilde{W}(t)=U_0^\dagger(t) W U_0(t)$.
Formal integration leads to 
\begin{equation}
  \label{eq:formalintegrationbornone}
  |\psi(t)\>_I=|\psi(0)\>-\imath \lambda \int_0^t \d \tau \tilde{W}(\tau)|\psi(\tau)\>_I.
\end{equation}
Solving the integral by iteration we obtain
\begin{multline}
  \label{eq:formalintegrationborn}
  |\psi(t)\>_I=\\ \left( \openone + (-\imath \lambda) \int_0^t \d \tau \tilde{W}(\tau)
    +(-\imath   \lambda)^2    \int_0^t   \d   \tau    \int_0^\tau   \d   \tau'    \tilde{W}(\tau')
    + \ldots \right) |\psi(0)\>. 
\end{multline}
Comparing \eqref{eq:defstateinteract}  and \eqref{eq:formalintegrationborn} we
obtain the Born expansion:
\begin{multline}
  \label{eq:bornexpansionAPP}
  U_0^\dagger (t) U_\lambda (t)= 
   \openone + (-\imath \lambda) \int_0^t \d \tau \tilde{W}(\tau)\\
    +(-\imath   \lambda)^2    \int_0^t   \d   \tau    \int_0^\tau   \d   \tau'    \tilde{W}(\tau)\tilde{W}(\tau')
    + \cdots. 
\end{multline}

\addcontentsline{toc}{chapter}{Bibliography}
\bibliographystyle{halpha}
\bibliography{paperdef,miblibliografia}
\end{document}